\documentclass[amsmath,amssymb,aps,floats,amsfonts,notitlepage,superscriptaddress,eqsecnum,nofootinbib,prd,a4paper,longbibliography]{revtex4-1}
\allowdisplaybreaks

\usepackage[utf8]{inputenc}
\usepackage[]{graphicx}
\usepackage{subcaption}
\usepackage{hyperref}
\usepackage{url}
\usepackage{color}
\usepackage{soul}
\usepackage[usenames,dvipsnames,svgnames,table]{xcolor}
\usepackage{multirow}

\newcommand{\MC}[1]{}

\newcommand{\Kill}{{\bf k}}

\newcommand{\tS}{\bar{t}}
\newcommand{\rS}{\bar{r}}
\newcommand{\phiS}{\bar{\theta}}

\newcommand{\rt}{r}
\newcommand{\tht}{\theta}
\newcommand{\tit}{t}

\newcommand{\be}{\begin{equation}}
\newcommand{\ee}{\end{equation}}
\newcommand{\bea}{\begin{eqnarray}}
\newcommand{\eea}{\end{eqnarray}}

\newcommand{\rset}[1]{\langle #1\rangle}
\newcommand{\rsetall}[2]{\langle {#2}| {#1}|{#2}\rangle_{ren}}

\newcommand{\s}{\sigma}

\newcommand{\dn}{d_n}
\newcommand{\Sp}{\sideset{}{'}\sum}

\newcommand{\Is}{I}

\newcommand{\rb}{\bar r}
\newcommand{\ob}{\bar \omega}
\newcommand{\mb}{\bar m}

\newcommand{\ind}{m\omega}

\begin{document}

\author{Marc Casals}
\email{mcasals@cbpf.br, marc.casals@ucd.ie}
\affiliation{Centro Brasileiro de Pesquisas F\'isicas (CBPF), Rio de Janeiro,  CEP 22290-180,  Brazil.}
\affiliation{School of Mathematics and Statistics, University College Dublin, Belfield, Dublin 4, Ireland.}

\author{Alessandro Fabbri}
\email{afabbri@ific.uv.es}
\affiliation{Departamento de F\'isica Te\'orica and IFIC, Universidad de Valencia-CSIC, C. Dr. Moliner 50, 46100 Burjassot, Spain.}

\author{Cristi\'an Mart\'{\i}nez}
\email{martinez@cecs.cl}
\author{ Jorge Zanelli}
\email{z@cecs.cl}
\affiliation{Centro de Estudios Cient\'{\i}ficos (CECs), Av. Arturo Prat 514, Valdivia 5110466, Chile.}

\title{Quantum-corrected  
black holes and naked singularities
in ($2+1$)-dimensions
}

\begin{abstract}
We analytically investigate the pertubative effects of a  quantum conformally-coupled scalar field on rotating (2+1)-dimensional black holes and naked singularities. In both cases we obtain the quantum-backreacted metric analytically. In the black hole case, we explore the quantum corrections on different regions of  relevance for a rotating black hole geometry. We find that the quantum effects lead to a growth of  both the event horizon and the ergosphere, as well as to a reduction of the angular velocity compared to their corresponding unperturbed values. Quantum corrections also give rise to the formation of a curvature singularity at the Cauchy horizon and show no  evidence of the appearance of a superradiant instability. In the naked singularity case, quantum effects lead to the formation of a horizon that hides the conical defect, thus turning it into a black hole. The fact that these effects occur not only for static but also for spinning geometries makes a strong case for the r\^ole of quantum mechanics as a cosmic censor in Nature.

\end{abstract}

\maketitle

\newpage
\tableofcontents
\newpage


\section{Introduction}

The quantum regime of gravitation has been one of the outstanding conundrums of theoretical physics for almost a century. Even the perturbative semiclassical framework, where the matter fields are quantized while the quantum nature of a background geometry is ignored, is a difficult problem, both technically and conceptually. Yet, important results have been shown within the semiclassical framework. For example, in the presence of a black hole (BH), it has been shown that quantum effects  give rise to Hawking radiation \cite{HawkingRad}. Such a semiclassical framework is possibly a good approximation for astronomical BHs, but probably too crude for a microscopic BH near the end of the evaporation process.

In this paper we  focus in particular on a different question of interest within the semiclassical framework: the fate of  timelike singularities as solutions of the classical Einstein field equations when quantum matter effects are taken into account. Timelike space-time singularities appear in various settings. For example, rotating BHs possess a hypersurface, called the Cauchy horizon,  inside the event horizon, beyond which there is a timelike singularity. Such a singularity, while not visible to observers outside the black hole, may be visible to observers that fall inside the BH. This can be seen in Fig.\ref{fig:Penrose}(b), where $r=0$, $r_-$ and $r_+$ are the radii of, respectively, the singularity, Cauchy horizon and event horizon. Non-rotating but electrically-charged black hole solutions also possess a Cauchy horizon with a timelike singularity lying beyond it. Another example is that of space-time solutions (rotating or not) possessing timelike singularities but no event horizon; such `naked' singularities (NSs) would thus be visible even to far-away observers. 

The presence of a generic (timelike) singularity is an undesirable feature from a physical point of view, since it signifies the breakdown of predictability: Cauchy data on an initial hypersurface does not have a unique evolution; heuristically: we do not know what may `come out' of such a singularity. Therefore, Penrose  formulated a Cosmic Censorship Hypothesis (CCH)\cite{penrose1999question}. The weak version of CCH~\cite{penrose2002golden,Wald} essentially states that if a singularity forms from the gravitational collapse of matter, then it will be surrounded by an event horizon -- thus, it will not be visible to far-away observers. In its turn, the strong version of CCH~\cite{penrose1979singularities} essentially states that if a singularity forms from the gravitational collapse of matter, then it will generically be  spacelike or null (not timelike) -- thus, 
 the singularity will not be visible to any observers at all 
(although they may crash into it!).

Given that there exist exact space-time solutions of the classical Einstein equations which contain timelike singularities, it is  important to investigate whether they {\it generically} form under gravitational collapse. Investigating whether singularities are stable under field perturbations  will help ascertain whether they are generic singularities or not. 

In $(3+1)$-dimensions, it has been shown that classical field perturbations lead to a curvature (non-timelike) singularity at the Cauchy horizon in the case of spherically-symmetric and electrically-charged (Reissner-Nordstr\"om) BHs, with~\cite{Costa:2017tjc,Cardoso:2017soq} or without~\cite{poisson1989inner,Ori:1991zz,dafermos2005interior,luk2017strong} a positive cosmological constant, as well as in the case of rotating (Kerr) BHs~\cite{ori1992structure,DafermosLuk2017}. These results are in support of strong CCH.\footnote{There are different versions of strong CCH. These results are in support of some version or other of strong CCH: they show varying degrees of ``irregularity" of the field perturbation on the Cauchy horizon depending on the specific physical setting, while the $C^0$ character is preserved in all settings studied.}
 In space-times with a number of dimensions other than four, on the other hand, violation of strong CCH has been found in, e.g.,~\cite{lehner2010black,Santos:2015iua}, as due to the Gregory-Laflamme instability~\cite{Gregory:1993vy}. 

As for weak CCH, recent work~\cite{sorce2017gedanken} has shown that a Kerr BH or a Kerr-Newman (i.e., electrically-charged Kerr)  BH cannot be turned into a NS by throwing matter into it, as long as its stress-energy tensor satisfies the null energy condition. However, in the specific case of ($3+1$)-D anti de Sitter (AdS) space-time (i.e., a Universe with a negative cosmological constant), Ref.~\cite{Crisford:2017zpi} has shown that weak CCH may be violated.

The above examples deal with the {\it classical} stability of space-times possessing timelike singularities. It is also important to investigate their stability properties  under {\it quantum} field perturbations. This can be achieved via the semiclassical Einstein equations, in  which 
the classical stress energy tensor  
is supplemented with the renormalized expectation value of the quantum stress-energy tensor (RSET) calculated on a fixed, classical background space-time.

In the quantum case, the results for timelike singularities in ($3+1$)-dimensions are very scarce. One of the very few results is the argument in~\cite{hiscock1980quantum,birrell1978falling,Ottewill:2000qh} that the RSET calculated on Reissner-Nordstr\"om or Kerr(-Newman) background space-time diverges on  
(at least a part of) the CH; there is also the recent~\cite{Lanir:2018vgb}, which contains an exact calculation of the renormalized expectation value of the square of the field on the Cauchy horizon of Reissner-Nordstr\"om and is found to be regular there, while the trace of the RSET diverges. We note, however, that the RSET was not obtained explicitly  in these works and, therefore,  the space-time resulting from the quantum perturbations of the  Reissner-Nordstr\"om or Kerr(-Newman) background could not be obtained. In order to understand the full structure of the backreacted space-time, resulting from quantum field perturbations, one should solve the semiclassical Einstein equations. To the best of our knowledge, this  has not been achieved exactly\footnote{See~\cite{york1985black}, where an {\it approximation} for the RSET was used in (3+1)-D Schwarzschild space-time.} 
for any $(3+1)$-D BH space-time. 
There already exist some works in the literature where  the quantum-backreacted metric has been obtained
in $(1+1)$-dimensions (see for example \cite{libro} and references therein) as well as in  $(2+1)$-dimensions. We next review quantum-backreaction results on a specific $(2+1)$-D case: the so-called Ba\~nados-Teitelboim-Zanelli (BTZ) geometries, which include both BHs~\cite{BTZ,BHTZ} and NSs~\cite{MiZ}.

Semiclassical backreaction on  {\it static} BTZ space-times has been studied in the following works. Refs.~\cite{Lifschytz:1993eb,MZ2} showed that the horizon of a static BTZ BH is ``pushed out" due to backreaction and that a curvature singularity forms at the centre of the BH (although this region where the curvature singularity forms is in principle beyond the regime of validity of the semiclassical approximation). Also in the case of a static BTZ BH,~\cite{shiraishi1994vacuum} found that the contribution of the backreaction to the gravitational force on a static particle may be positive or negative depending on the radius.

These works are for the case that the background space-time is that of a static BTZ BH, which does not possess a timelike singularity. In the case of a static (timelike) BTZ NS,  we showed in~\cite{CFMZ1} that backreaction  creates an event horizon and forms a curvature singularity at its centre (although, again, this region inside the BH in principle lies beyond the regime of validity of the semiclassical approximation).

In the important case of nonzero rotation, to the best of our knowledge, the only work up until recently which aimed at investigating quantum-backreaction was that of Steif in~\cite{Steif}.  Steif found that, in the case of a rotating BTZ BH,  the RSET diverges as the inner horizon is approached from its inside. In the Letter~\cite{CFMZ2} we went further and we presented results for the backreacted metric, both in the case of a rotating BTZ BH and a rotating BTZ NS. In this paper we provide the full details of the calculation presented in~\cite{CFMZ2}. 
We analytically obtain the quantum-backreacted metric everywhere for these  two  background space-times. This  enables us to thoroughly study the effect of quantum corrections on rotating geometries describing both BHs and naked conical singularities in 2+1 dimensions. In particular, we study the quantum stability of such space-times in relation to CCH. We  also investigate the effects of quantum backreaction on other interesting regions of the space-times. For example, in the case of the rotating BH space-time, we  determine the quantum backreaction on  the event horizon and on the ergosphere (region outside the rotating horizon where observers cannot remain static). Our results show that, in the BH case, the event horizon is pushed out (as in the static case) and the inner horizon develops a {\it curvature} singularity.
This singularity in the backreacted spacetime may be spacelike or timelike, depending on the values of the mass and angular momentum of the black hole; when it is spacelike, strong CCH is enforced.
In the NS case, we find that an event horizon forms and shields the singularity, which becomes a spacelike curvature singularity (as in the static case of~\cite{CFMZ1}).
Quantum effects on the NS thus act to enforce strong CCH.

There is an
issue worth mentioning regarding our space-time setting and evolution of initial data. 
Our BTZ geometries are asymptotically AdS. Therefore, they are not globally-hyperbolic and the Cauchy value problem is, in principle, not well-posed. It is known, however, that this issue may be resolved by imposing specific boundary conditions for the matter field on the AdS boundary~\cite{Avis1978quantum} -- see $r=\infty$ in Fig.\ref{fig:Penrose}.
We
specifically
impose the so-called transparent boundary conditions~\cite{Avis1978quantum} on the AdS boundary.
Furthermore,
we are dealing with regions of space-time which possess a timelike singularity. This is true, of course, for the NS case, but also for the region inside the Cauchy horizon of the rotating BH case (which is the region that we need to deal with in order to find the instability of the Cauchy horizon).
Similarly to the AdS boundary, 
the field effectively satisfies some specific boundary conditions 
on the timelike singularity, so that unique evolution of initial data is restored.

Another point worth mentioning is that the singularity on the Cauchy horizon that we find
appears in the limit as we approach the Cauchy horizon from its {\it inside}. 
However, as opposed to Kerr, in the rotating BTZ geometry there exist no closed timelike curves. Therefore, we are not  faced with the issues that such curves cause in relation to the initial value problem 
in the region inside the Cauchy horizon in Kerr.

An important point of our results is that they show that the quantum effects on black holes  and naked singularities found in the static case \cite{Lifschytz:1993eb,MZ2,CFMZ1} are rather generic. They do not require the geometry to be static, but they are also present in many of the spinning cases.

Finally, we note that, since three-dimensional gravity has no local dynamical degrees of freedom, the quantum effects can only be due to the quantized matter source, which in our case is provided by a (conformally coupled\footnote{The choice of conformal coupling is motivated by simplicity: because AdS space-time is conformal to Minkowski space-time,  the quantum propagator in AdS is then obtained directly from its expression in flat space-time.}) scalar field. As mentioned, the quantum fluctuations of the scalar field vacuum on a fixed background geometry give rise to a RSET which is of $O(\hbar)$ and acts as a source of Einstein's equations. These corrected equations give rise to a one-loop correction on the geometry (backreaction). In principle, one could go on to compute the second order correction to the RSET by recalculating it, this time, on the backreacted geometry. However, those would in principle be corrections of $O(\hbar^2)$ and we choose not to continue in this direction.

The paper is organized as follows.
In Sec.\ref{sec:review} we review the classical rotating BTZ geometries, both for black holes and for naked singularities. In that section we also review an exact black hole solution of the Einstein equations with a source given by a particular classical scalar field configuration.
In Sec.\ref{sec:QFT} we consider a quantum scalar field on a rotating BTZ geometry and calculate the two-point function and the RSET.
We analytically solve the semiclassical Einstein equations in Sec.\ref{sec:semicl sln}.
We analyse in depth the physical features of these quantum-backreacted geometries in Sec.\ref{sec:analysis}.
We finish the main body of the paper with a discussion in Sec.\ref{sec:discussion}, where we summarize our results and point to open questions.
After the main body  there are three appendixes:
in App.\ref{sec:BTZ as ids} we present the background BTZ geometries as the result of identifying points in the embedding space $\mathbb{R}^{2,2}$; in App.\ref{sec:G_A}
we review the two-point function in (the covering space of) $AdS_3$; in the last appendix, \ref{sec:GF in BTZ modesum},
we (re-)derive the two-point function in a static naked singularity space-time via the alternative method of mode sums.

We use units such that the cosmological constant is $\Lambda = -\ell^{-2}$
and the Planck length is $l_P=\hbar \kappa/(8\pi)$, where $\ell$ is the radius of curvature and $\kappa$ is the ($2+1$)-dimensional gravitational constant. We choose metric signature $(-++)$.

\section{Review of BTZ geometries: black holes and conical singularities}\label{sec:review}  

Three-dimensional BTZ BH and NS space-times are exact solutions of the vacuum Einstein field equations with a negative cosmological constant ``$-\ell^{-2}$", described by the line element
\begin{equation} \label{BTZ}
ds^2= \left(M-\frac{r^2}{\ell^2}\right)dt^2 -Jdt d\theta + \left(\frac{r^2}{\ell^2}-M+\frac{J^2}{4r^2}\right)^{-1}dr^2+r^2 d\theta^2,
\end{equation}
where $-\infty < t < +\infty$, $0 < r < \infty$, $0 \leq \theta < 2\pi$ (periodic). The constants $M$ and $J$ are, respectively, the  mass\footnote{The Hamiltonian mass and angular momentum of the BTZ space-time are, in fact, $M\pi/\kappa$ and $J\pi/\kappa$, respectively, but we shall just refer to $M$ and $J$ as the mass and angular momentum.} and angular momentum of these space-times.
In this section we review  in some detail these classical solutions. For further details, we refer the reader to the original papers \cite{BTZ,BHTZ} in the BH case, and \cite{MiZ} in the NS case.

\vspace{0.5cm}
\textbf{1. Black hole} \\

The metric \eqref{BTZ} describes a spinning black hole provided $M \ell \ge |J| > 0$. In this case,
the space-time possesses a Cauchy horizon at $r=r_-> 0$
and an event horizon at $r=r_+\ge r_-$,
where
\begin{equation} \label{alfa+-}
r_{\pm}\equiv \frac{\ell |\alpha_{\pm}|}{2},\quad \alpha_{\pm} \equiv \sqrt{M+\frac{J}{\ell}}\pm\sqrt{M-\frac{J}{\ell}}.
\end{equation}
Note that
\begin{equation} \label{M and J BH}
 M=\frac{\alpha_{+}^2+\alpha_{-}^2}{4} >0  \quad \mbox{and} \quad J=\frac{\ell\alpha_{+} \alpha_{-}}{2},   
\end{equation}
with $\alpha_{+}>0 $, $\alpha_{+}\geq\alpha_{-}$, and  $\alpha_{+}^2 - \alpha_{-}^2=4\sqrt{M^2-J^2/\ell^2}$. The static BH is obtained for $J=0$, where  $\alpha_{+}=2\sqrt{M}>0$, $\alpha_{-}=0$, and there is no Cauchy horizon.

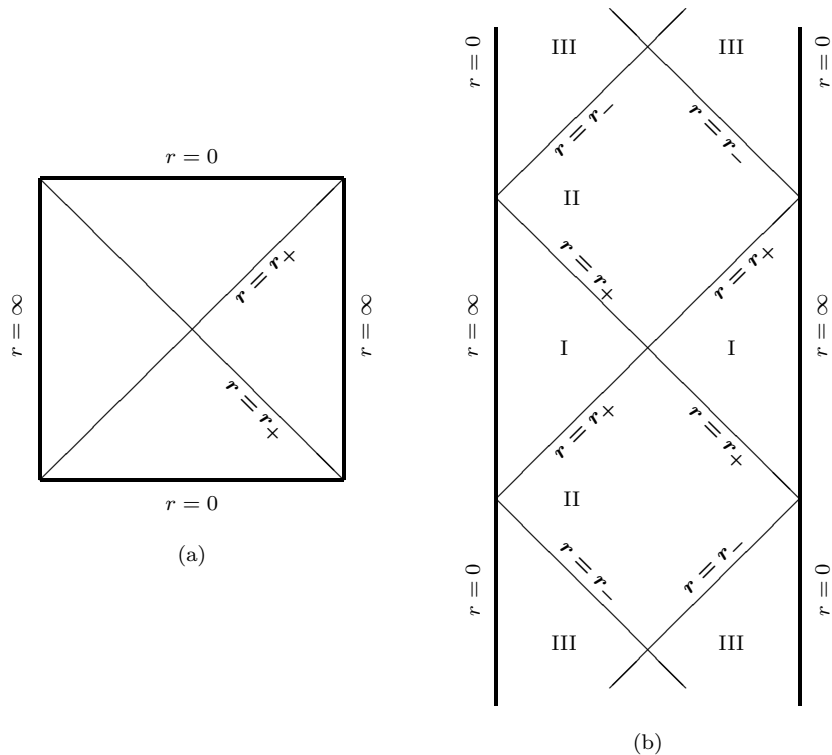
\begin{figure}[h!]
\begin{center}
\setlength{\unitlength}{1mm}
\begin{picture}(100,100)
\footnotesize
\put(0,40){\linethickness{0.4mm}\line(0,1){40}}
\put(40,40){\linethickness{0.4mm}\line(0,1){40}}
\put(0,40){\linethickness{0.4mm}\line(1,0){40}}
\put(0,80){\linethickness{0.4mm}\line(1,0){40}}
\put(-3,60,5){\makebox(0,0){\rotatebox{90}{$r=\infty$}}}
\put(43,60,5){\makebox(0,0){\rotatebox{90}{$r=\infty$}}}
\put(20,37){\makebox(0,0){\rotatebox{0}{$r=0$}}}
\put(20,83){\makebox(0,0){\rotatebox{0}{$r=0$}}}
\put(0,40){\linethickness{0.6mm}\line(1,1){40}}
\put(0,80){\linethickness{0.6mm}\line(1,-1){40}}
\put(30,67){\makebox(0,0){\rotatebox{45}{$\boldsymbol{r=r_+}$}}}
\put(28,49){\makebox(0,0){\rotatebox{-45}{$\boldsymbol{r=r_+}$}}}
\put(20,30){\makebox(0,0){\rotatebox{0}{(a)}}}
\put(60,10){\linethickness{0.4mm}\line(0,1){90}}
\put(100,10){\linethickness{0.4mm}\line(0,1){90}}
\put(60,37.5){\linethickness{0.6mm}\line(1,1){40}}
\put(60,77.5){\linethickness{0.6mm}\line(1,1){25}}
\put(100,37.5){\linethickness{0.6mm}\line(-1,-1){25}}
\put(60,37.5){\linethickness{0.6mm}\line(1,-1){25}}
\put(60,77.5){\linethickness{0.6mm}\line(1,-1){40}}
\put(100,77.5){\linethickness{0.6mm}\line(-1,1){25}}
\put(57,25,5){\makebox(0,0){\rotatebox{90}{$r=0$}}}
\put(57,60,5){\makebox(0,0){\rotatebox{90}{$r=\infty$}}}
\put(57,95,5){\makebox(0,0){\rotatebox{90}{$r=0$}}}
\put(103,25,5){\makebox(0,0){\rotatebox{90}{$r=0$}}}
\put(103,60,5){\makebox(0,0){\rotatebox{90}{$r=\infty$}}}
\put(103,95,5){\makebox(0,0){\rotatebox{90}{$r=0$}}}
\put(93,67,5){\makebox(0,0){\rotatebox{45}{$\boldsymbol{r=r_+}$}}}
\put(89,28,5){\makebox(0,0){\rotatebox{45}{$\boldsymbol{r=r_-}$}}}
\put(89,45,5){\makebox(0,0){\rotatebox{-45}{$\boldsymbol{r=r_+}$}}}
\put(89,85,5){\makebox(0,0){\rotatebox{-45}{$\boldsymbol{r=r_-}$}}}
\put(72,46,5){\makebox(0,0){\rotatebox{45}{$\boldsymbol{r=r_{+}}$}}}
\put(72,86,5){\makebox(0,0){\rotatebox{45}{$\boldsymbol{r=r_-}$}}}
\put(72,67,5){\makebox(0,0){\rotatebox{-45}{$\boldsymbol{r=r_+}$}}}
\put(72,27,5){\makebox(0,0){\rotatebox{-45}{$\boldsymbol{r=r_-}$}}}
\put(69,18,5){\makebox(0,0){III}}
\put(91,18,5){\makebox(0,0){III}}
\put(69,57,5){\makebox(0,0){I}}
\put(91,57,5){\makebox(0,0){I}}
\put(70,37,5){\makebox(0,0){II}}
\put(70,77,5){\makebox(0,0){II}}
\put(69,97,5){\makebox(0,0){III}}
\put(91,97,5){\makebox(0,0){III}}
\put(80,5){\makebox(0,0){\rotatebox{0}{(b)}}}
\end{picture}
\end{center}
\caption{Penrose diagrams for BTZ black holes: static black hole in panel (a) and rotating, non-extremal black hole in panel (b).} 
\label{fig:Penrose}
\end{figure}

The coordinates in Eq.\eqref{BTZ} do not cover the maximal analytical extension of the rotating BTZ BH space-time. The maximal analytical  extension  is represented  in Fig.\ref{fig:Penrose} by means of a Carter-Penrose diagram.

Clearly, the extremal (i.e., maximally-rotating) BH corresponds to $M\ell = |J|$. See Eq.\eqref{BTZline} for an expression of  the sub-extremal line-element \eqref{BTZ} in terms of $\alpha_{\pm}$
and
Eq.\eqref{mext} for the line-element for the extremal BTZ BH. 

The inner horizon is classically unstable \cite{Chan:1994rs,husain1994radiation}  in a similar manner to that of Kerr or Reissner-Nordstr\"om space-times \cite{poisson1989inner,brady1995nonlinear,ori1992structure}. 
Unlike the ($3+1$)-D Kerr geometry, however, the ($2+1$)-D BH possesses no curvature singularities -- instead, it  possesses a {\it causal} singularity
at $r=0$\footnote{In a slight abuse of language, we refer to  $r=0$ although, this singularity is, strictly speaking,  not a point of the space-time.}: there exist inextendible incomplete geodesics that hit $r=0$~\cite{BTZ,BHTZ}. 
Like the singularity  in Kerr, the singularity 
of  the BTZ BH is timelike.
The past boundary of the causal future of the timelike singularity is the (future) Cauchy horizon. The name of ``Cauchy" given to this horizon is because the Cauchy  problem  the \footnote{The Cauchy  problem is the initial value problem when the field data is given on a certain constant-coordinate hypersurface.} is not well-posed to its future.
In Kerr, the situation is even worse since there exist closed timelike curves near its singularity~\cite{carrollbook}.
In the rotating BTZ space-time, on the other hand, there exist no closed timelike curves {\it by construction} of the space-time.

Conformal infinity  $\mathcal{I}$
for null geodesics
corresponds to the so-called
$AdS$ boundary at $r=\infty$. This boundary is a {\it timelike} hypersurface and so
the space-time is not globally hyperbolic.
Fig.\ref{fig:Penrose}  shows the causal structure that gives the
defining characters to the
event and Cauchy horizons,
as well as to the $AdS$ boundary.

The metric in Eq.\eqref{BTZ} is stationary and axially symmetric, with associated Killing vectors $\partial/\partial t$ and $\partial/\partial\theta$, respectively.
The Killing vector $\partial/\partial t$ is timelike for $r> r_{SL}\equiv\sqrt{M}\ell$, it is null at $r= r_{SL} $ and it is spacelike for $r_+ < r <r_{SL}$. This means that no static observers can lie in the region $r<r_{SL}$. The hypersurface $r=r_{SL}$ is called the static limit surface and the region $r \in (r_+, r_{SL})$ is called the ergosphere. The existence of an ergosphere allows for the Penrose process, whereby particles (only massless ones in the BTZ case) can extract rotational energy from the BH (see~\cite{penrose2002golden} in Kerr and~\cite{cruz1994geodesic} in rotating BTZ). The ergosphere also allows for the wave-equivalent of the Penrose process, the so-called phenomenon of superradiance, whereby boson field waves can extract rotational energy from the BH. For superradiance, see~\cite{starobinskii1973amplification,zel1971generation} in Kerr and~\cite{winstanley2001classical} in asymptotically-AdS Kerr. In BTZ, on the other hand, a massless scalar field obeying Dirichlet boundary conditions does not exhibit
superradiance~\cite{ortiz2012no}, although the specific case of a massive scalar field obeying certain Robin boundary conditions does exhibit superradiance~\cite{dappiaggi2018superradiance}.

In its turn, the Killing vector $\chi\equiv \partial/\partial t+ \Omega \partial/\partial\theta $, where $\Omega\equiv J/(2r_+^2)$ is the angular velocity of the event horizon, is the  generator of the event horizon. The vector $\chi$ is null at the event horizon and, in the nonextremal case, it is timelike for $r>r_+$. This means that, in the nonextremal case, timelike observers that rigidly rotate at the angular velocity of the BH can lie anywhere outside the event horizon, i.e., there is no speed-of-light surface as in Kerr. In the extremal case, on the other hand, the  Killing vector $\chi$ is null everywhere. 



\begin{figure}[h!]
\begin{center}
  \includegraphics[width=12cm]  {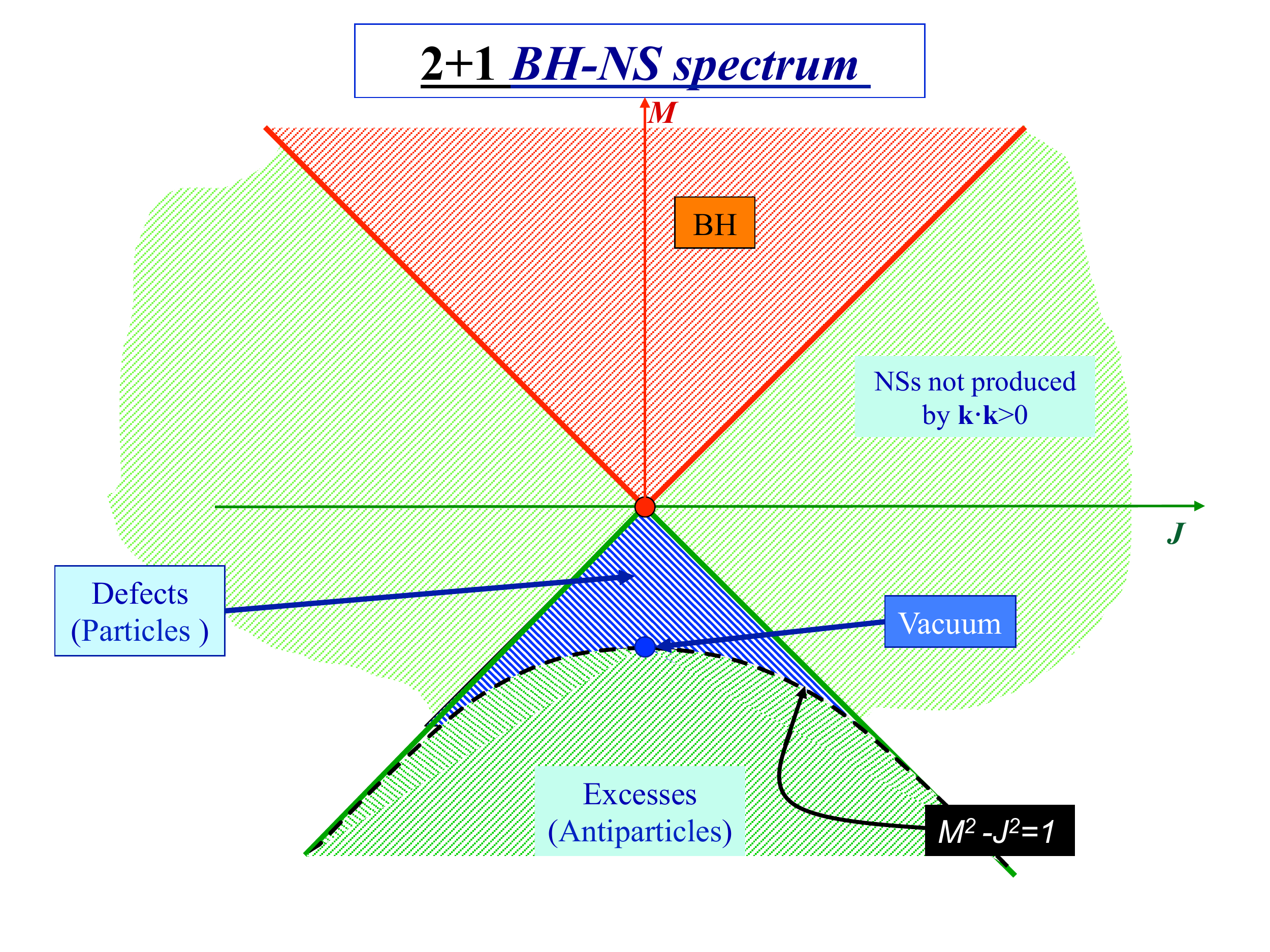}
  \end{center}
\caption{Schematic representation of BTZ black hole and naked singularity states for different values of $M$ and $J$. }
\label{fig:M-J} 
\end{figure}

Spinning BHs can also be obtained by boosting a static BH of a given mass $M_0$, yielding a new BH state of mass $M$ and angular momentum $J$, with 
\begin{equation}
M=\frac{M_0 (1+\omega^2)}{(1-\omega^2)}\, , \, \, J=\frac{2\omega M_0 \ell}{(1-\omega^2)}\, ,
\end{equation}
where $\omega$ is the boost parameter in the Lorentz transformation and it satisfies $|\omega| <  1$. In this way, all BH states with $M$ and $J$ lying on the hyperbola 
$M^2-J^2/\ell^2=const.$
on the $M$-$J$ plane -- see Fig.\ref{fig:M-J}  -- are connected by boosts \cite{MTZ}.\\

\textbf{2. Naked singularity}\\ 

If the mass in the BTZ metric \eqref{BTZ} is continued to negative values, the geometry then becomes a conical NS (there is a curvature singularity at $r=0$)~\cite{MiZ}, with the single exception of non-rotating AdS$_3$ space-time ($M=-1,J=0$). For  $-M \ell > |J|$, we define
\begin{equation} \label{betas}
\beta_{\pm}\equiv \sqrt{-M+J/\ell}\pm\sqrt{-M-J/\ell}\in\mathbb{R},
\end{equation}
so that
\begin{equation} \label{M and J NS}
 M=-\frac{\beta_{+}^2+\beta_{-}^2}{4}<0 \quad \mbox{and} \quad J=\frac{\ell\beta_{+} \beta_{-}}{2},   
\end{equation}
with $\beta_{+}\ge |\beta_{-}|\geq 0$ and  $\beta_{+}^2 - \beta_{-}^2=4\sqrt{M^2-J^2/\ell^2}$.

It follows from the line element \eqref{BTZ} (see Eq.\eqref{NSline} for the NS line element in terms of $\beta_{\pm}$) with mass $M<0$ that its metric components  are well-defined everywhere for $r\neq 0$, which means that there is no horizon and it therefore describes a NS. Its conformal structure at infinity is as in the BH case and so it also possesses a (timelike) $AdS$ boundary at $r=\infty$.

The spinless (i.e., $J=0$) states in the range $-1< M<0$ correspond to conical space-times with angular  defects $\Delta= 2 \pi(1-\sqrt{-M})$ (particles), while those with $M<-1$ are conical  excesses (antiparticles). The dividing case, $M = -1$, corresponds to AdS$_3$  vacuum space-time. The static conical singularities can also be boosted to obtain spinning (anti-)particles, in the same manner as for BHs. All of these states are described by the same BTZ metric, Eq.\eqref{BTZ}, with $M\ell \leq -|J|$. 

Like the BH metric, the NS metric is also stationary and axially symmetric, with the same associated Killing vectors 
$\partial/\partial t$ and $\partial/\partial\theta$, respectively. 
In this geometry, 
$\partial/\partial t$ is always timelike and there is no ergosphere. The extremal NS case corresponds to maximal rotation,  $M\ell=-|J|$,
and its metric is given in 
Eq.\eqref{Extreme-g}.

The spectrum of BHs and NSs can thus be summarized as follows:
\begin{eqnarray}
M>0 
\quad 
\& 
\quad
&
0 \leq M^2\ell^2 - J^2 <\infty		& \mbox{: Black holes}
\nonumber
\\
M<0
\quad 
\& 
\quad
&
0\le M^2\ell^2 - J^2 <1				& 	\mbox{: Particles}
\nonumber
\\
M<0
\quad
\&  
\quad
&
J\neq 0\ \& \
1=M^2\ell^2 - J^2 					
& \mbox{: Rotating $AdS_3$}
\nonumber
\\
M=-1
\quad
\&  
\quad
&
J=0 					& \mbox{: Non-rotating $AdS_3$ (vacuum)}
\nonumber
\\
M<0
\quad
\& 
\quad
&
1< M^2\ell^2 - J^2<\infty		& \mbox{: Antiparticles}
\nonumber
\end{eqnarray}
 This spectrum is represented schematically 
 in the $M$-$J$ plane
 in Fig.\ref{fig:M-J}. 
The case $M=J=0$ is known as the 
 ``zero-mass black hole" or the
 maximum-deficit conical singularity.

\vspace{0.5cm}
\textbf{3. Construction of the classical BTZ geometries} \\

In order to construct the BTZ geometries, we first  consider flat $\mathbb{R}^{(2,2)}$ with 
coordinates $X^0,X^1,X^2,X^3\in\mathbb{R}$  and
metric
\begin{equation}\label{eq:R22}
ds^2 = -\left(dX^0\right)^2+\left(dX^1\right)^2+\left(dX^2\right)^2-\left(dX^3\right)^2.
\end{equation}
We can then think of AdS$_3$ as the pseudosphere 
 \begin{equation}\label{eq:pseudo}
 -(X^0)^2+(X^1)^2+(X^2)^2- (X^3)^2=-\ell^2
 \end{equation}
  embedded in $\mathbb{R}^{(2,2)}$.
However, the 
  topology of AdS$_3$ is $S^1\text{(time)}\times \mathbb{R}^2\text{(space)}$
  and so the space-time contains closed timelike curves.
  The covering of AdS$_3$, denoted by CAdS$_3$, is obtained by ``unwrapping" the 
  $S^1$,
  so that the resulting space-time does not 
contain closed timelike curves.
We can now obtain BHs and NSs in ($2+1$)-dimensions as locally negative constant curvature 
geometries 
by identifying points in the covering
space CAdS$_3$,
which is represented by its embedding in flat $\mathbb{R}^{2,2}$. The identification is a quotient of CAdS$_3$ by a Killing vector $\Kill$ in the algebra $ so(2,2)$ of global isometries of the pseudosphere. 

In the \textbf{BH} case ($M \geq|J|\geq 0$), $\Kill$ is a Killing vector that acts transitively, that is, leaving no fixed points on CAdS$_3$. The region where the Killing vector is spacelike ($\Kill^2>0$) is identified as $r>0$ in the resulting manifold, while the region where $\Kill$ is timelike ($\Kill^2=r^2<0$) is removed in order to avoid traversable closed timelike curves \cite{BHTZ}. 
Thus $r=0$ is a causal singularity.
The specific form of $\Kill$ depends on the mass $M$ and angular momentum $J$ of the BH, with $M\geq |J|/\ell$.

In the \textbf{NS} case, the Killing vector for the identification is a spacelike rotation that keeps $r=0$ fixed. The manifold CAdS$_3/\Kill(M,J)$ then has a  conical NS at the fixed point of $\Kill$ (i.e., $r=0$), where the curvature has a Dirac-$\delta$ singularity. The corresponding identification is along the compact coordinate $\theta$.

These identifications can also be expressed as the action of a matrix
$H(\Kill)$ 
that maps every point in $\mathbb{R}^{2,2}$ to its image under $\Kill$, given in Table 1. The identification matrices in $\mathbb{R}^{2,2}$ corresponding to the different 
BHs and conical singularities are given explicitly in Appendix \ref{identifications}.

\begin{table}[h!] \label{Table1}
\caption{Identification vector $\Kill$ for the nonextremal ($M^2\ell^2 >J^2$) and extremal ($M^2\ell^2 =J^2$) BH and NS geometries in terms of the $ so(2,2)$ generators $J_{ab}$ (see Appendix~\ref{sec:BTZ as ids}).}
\begin{ruledtabular}
\begin{tabular}{lccl}
 & $\qquad M^2\ell^2 >J^2$ \qquad &   $M^2\ell^2 =J^2$  &  Type of Killing vector $\Kill$ \\[1mm]
\hline
$M>0$ & $\frac{1}{2} ( \alpha_+ J_{0 1}+\alpha_-J_{23})$	& $\alpha (J_{01}+J_{23})+\frac{1}{2} (J_{02}+J_{03}+J_{12}+J_{13})$ & Spacelike, no fixed points  \\[1mm]
 $M<0$& $\frac{1}{2}(\beta_+ J_{2 1}+\beta_- J_{30})$ &	$\beta(J_{03}-J_{12})-\frac{1}{2} (J_{01}+J_{03}+J_{12}-J_{23})$ & Spacelike, $r=0$ fixed point \\[0.5mm]
\end{tabular}
\end{ruledtabular}
\end{table}

\textbf{4. Black hole solutions with a scalar field}\\
\label{BHS}

We complete the discussion of the classical system by reviewing an exact solution of Einstein equations in the presence of a source given by a massless and conformally-coupled real scalar field $\phi$ \cite{MZ1}. The action in three space-time dimensions reads
\begin{equation} \label{action}
I= \int d^3 x \sqrt{-g} \left[ \frac{R+2 \ell^{-2}}{2\kappa}
-\frac{1}{2}g^{\mu \nu}\nabla_{\mu}\phi\nabla_{\nu}\phi
-\frac{1}{16}R
\,\phi^2 \right],
\end{equation}
which provides the following field equations:
\begin{equation} \label{Einsteineqs}
G_{\mu \nu}- \ell^{-2}g_{\mu \nu}= \kappa T_{\mu \nu}, \quad \Box \phi-\frac{1}{8}R \phi=0,
\end{equation}
where the stress-energy tensor is given by
\begin{equation} 
T_{\mu\nu}=\nabla_{\mu}\phi\nabla_{\nu}\phi-\frac{1}{2}g_{\mu\nu} 
 g^{\alpha \beta} \nabla_{\alpha}\phi\nabla_{\beta}\phi 
+\frac{1}{8}\left(g_{\mu \nu} \Box -
\nabla_{\mu}\nabla_{\nu} +G_{\mu \nu}\right)\phi^2. \label{ClassicalTuv}
\end{equation}
It is straightforward to check that this stress-energy tensor is conserved and traceless, which in turn implies that the
geometry has a constant Ricci scalar, 
\begin{equation} \label{R}
R=-6 \ell^{-2} .
\end{equation}

An exact  static, circularly-symmetric solution was found in \cite{MZ1}. Its line element is given by
\begin{equation} \label{ds}
ds^2= -f(r)dt^2+f^{-1}(r)dr^2+r^2 d\theta^2,
\end{equation}
where 
\begin{equation} \label{Fp}
f(r)\equiv \frac{1}{ \ell^2}\left(r^2-3C^2- \frac{2C^3}{r}\right)=
\frac{(r+C)^2(r-2C)}{ \ell^2 r},
\end{equation}
is the lapse function, $C$ is an arbitrary integration constant and the corresponding scalar field is given by
\begin{equation} \label{psie}
\phi(r)=\sqrt{\frac{8C}{\kappa (r+C)}}.
\end{equation}
This exact solution  describes a BH with an event horizon at $r_+=2C$ provided $C>0$. In that case, the event horizon surrounds a single curvature singularity at $r=0$,  as can be shown by calculating the Kretschmann scalar,
\begin{equation} \label{KR}
R^{\mu \nu \lambda \rho}R_{\mu \nu \lambda \rho}=\frac{
12(r^6+2C^6)}{ \ell^4 r^6}.
\end{equation}
For $C=0$ the solution reduces to the massless BTZ spacetime with a vanishing scalar scalar field. The on-shell stress-energy tensor is given by 
\begin{equation} \label{eq:SET scalar}
T^{\mu}{}_{\nu}=\frac{C^3}{\kappa   \ell^2 r^3}\textrm{diag}(1,1,-2),
\end{equation}
which is consistently traceless. It should be noted that, except for the constant factor $C^3/(\kappa \ell^2)$, the rest in the expression in \eqref{eq:SET scalar} coincides exactly with the renormalized stress-energy tensor (\ref{eq:RSET BH'}) to be presented in the next section.

\section{Quantum scalar field }\label{sec:QFT}

The semiclassical Einstein equations are obtained by replacing the classical stress-energy of the matter field(s) by the renormalized expectation value of the quantum stress-energy tensor operator (RSET). In the presence of a  cosmological constant  $\Lambda = -\ell^{-2}$, the semiclassical Einstein equations are
\begin{equation}\label{eq:EFE}
G_{\mu\nu}-\frac{g_{\mu\nu}}{\ell^2}= \kappa \rsetall{T_{\mu\nu}}{\psi},
\end{equation}
where $\rsetall{T_{\mu\nu}}{\psi}$ is the RSET for a quantum field in a state $|\psi\rangle$. For ease of notation, we henceforth drop the subindex `$ren$' as well as the symbol for the quantum state in the RSET, and we thus denote it by $\rset{T_{\mu\nu}}$.

\subsection{Two-point functions}\label{sec:two-pt func}

From now on we shall consider a massless, conformally-coupled scalar  field $\phi$ (conformal coupling in three dimensions corresponds to a coupling constant $\xi=1/8$~\cite{Birrell:Davies}).
In this case, the (Klein-Gordon) field equation is 
\begin{equation}\label{wave-eq} 
\left(\Box +\frac{3}{4\ell^2}\right)\phi(x) = 0.
\end{equation}
As opposed to Eq.\eqref{Einsteineqs}, the d'Alembertian $\Box=g^{\mu \nu}\nabla_{\mu}\nabla_{\nu}$ here is with respect to a {\it background} metric $g_{\mu\nu}$ (i.e., it is a solution of the classical {\it vacuum} Einstein equations) which, in our case, we shall take to be a BTZ geometry.

The RSET for the quantum scalar field $\phi$ in a state $|\psi\rangle$ is typically constructed from a geometric differential operator acting on the Hadamard elementary two-point function, which is the anti-commutator $G^{(1)}(x,x')=\langle\psi | \left\{\phi(x),\phi(x')\right\}|\psi\rangle$~\cite{Christensen:1978}, where $x$ and $x'$ are space-time points. The anti-commutator is related to the Feynman Green function $G_F(x,x')$ and to the Wightman function $G^+(x,x')=\langle\psi | \phi(x)\phi(x')|\psi\rangle$ as~\cite{DeWitt:1965,Birrell:Davies}:
\begin{equation}\label{eq:elem func}
G^{(1)}(x,x') =
2\ \text{Im} \left(G_F(x,x')\right) = 2\ \text{Re} \left(G^+(x,x')\right).
\end{equation}
Clearly from their definitions, both the anti-commutator and the Wightman function satisfy  (with respect to either $x$ or $x'$) the homogeneous scalar field  equation \eqref{wave-eq}.
In its turn, the Feynman Green function satisfies the Green function equation
\begin{equation}\label{eq:GF eq} 
\left(\Box +\frac{3}{4\ell^2}\right)G_F(x,x') = -\frac{\delta^{(3)}(x-x')}{\sqrt{-g}},
\end{equation}
where $g\equiv det(g_{\mu\nu})$ and $\delta^{(3)}$ is the  Dirac-$\delta$ distribution in three dimensions.

\vspace{0.5cm}
\textbf{ 1. Locally AdS$_3$ space-time}
\\

In principle, there are two possible approaches to compute the two-point function in the BTZ geometries. The first one is to  expand 
this function 
in terms of elementary modes
of the wave equation (\ref{wave-eq}) 
satisfying  appropriate boundary conditions. The second approach is to use the fact that these geometries can be obtained by an appropriate identification in the covering AdS$_3$ geometry. This second approach is the one followed by~\cite{Steif,Lifschytz:1993eb,Shiraishi,MZ2}  and the one that we shall follow here - except in App.\ref{sec:GF in BTZ modesum}, where we follow the first approach. 
 
Within the second approach, the two-point function in BTZ can be readily obtained from the two-point function in the embedding space CAdS$_3$~\cite{Lifschytz:1993eb,Steif}. 
As mentioned in Sec.\ref{sec:review}, the BTZ space-times are not globally hyperbolic.
For the
Cauchy problem to be well-defined in these space-times, one must impose boundary conditions
on the timelike  AdS boundary~\cite{Avis1978quantum} (as well as on the timelike singularity in the NS case).
The field may obey different boundary conditions on the AdS boundary. We choose transparent boundary conditions, which correspond to defining the field modes that are smooth on the entire Einstein Static Universe to which the AdS geometry can be conformally mapped \cite{Lifschytz:1993eb,Avis1978quantum}.
Taking advantage of the fact that AdS$_3$ is a maximally-symmetric space-time, the anti-commutator in CAdS$_3$ corresponding to these boundary conditions can be found to be \cite{Avis1978quantum,Shiraishi,Shiraishi:1993ti,Decanini:Folacci:2005a,Steif}
\begin{equation}\label{eq:G^1A as sigma_M}
G^{(1)}_A(x,x')=
\frac{1}{2\sqrt{2}\pi}
\frac{\Theta\left(\s(x,x')\right)}{
\sqrt{\s(x,x')}
},
\end{equation}
where
$\Theta$ is the Heaviside step function,
\begin{equation}\label{eq:world func R22}
\s(x,x')\equiv
\left( -(X^0-X'^0)^2+(X^1-X'^1)^2+(X^2-X'^2)^2- (X^3-X'^3)^2\right)/2,
\end{equation}
and $x$ and $x'$ are points in AdS$_3$.
Here, 
$X^a$ and $X'^a$, with $a=0,1,2,3$,
are the coordinates in the embedding space $\mathbb{R}^{(2,2)}$ of the points $x$ and $x'$, respectively.
We note that $\s(x,x')$ 
is equal to one-half of the square of the
geodesic distance  
between the two points 
$X^a$ and $X'^a$
in flat $\mathbb{R}^{(2,2)}$
(this is Synge's
world function
in $\mathbb{R}^{(2,2)}$, not in CAdS$_3$). Since $X^a$ and $X'^a$ belong to the pseudosphere, $\sigma(x,x')$ is the chordal distance between $x$ and $x'$.
Throughout the paper, 
we use 
Latin letters (such as $a$ and $b$) for indices of coordinates of points in $\mathbb{R}^{2,2}$
and Greek letters 
(such as $\mu$ and $\nu$)
for indices of coordinates of points in CAdS$_3$ and BTZ geometries.
See App.\ref{sec:G_A} for further details and an explicit coordinate expression for $G^{(1)}_A(x,x')$.

\vspace{0.5cm}
\textbf{2. Multiply connected spaces}
\\

Let us now turn to the calculation of the two-point function and the RSET specifically in the BTZ geometries. Applying the method of images --according to which one must sum over all distinct images of a point obtained by the identification in the embedding space--, it readily follows that the anti-commutator both for the BH and NS geometries reads 
\begin{equation} \label{G^1BTZgeometries}
G^{(1)}(x,x')= \sum_{n\in I} G^{(1)}_A(x,H^n x'),
\end{equation}
where $H$ is the identification matrix in $\mathbb{R}^{2,2}$ introduced in Sec.\ref{sec:review}.3\footnote{Strictly speaking, $H$ is meant to act on a point in $\mathbb{R}^{2,2}$.
As a slight abuse of notation, by $H^nx$ we shall mean $H^n$ acting on the point on the pseudosphere in $\mathbb{R}^{2,2}$ that corresponds to the point $x$ in the BTZ space-time.} and the range $I$ is decribed below.
In the case of transparent boundary conditions, the two-point function can be written as
\begin{equation} \label{G^1BTZgeometriesTC}
G^{(1)}(x,x')= 
\frac{1}{2\sqrt{2}\pi} \sum_{n\in I} \frac{\Theta(\s(x,H^n x'))}{\sqrt{\s(x,H^n x')}}.
\end{equation}
This expression
applies to the case that the field obeys specific boundary conditions on the AdS boundary ($r=\infty$) and, if the spacetime possesses one (which is the case for all BTZ geometries except for the static BH), also on the timelike singularity ($r=0$).
In the case of a static NS,
we (re-)derive the 
expression \eqref{G^1BTZgeometriesTC}
in Appendix~\ref{sec:GF in BTZ modesum}
using the alternative method of mode sums,
and we see explicitly that the boundary conditions satisfied at the timelike singularity are square-integrability.

In the expressions above,
 $I\subset \mathbb{Z}$ is a summation range over all the various distinct images (see Appendix~\ref{sec:GF in BTZ modesum}
 where,
 in the static NS case,
 the ``sum over images" arises as a ``sum over caustics"). The identification matrices for the BH and NS cases are different and we give them explicitly in Appendix \ref{identifications}; the ranges $I$ are also  different in each case and we describe them next. 

\vspace{0.5cm}
\textit{1. Black hole}
\vspace{0.3cm}

The Green function for the three-dimensional BTZ BH was discussed in \cite{Lifschytz:1993eb,Shiraishi,Steif}. Since the identification matrix $H$ acts transitively on $\mathbb{R}^{2,2}$, the sum in Eq.\eqref{G^1BTZgeometries} includes an infinite countable number of images: $n \in I= \mathbb{Z}$. As is shown in Appendix A,  the $H$ matrix for the rotating black hole is given by
\begin{equation}\label{HBTZrot}
H=
 \left( \begin{array}{cccc}
\cosh(\pi\alpha_{+})  & \sinh(\pi\alpha_{+})  & 0 & 0 \\
\sinh(\pi\alpha_{+})  & \cosh(\pi\alpha_{+})  & 0 & 0 \\
0 & 0 & \cosh(\pi\alpha_{-})  & -\sinh(\pi\alpha_{-})  \\
0 & 0 &-\sinh(\pi\alpha_{-})  & \cosh(\pi\alpha_{-})
 \end{array} \right).
\end{equation}

\vspace{0.5cm}
\textit{2. Conical singularity}
\vspace{0.3cm}

In the case of a conical singularity, the method of images does not reproduce the mode expansion for the two-point function for {\it arbitrary} values of $M$ and $J$. Let us for now focus on the static case. If the deficit angle $\Delta$ is of the form $2\pi (k-1)/k$, $k\in\mathbb{Z}^+$, the angular identification produces a finite number of  images\footnote{A finite number of images is also obtained for rational values of $k$.}. On the other hand, for arbitrary real values of $\Delta$  the sum in Eq.(\ref{G^1BTZgeometries}) must be replaced by an integral since the associated eigenfunctions acquire a continuous degree and order \cite{Hobson}. The integral expressions, however, interpolate between the discrete sums that occur for consecutive deficit angles, $2\pi (k-1)/k$ and $2\pi k/(k+1)$.

The rationale that explains the difference between the BH case  NS cases is as in electrostatics: the method of images between two parallel conducting plates generates a countable but infinite number of images regardless of the distance between the plates. On the other hand, if the plates form an angle $\theta=2\pi/k$,  a finite number of images is produced, for $k\in\mathbb{Z}$, whereas a dense distribution of images are generated for a generic $k$. In  the case of angular excesses (negative angular deficit and $M < -1$) the geometry is also described by Eq.~\eqref{BTZ}, but the method of images is inadequate.
Therefore, from now on, for  NS geometries (whether rotating or not), we restrict ourselves to the case $M^2\ell^2-J^2<1$ (and $M<0$).

The identification matrix $H$  is that in Eq.\eqref{HNS} for $\beta\equiv\beta_+=2 \sqrt{-M}$, $\beta_-=0$, namely
\begin{equation}\label{HNSstatic}
H = 
 \left( \begin{array}{cccc}
1  & 0 & 0 & 0  \\
0 &\cos (\pi  \beta ) &  -\sin (\pi  \beta )  & 0 \\
0 & \sin (\pi  \beta ) & \cos (\pi  \beta )   & 0  \\
0 & 0 & 0 & 1
 \end{array} \right).
\end{equation}

The number of terms in the sum in Eq.\eqref{G^1BTZgeometries} is given by the number of distinct images produced by the action of the identification matrix $H$, which in this case is $N-1$, where $N$ is the smallest positive integer such that $H^{N}=1$. 
The condition that such a number $N$ exists
implies that $\beta$ is a rational number. In~\cite{Souradeep-Sahni} and in asymptotically flat (instead of AdS) space-time, the method of images was applied specifically to the case $\beta=2/N$, with $N$ a positive integer. Furthermore, in App.\ref{sec:GF in BTZ modesum} we obtain the two-point function for this $\beta$
using the method of mode sums, without relying on the method of images. Therefore, henceforth we shall consider only the case $\beta=2/N$, $N\in\mathbb{Z}^+$, for static NSs. Both from the method of images and from the independent mode-sum calculation of App.\ref{sec:GF in BTZ modesum}, it follows that in Eq.\eqref{G^1BTZgeometries} the sum over the images yields
\begin{align}  \label{Green-NS2}
G^{(1)}_{NS}(x,x')=
\sum_{n=0}^{N-1} G^{(1)}_A(x, H^n x')=
 \frac{1}{2\sqrt{2}\pi} \sum_{n=1}^{N-1} \frac{\Theta(\s(x,H^nx')}{\sqrt{\s(x,H^nx')}}.
\end{align}

The mode expansion in \cite{Cheeger,Souradeep-Sahni} for a conical space-time without a cosmological constant can possibly be extended to the asymptotically AdS$_3$ case by replacing Bessel functions by Legendre functions in the homogeneous solutions -- see Eq.\eqref{eq:slns radial ODE static NS}.

Let us now turn to  the rotating case. In this case, the identification matrix (given in Eq.\eqref{HNS}) depends on two parameters, $\beta_+$ and $\beta_-$, 
\begin{equation}\label{HrotNS}
H = 
 \left( \begin{array}{cccc}
\cos \left(\pi  \beta _-\right)   & 0 & 0 &  -\sin \left(\pi  \beta _-\right)  \\
0 &\cos \left(\pi  \beta _+\right) &  -\sin \left(\pi  \beta _+\right)  & 0 \\
0 & \sin \left(\pi  \beta _+\right) & \cos \left(\pi  \beta _+\right)   & 0  \\
\sin \left(\pi  \beta _-\right)  & 0 & 0 & \cos \left(\pi  \beta _-\right) 
 \end{array} \right).
\end{equation}
Again, the number of terms in the sum in Eq.\eqref{G^1BTZgeometries} is given by the number of distinct images produced by the action of the identification matrix $H$. We shall henceforth consider only the case $\beta_{\pm}=2/N_{\pm}$, $N_{+}\in\mathbb{N}$, $|N_{-}|\in\mathbb{N}$ for rotating NSs, where $|N_-| > N_+$. The smallest $N$ for which $H^N=1$ occurs when $N$ is the least common multiple of $N_+$ and $N_-$. This means that the number of images in the sum in Eq.\eqref{G^1BTZgeometries} is $N-1$ and the expression for the two-point function is
formally the same as in Eq.\eqref{Green-NS2}.

\subsection{Renormalized stress-energy tensor}\label{sec:RSET}

Equipped with the two-point function, we now turn to the calculation of the RSET. As mentioned above, the quantum stress-energy tensor would in principle be calculated by applying a certain geometric differential operator on the two-point function $G^{(1)}(x,x')$. However, as is well known, the two-point function typically diverges at coincidence ($x=x'$) -- this can readily be seen in the BTZ case from Eq.\eqref{G^1BTZgeometriesTC} and the fact that $\s(x,x)=0$.
Therefore, in order to obtain the RSET, one must first renormalize the two-point function by subtracting from it an appropriate  bitensor $G^{(1)}_{div}(x,x')$ which is purely geometric. The RSET for the conformally-coupled scalar field can thus be obtained from the Hadamard elementary function as~\cite{Birrell:Davies,Steif}\footnote{The operator in Eq.\eqref{T-munu} is $1/2$ times the corresponding operator in~\cite{CFMZ1,casals2018quantum}. The reason is that the definition of the anticommutator $G^{(1)}(x,x')$ here is $2$ times the definition used in~\cite{CFMZ1,casals2018quantum}, so that all the results in here and in~\cite{CFMZ1,casals2018quantum}  agree.}:
\begin{align}\label{T-munu}
\kappa\rset{T_{\mu\nu}(x)}= & 
\pi l_P
\lim_{x' \rightarrow x}
\left( 3\nabla^x_\mu \nabla^{x'}_\nu - g_{\mu \nu} g^{\alpha \beta} \nabla^x_\alpha \nabla^{x'}_\beta 
-\nabla^x_\mu \nabla^x_\nu - \frac{1}{4\ell^2}g_{\mu \nu} \right) \left(G^{(1)}(x,x')
-G^{(1)}_{div}(x,x')
\right).
\end{align}
We note that the Heaviside step function in Eq.\eqref{G^1BTZgeometries} does not actually appear in ~\cite{Lifschytz:1993eb,Shiraishi,Shiraishi:1993ti,Steif}. The reason is that  these references calculate either a two-point function different from the anti-commutator or else the anti-commutator only in the static case. In the static case (whether BH or NS), $\s(x,H^nx)$ is non-negative
and so the step function is redundant in this case. However, in the rotating case (whether BH or NS), $\s(x,H^nx)$ can be negative and so it is important to include the step function. 

Let us here note some properties of the RSET. Firstly, since we are dealing with a massless and conformally-coupled scalar field, the trace of its classical stress-energy tensor must be zero. Furthermore, since we are dealing with a three-dimensional space-time, the trace of the RSET must also be zero (i.e., there is no trace anomaly)~\cite{Birrell:Davies}. Secondly, the divergent term $G^{(1)}_{div}$ is constructed in a way so that the RSET is also conserved with respect to the classical background metric. In the BTZ case, the subtraction of $G^{(1)}_{div}(x,x')$ corresponds to simply removing the $n=0$ term from the $n$-sum in Eq.(\ref{G^1BTZgeometries}) ~\cite{PhysRevD.40.948,Souradeep-Sahni,Shiraishi}. Therefore, the $n$-sums for the RSET that follow from Eqs.\eqref{G^1BTZgeometries} and \eqref{T-munu} will be over the summation range $I\setminus \{0\}$, instead of the range $I$ which we described in Sec.\ref{sec:two-pt func}.2 for the various space-time settings. 

Furthermore, as follows from from Eqs.\eqref{G^1BTZgeometries} and \eqref{T-munu}, the $n$-summands in the RSET will contain the quantity
\begin{equation}\label{eq:dn}
\dn\equiv 2\s(x,H^nx)
\end{equation}
as well as $\Theta(\dn)$. Therefore, in order to facilitate the notation for the $n$-sums we define a new summation symbol, 
\begin{equation}\label{eq:sum theta}
\Sp_{n} s_n\equiv
\sum_{n}
\Theta(\dn)\, s_n,
\end{equation}
for some summand $s_n$ and some summation range.

It follows from~\cite{Steif} that,
by inserting the general form Eq.\eqref{G^1BTZgeometries} for the two-point function into
Eq.\eqref{T-munu}, and using Eq.\eqref{eq:world func R22} for $\s$,
the RSET for a conformal scalar field satisfying transparent boundary conditions
on a BTZ geometry takes the  form 
\begin{align}\label{eq:RSET S}
\kappa\rset{T_{\mu\nu}}=
&   
\frac{3 l_P}{2 } 
\Sp_{n\in I\setminus \{0\}}
\left(  S^n_{\mu\nu} - \frac{1}{ 3} g_{\mu \nu} g^{\lambda\rho}
S^n_{\lambda\rho} \right), 
\end{align}
where $ S^n_{\mu\nu} \equiv \partial_{\mu}X^a \partial_{\nu}X^b  S^n_{ab}$ 
is the pull back to AdS$_{3}$ of  
\begin{equation}\label{eq:S}
 S^n_{ab}  \equiv  
 \frac{(H^n)_{ab}}{\dn^{3/2}}  +  \frac{3 (H^n)_{ac} X^{c}
(H^{-n})_{bd} X^{d} - (H^n)_{ac} X^{c}
(H^n)_{bd} X^{d}}{\dn^{5/2} }.
\end{equation}
Even though this expression for the RSET was given in~\cite{Steif} for the BH case, it also applies to the NS with the appropriate summation range $I$.

We now proceed to give explicit  expressions for the RSET and describe its main physical features, separately for the BH and NS cases. We will make use of the fact that the summand in Eq.\eqref{eq:RSET S} is either symmetric or antisymmetric -- depending on the specific component -- under $n\to -n$.

\vspace{0.5cm}
\textbf{1. Black hole} 
\vspace{0.5cm}

Here we give the RSET in the BH geometries.
Using the symmetries under $n\to -n$ mentioned above and the fact that $\Is=\mathbb{Z}$ is symmetric with respect to $n=0$, the explicit expressions for the RSET that we shall give will contain $n$-sums involving only $n>0$.

We first summarize the RSET result in~\cite{MZ2} in the static case.
We then re-derive (and make a slight correction to) the RSET in~\cite{Steif} in the non-extremal rotating case and plot its components. We finally derive the RSET in the extremal case.
For the rotating BH cases, 
we also give the specific radii  inside the Cauchy horizon at which the RSET diverges.

\vspace{0.5cm}
\textbf{\textit{1A. RSET for the static BTZ black hole}} \\

\begin{figure}[h!]
\begin{center}
  \includegraphics[width=12cm]  {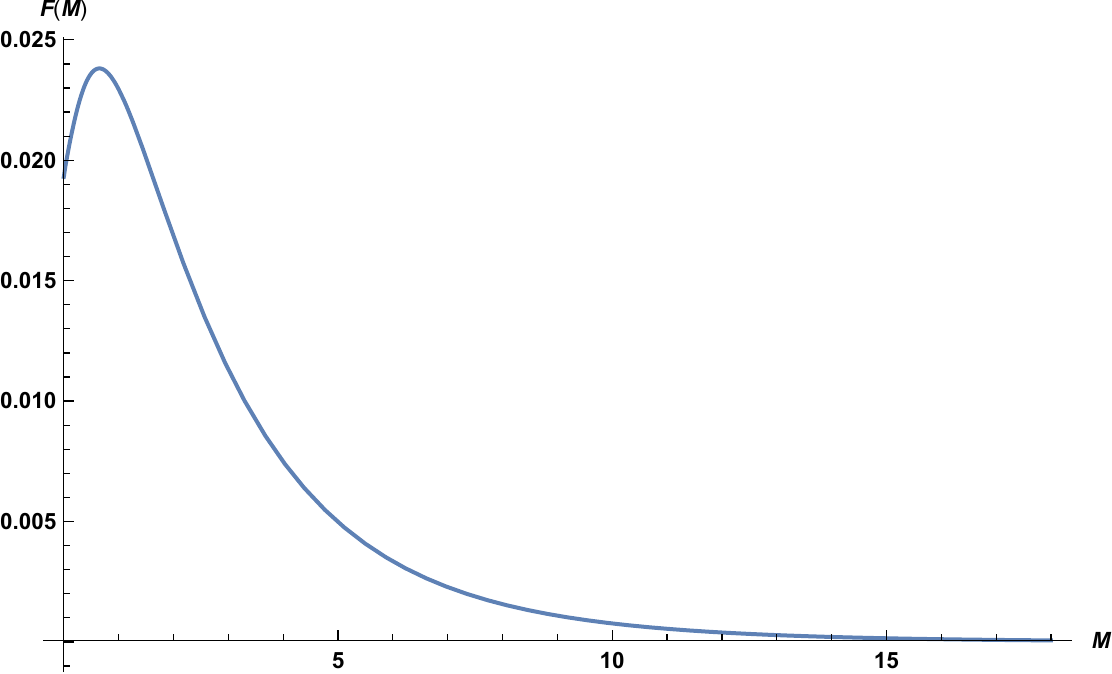}
  \end{center}
\caption{The function $F(M)$ of Eq. \eqref{eq:FBH} that defines the RSET profile for a static BH. This function has a maximum at $M\approx 0.648876$, decays exponentially for large $M$ and $F(0^+)=\zeta(3)/(2 \pi^3)\approx 0.0193841$.}
\label{fig:FBH} 
\end{figure}

The RSET in the static BTZ BH is obtained from Eqs.\eqref{eq:RSET S}, \eqref{regionI}, \eqref{regionII},
 and \eqref{HBTZ} with ($\alpha_-=0$),
 and the summation range from $-\infty < n < \infty$ in  
 Eq.\eqref{G^1BTZgeometriesTC}. In this setting, it is $d_n>0$, $\forall n>0$, at any space-time point, and so $\Theta(d_n)=1$ in Eq.\eqref{eq:sum theta}. The RSET
 in this case is
\begin{equation}\label{eq:RSET static BH}
\kappa\, \rset{T^{ \mu}{}_{ \nu}(x)} =\frac{l_P}{ r^3}  F(M)\, \text{diag}(1,1,-2),
\end{equation}
in $\{t,r,\theta\}$ coordinates, where
\begin{equation}\label{eq:FBH}
F(M)\equiv \frac{M^{3/2}}{2\sqrt{2}} \sum_{n=1}^{\infty} \frac{\cosh (2n\pi \sqrt{M})+3}{\left(\cosh (2n\pi \sqrt{M}\right)-1)^{3/2}}.
\end{equation}
We plot the function $F(M)$ in Fig.\ref{fig:FBH}.
Also, we note that we obtain the same expression for the RSET regardless of which region of the space-time,
$r>r_+$ (Eq.\eqref{regionI}) or
$0<r<r_+$ (Eq.\eqref{regionII}),
we calculate it in. The result \eqref{eq:FBH} was previously found in \cite{Lifschytz:1993eb} and \cite{Steif}.

\vspace{0.5cm}
\textbf{\textit{1B. RSET for the rotating nonextremal BTZ black hole}} \\%

Let us now include (non-extremal) rotation to the BH. 
From Eqs.\eqref{eq:RSET S}, \eqref{regionI}, \eqref{regionII},
\eqref{regionIII} and \eqref{HBTZ} 
we obtain the RSET
in the nonextremal BH case:
\begin{align}
\label{eq:RSET subext BH tt}
\kappa \langle T^{t}\,_{t} \rangle=& \frac{ 2 l_{P}}{(\alpha_{+}^2-\alpha_{-}^2)^{2}}\Sp_{n=1}^{\infty} \frac{2\left(2 r^2 \left(3 {a}_{n}+\left(\alpha _+^2-\alpha _-^2\right) {b}_{n}\right)-\ell^2 g_n\right) {c}_{n} +3 \alpha_{+}\alpha_{-}{e}_{n}\left(8 r^2-\alpha _-^2 \ell^2+\alpha _+^2 \ell^2\right)}{{d}_{n}^{5/2}},
\\
\kappa\langle T^{r}\,_{r} \rangle=& l_{P}\Sp_{n=1}^{\infty} \frac{{c}_{n}}{{d}_{n}^{3/2}},
\\
\kappa \langle T^{\theta}\,_{\theta} \rangle=& -\frac{ 2 l_{P}}{(\alpha_{+}^2-\alpha_{-}^2)^{2}}\Sp_{n=1}^{\infty} \frac{2\left(2 r^2 \left(3 \bar{a}_{n}-\left(\alpha _+^2-\alpha _-^2\right) {b}_{n}\right)-\ell^2 \bar{g}_n\right) {c}_{n}+3  \alpha_{+}\alpha_{-}{e}_{n}\left(8 r^2-\alpha _-^2 \ell^2+\alpha _+^2 \ell^2\right)}{{d}_{n}^{5/2}},
\\
\kappa\langle T^{t}\,_{\theta} \rangle=& -\frac{6  l_{P}\ell}{ (\alpha_{+}^2-\alpha_{-}^2)^{2}}  \Sp_{n=1}^{\infty} \frac{\left(4 ({c}_{n}-4) r^2-{a}_{n} \ell^2\right){c}_{n}\alpha _+\alpha _- +{e}_{n} \left(-4 r^2\left(\alpha _-^2+\alpha _+^2\right)+2\alpha_{-}^2 \alpha _+^2 \ell^2 \right)}{{d}_{n}^{5/2}},
\\
\kappa\langle T^{\theta}\,_{t} \rangle=&  \frac{6  l_{P}}{ \ell(\alpha_{+}^2-\alpha_{-}^2)^{2}} \Sp_{n=1}^{\infty} \frac{\left(4 ({c}_{n}-4) r^2-{a}_{n}\ell^2\right){c}_{n}\alpha _+\alpha _-   +{e}_{n}\left(-4 r^2\left(\alpha _-^2+\alpha _+^2\right)+\ell^2(\alpha _-^4 +\alpha _+^4 )\right)}{{d}_{n}^{5/2}},
\label{eq:RSET subext BH tht}
\end{align}
with
\begin{align}
a_{n}\equiv &2\alpha_{+}^2 \sinh ^2\left(\frac{n\pi  \alpha _- }{2} \right)+2\alpha_{-}^2 \sinh ^2\left(\frac{n\pi  \alpha _+ }{2} \right), \label{anBH}
\\
\bar{a}_{n}\equiv &2\alpha_{+}^2 \sinh ^2\left(\frac{n\pi  \alpha _+ }{2} \right)+2\alpha_{-}^2 \sinh ^2\left(\frac{n\pi  \alpha _- }{2} \right),
\\
{b}_n\equiv &\cosh \left(\pi  n \alpha_{+}\right)-\cosh \left(\pi  n \alpha_{-}\right)=2\left( \sinh ^2\left(\frac{\pi n \alpha _+ }{2} \right)-\sinh ^2\left(\frac{\pi  n \alpha _- }{2} \right) \right),
\label{bnBH}
\\
{c}_n\equiv & \cosh \left(\pi  n \alpha_{+}\right)+\cosh \left(\pi  n \alpha_{-}\right)+2, \label{cnBH}
\\
{e}_n\equiv & 2 \sinh \left(\pi  n \alpha_{+}\right) \sinh \left(\pi  n \alpha_{-}\right), \label{enBH}
\\
 {g}_n\equiv &\alpha _-^2 \left(\alpha _+^2+2 \alpha _-^2\right) \sinh ^2\left(\frac{n\pi  \alpha _+ }{2} \right)+\alpha _+^2 \left(\alpha _-^2+2 \alpha _+^2\right) \sinh ^2\left(\frac{n\pi  \alpha _- }{2} \right), \\
\bar{g}_n\equiv &\alpha _-^2\left(\alpha _-^2+2 \alpha _+^2 \right) \sinh ^2\left(\frac{n\pi  \alpha _+ }{2} \right)+\alpha _+^2\left(\alpha _+^2+2 \alpha _-^2 \right) \sinh ^2\left(\frac{n\pi  \alpha _- }{2} \right),
\end{align}
and, as per Eq.\eqref{eq:dn},
\begin{equation} \label{dnBH}
d_{n}
= 2\s(x,H^nx)
=4 \ell^2 \frac{\alpha_{+}^2 \sinh ^2\left(\frac{\pi n \alpha _- }{2} \right)-\alpha_{-}^2 \sinh ^2\left(\frac{\pi n \alpha _+ }{2} \right) +2 r^2\ell^{-2} {b}_n}{\alpha_{+}^2-\alpha_{-}^2}.
\end{equation}

In this setting, it is $d_n>0$ for all $n>0$ and any space-time point with $r> \ell |\alpha_-|/2$ (the Cauchy horizon of the BTZ background). Therefore, in general, the $\Theta(d_n)$ of Eq.\eqref{eq:sum theta} must be kept in the above equations.
Also, we note that we obtain the same expression for the RSET regardless of which region of the space-time,
$r>r_+$ (Eq.\eqref{regionI}), 
$r_-<r<r_+$ (Eq.\eqref{regionII}),
 or $0<r<r_-$ (Eq.\eqref{regionIII}),
we calculate it in.

An important issue appears in the region $r<r_-$: in this region, $d_n$ takes negative values and it vanishes at the  radii given by
\begin{equation} \label{eq:rn subext BH}
r^2=r_n^2\equiv \ell^2 \frac{\alpha_{-}^2 \sinh ^2\left(\frac{\pi n \alpha _+ }{2} \right)-\alpha_{+}^2 \sinh ^2\left(\frac{\pi n \alpha _- }{2} \right)}{ 2 {b}_n}, \quad
n\in \mathbb{Z}^+.
\end{equation}
Consequently, all components of $\langle  T^{\mu}\,_{\nu}  \rangle$ diverge at these various radii $r=r_n<r_-$. Moreover, $r_n^2\to r_-^2$ as $n\to \infty$, and therefore, $r_-$ is an accumulation point of singularities from the left.

We note that our RSET expressions in Eqs.\eqref{eq:RSET subext BH tt}--\eqref{eq:RSET subext BH tht} agree with those in Eq.19 in~\cite{Steif} except for a factor in one component.
Eq.19 in~\cite{Steif} is in a different set of coordinates, which are defined in Eq.(6) in~\cite{Steif} and which we  denote here by $\{\tS,\rS,\phiS\}$.
If we transform our Eqs.\eqref{eq:RSET subext BH tt}--\eqref{eq:RSET subext BH tht} to $\{\tS,\rS,\phiS\}$ coordinates, our result  is equal to that in Eq.(19) in~\cite{Steif} but with an extra factor ``$-2$" in the
$\langle T^{\phiS}{}_{\tS} \rangle$ component.

\begin{figure}
\begin{tabular}{cc}
  \includegraphics[width=8cm] {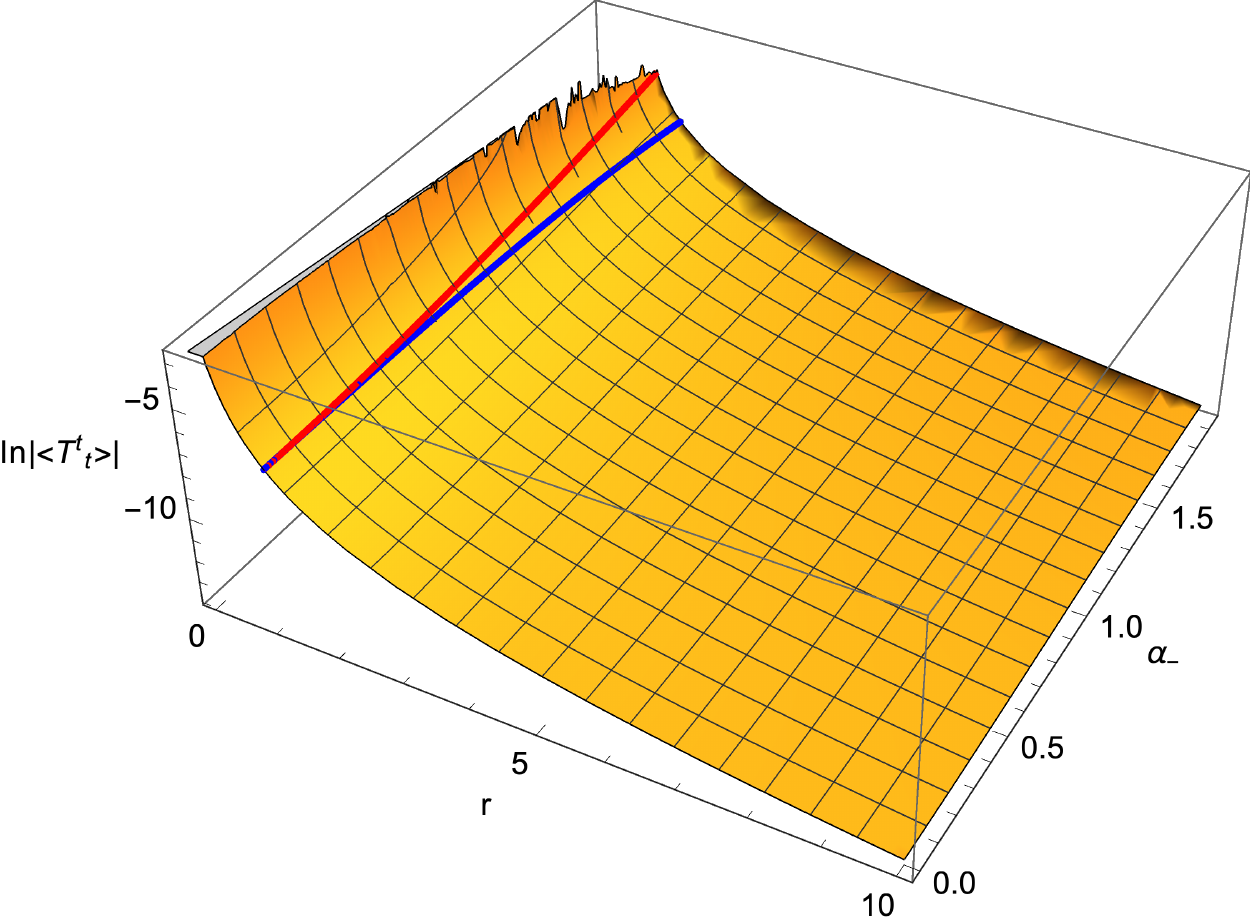}
        \includegraphics[width=8cm] {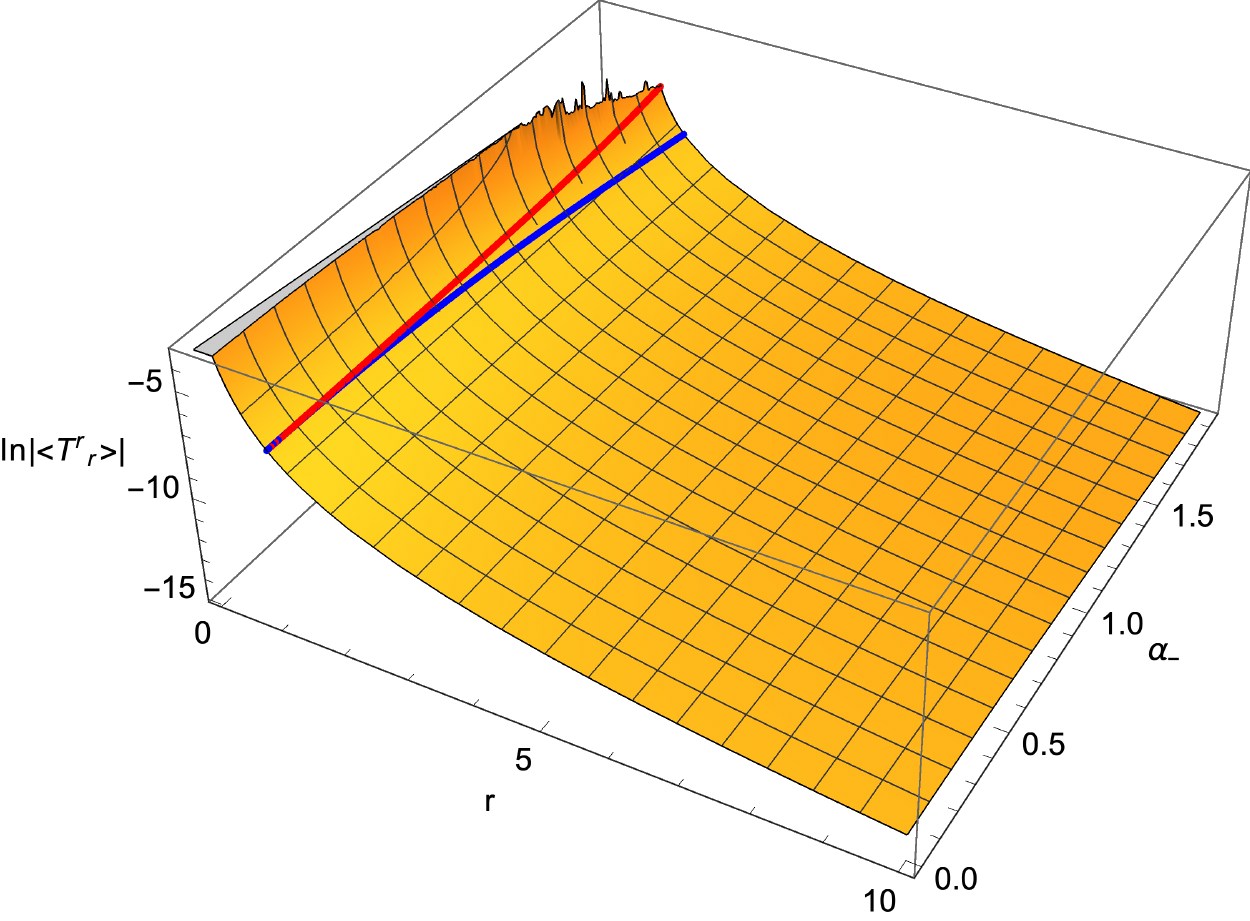}
  \end{tabular}
\caption{
Plot of the log of the absolute value of the RSET components
$\langle T^{\mu}{}_{\nu} \rangle$
as functions of $r\in (r_-,10]$ and $\alpha_-\in [0,\alpha_+]$ for the specific values of $\alpha_+=(\sqrt{3}+1)/\sqrt{2}$, $\ell=1$, $\kappa=8\pi$ and $l_P=1$.
Left: $\langle T^{t}{}_{t} \rangle$;
right: $\langle T^{r}{}_{r} \rangle$.
The continuous red and blue lines correspond
to, respectively, $r_+$ and $r_{SL}$. 
The vertical axis has been capped at a fixed value.
}
\label{fig:Tmunu 3D} 
\end{figure}

\begin{figure}
\begin{tabular}{cc}
             \includegraphics[width=8cm] 
      {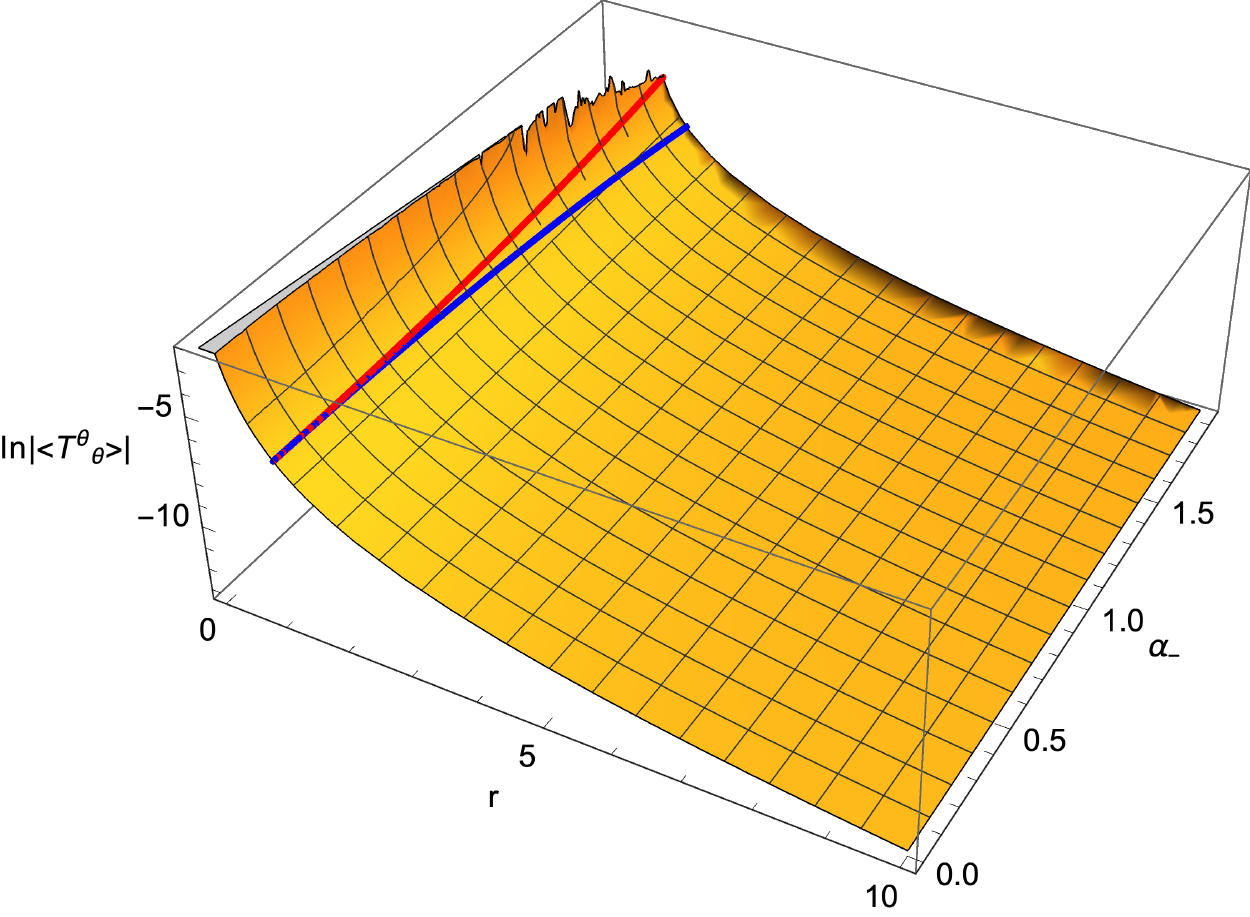}
  \end{tabular}
\caption{
Same as Fig.\ref{fig:Tmunu 3D} 
but for the component
$\langle T^{\theta}{}_{\theta} \rangle$. 
}
\label{fig:Tmunu 3D 2} 
\end{figure}

\begin{figure}
\begin{center}
           \includegraphics[width=8cm] {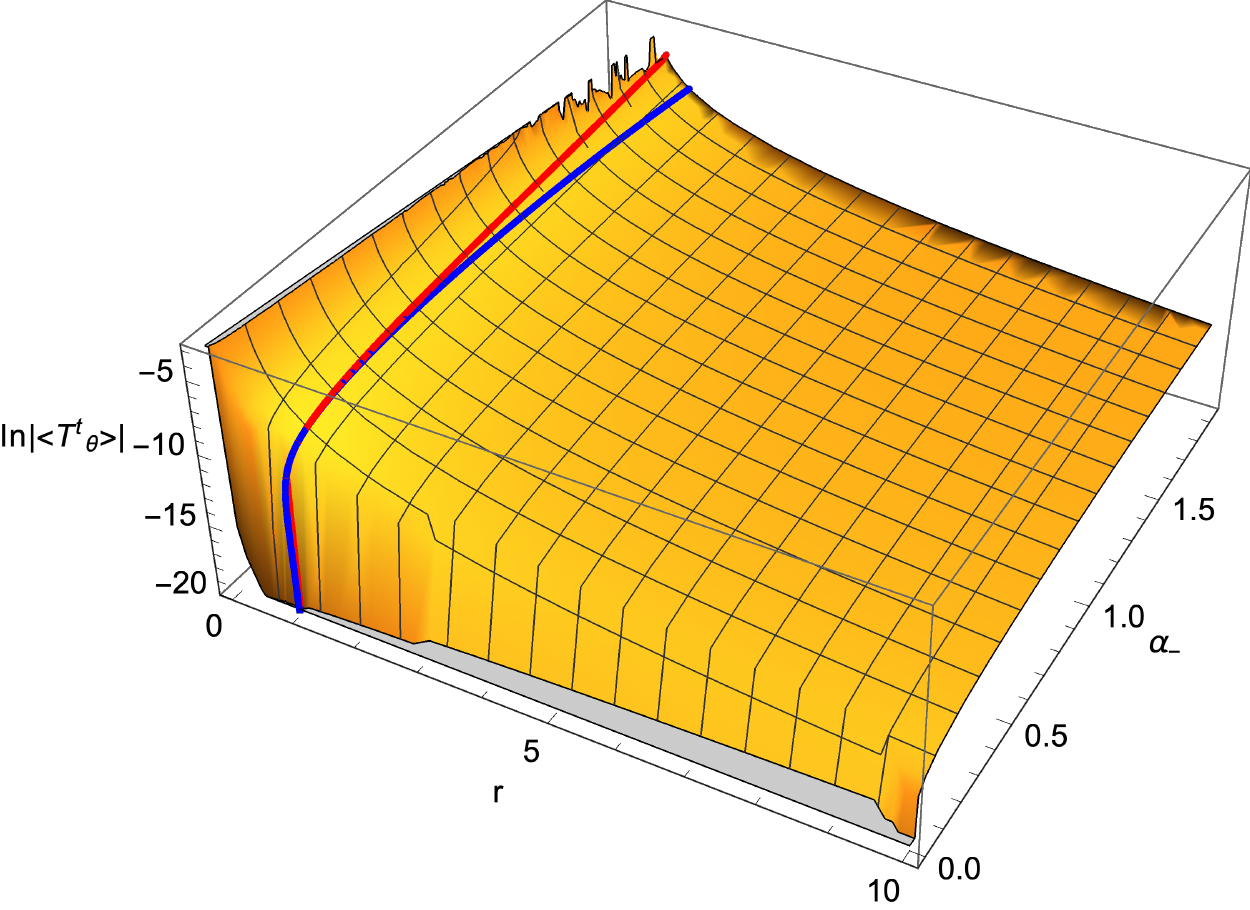}
     \includegraphics[width=8cm] {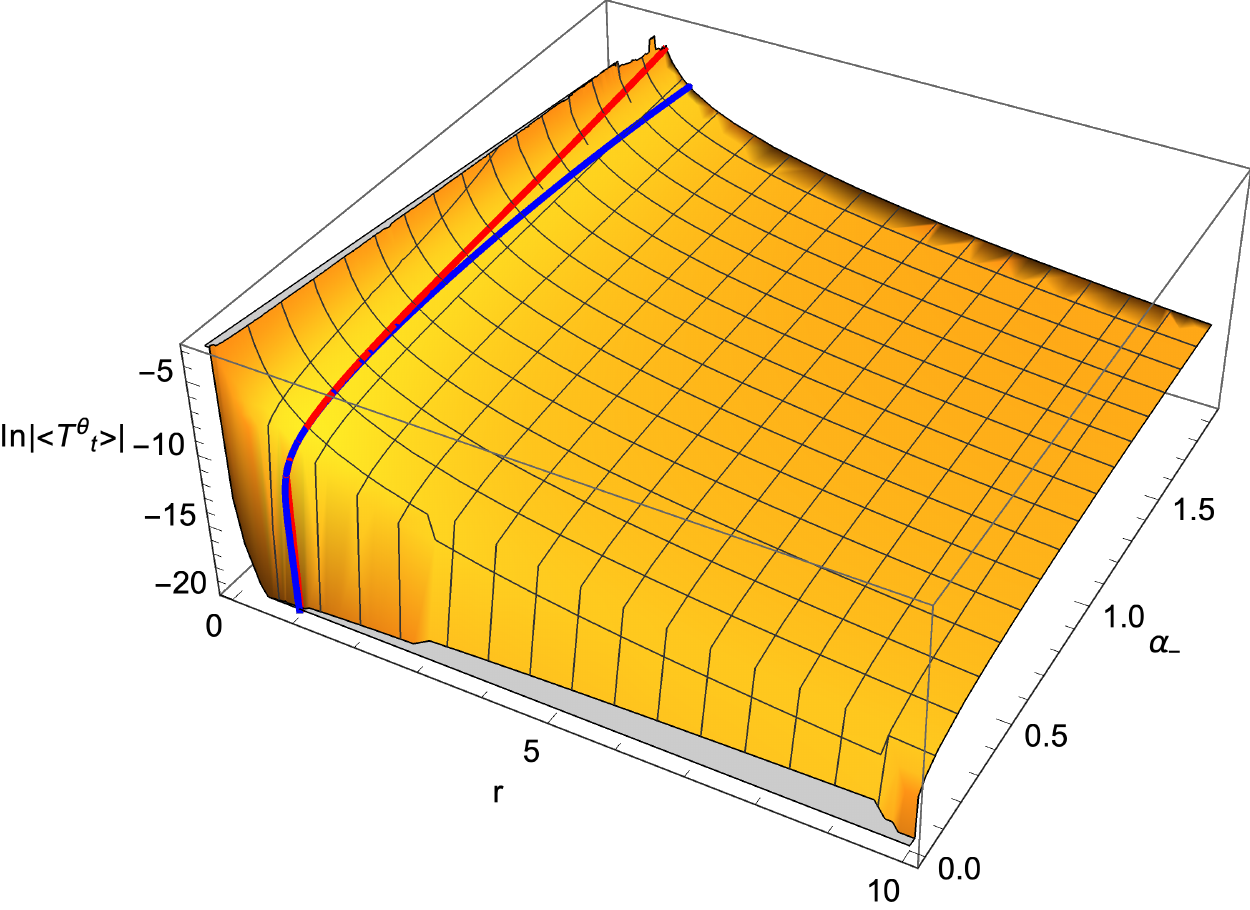}
  \end{center}
\caption{
Same as Fig.\ref{fig:Tmunu 3D} 
but for the (non-diagonal) components
$\langle T^{t}{}_{\theta} \rangle$ 
(left) and
$\langle T^{\theta}{}_{t} \rangle$ 
(right).
}
\label{fig:Tmunu 3D 3} 
\end{figure}

In Figs.\ref{fig:Tmunu 3D}--\ref{fig:Tmunu 3D 3}  we plot
the RSET components
$\langle T^{\mu}{}_{\nu} \rangle$
as functions of $r$ and $\alpha_-$ for a fixed
value of $\alpha_+$.
It can be observed that they all diverge as $r\to r_-$ as expected.
For comparison with different boundary conditions, we note that Ref.~\cite{Kothawala} plotted the RSET for the case of {\it Dirichlet} boundary conditions --instead of transparent boundary conditions, as in our case -- and explicit analytical expressions for the RSET in Dirichlet,   Neumann and Robin boundary conditions are given in~\cite{WordenPhD,WinstanleyWorden}.


\vspace{0.5cm}
\textbf{\textit{1C. RSET for the extremal BTZ black hole}} \\

The angular momentum in the extreme BTZ BH of mass $M$ is $J= \gamma M \ell$ with $\gamma=\pm 1$ (i.e., $\alpha_+= \gamma \alpha_-$). Here we define $\alpha\equiv r_+/\ell =\sqrt{M/2}>0$.
From Eqs.\eqref{eq:RSET S}, \eqref{eq:X ext BH,r>rH}, \eqref{eq:X ext BH,r<rH} and
\eqref{eq:Hn ext BH,r>rH} 
we then find the following expression for the RSET valid everywhere ($\forall r>0$):
\begin{align}\label{eq:RSET ext BH tt}
\kappa \langle T^{t}\,_{t} \rangle= & \frac{ \ell^2 l_P}{2 \alpha} \Sp_{n=1}^{\infty}\frac{ 1}{ d_n^{5/2}}\left[3 A(r) \left(6 \pi ^2 \alpha ^2 n^2+\left(2 \pi ^2 \alpha ^2 n^2+1\right) \cosh (4 \pi  \alpha  n)-1\right) \right. \nonumber \\& \left. +16 \pi  \alpha  n \left(3 \alpha -A(r)\right) \sinh (\pi  \alpha  n) \cosh ^3(\pi  \alpha  n)-4 \alpha  \sinh ^2(2 \pi  \alpha  n)\right],
\\
\kappa \langle T^{t}\,_{\theta} \rangle=&-\frac{3\gamma \ell^3 l_P}{2 \alpha}\Sp_{n=1}^{\infty}\frac{ 1}{ d_n ^{5/2}}\left[A(r) \left(6 \pi ^2 \alpha ^2 n^2+\left(2 \pi ^2 \alpha ^2 n^2-1\right) \cosh (4 \pi  \alpha  n)+1\right) \right. \nonumber \\& \left. +16 \alpha  \sinh (\pi  \alpha  n) \cosh ^2(\pi  \alpha  n) (\pi  \alpha  n \cosh (\pi  \alpha  n)-\sinh (\pi  \alpha  n))\right],
\\
\kappa \langle T^{r}\,_{r} \rangle=& \, 4 l_P \Sp_{n=1}^{\infty}\frac{\cosh ^2(n \pi  \alpha )}{ d_n^{3/2} },
\\
\kappa \langle T^{\theta}\,_{t} \rangle=&\frac{3\gamma \ell l_P}{2 \alpha}\Sp_{n=1}^{\infty}\frac{ 1}{ d_n^{5/2}}\left[A(r) \left(6 \pi ^2 \alpha ^2 n^2+\left(2 \pi ^2 \alpha ^2 n^2-1\right) \cosh (4 \pi  \alpha  n)+1\right)\right. \nonumber \\& \left. +16 \alpha  \sinh (\pi  \alpha  n) \cosh ^2(\pi  \alpha  n) (\pi  \alpha  n \cosh (\pi  \alpha  n)+\sinh (\pi  \alpha  n))\right],
\\
\kappa \langle T^{\theta}\,_{\theta} \rangle=& - \frac{ \ell^2 l_P}{2 \alpha} \Sp_{n=1}^{\infty}\frac{ 1}{ d_n^{5/2}}\left[3 A(r) \left(6 \pi ^2 \alpha ^2 n^2+\left(2 \pi ^2 \alpha ^2 n^2+1\right) \cosh (4 \pi  \alpha  n)-1\right) \right. \nonumber \\& \left. +16 \pi  \alpha  n \left(3 \alpha +A(r)\right) \sinh (\pi  \alpha  n) \cosh ^3(\pi  \alpha  n)+4 \alpha  \sinh ^2(2 \pi  \alpha  n)\right],
\label{eq:RSET ext BH tht}
\end{align}
with
\begin{equation}\label{eq:dn ext BH}
d_n=
 2\s(x,H^nx)=
4 \ell^2 \sinh (\pi  \alpha  n) \left(\pi  n A(r) \cosh (\pi  \alpha  n)+\sinh (\pi  \alpha  n)\right),
\end{equation}
and 
\begin{equation}
A(r)\equiv \frac{r^2-\ell^2\alpha^2}{\ell^2 \alpha }.
\end{equation}

We note that the RSET in the extremal BH case in Eqs.\eqref{eq:RSET ext BH tt}--\eqref{eq:RSET ext BH tht} is actually equal to the RSET in the sub-extremal BH case in Eqs.\eqref{eq:RSET subext BH tt}--\eqref{eq:RSET subext BH tht} when taking the extremal limit $r_-\to r_+$. 

Similarly to the non-extremal BH case, $d_n$ is zero at certain values $r_n < r_+$, with
\begin{equation} 
r_n^2\equiv r_+^2 \left(1 -\frac{\tanh \left(n \pi \alpha \right)}{ n \pi \alpha}\right),\quad n\in\mathbb{Z}^+.
\end{equation}
This implies that the $n$-th term in the series for $\langle  T^{\mu}\,_{\nu}  \rangle$ diverges at these radii. Moreover, since $r_n \to r_+$ as $n \to \infty$, $r_+$ becomes an accumulation point of singularities from the left.
 
\vspace{0.5cm}
\textbf{2. Naked singularity}
\vspace{0.5cm}

Here we give the RSET in the NS geometries.
Here we shall make use of the symmetry $S^n_{\mu \nu}(x)=S^{n-N}_{\mu \nu}(x)$, a consequence of the property $H^N=1$ that allows to symmetrize the sum over positive and negative $n$ in Eq.\eqref{eq:RSET S} as
\begin{equation}\label{eq:symm summand NS}
\sum_{n=1}^{N-1} f_n =\frac{1}{2}\sum_{n=1}^{N-1} (f_n+f_{n-N})= \frac{1}{2}\sum_{n=1}^{N-1} (f_n+f_{-n}),
\end{equation}
where $f_n$ is the summand in
\eqref{eq:RSET S}.
Depending on the specific component of the RSET, we have $f_n=f_{-n}$ or $f_n=-f_{-n}$.

We first review the RSET in the static case obtained in~\cite{CFMZ1} and afterwards give our new RSET results in the rotating case
(we do not consider the extremal NS because it involves an infinite sum whose convergence would need to be addressed separately). \\

\textbf{\textit{2A. RSET for the static NS}} \\

\begin{figure}[h!]
\begin{center}\includegraphics[width=10cm]  {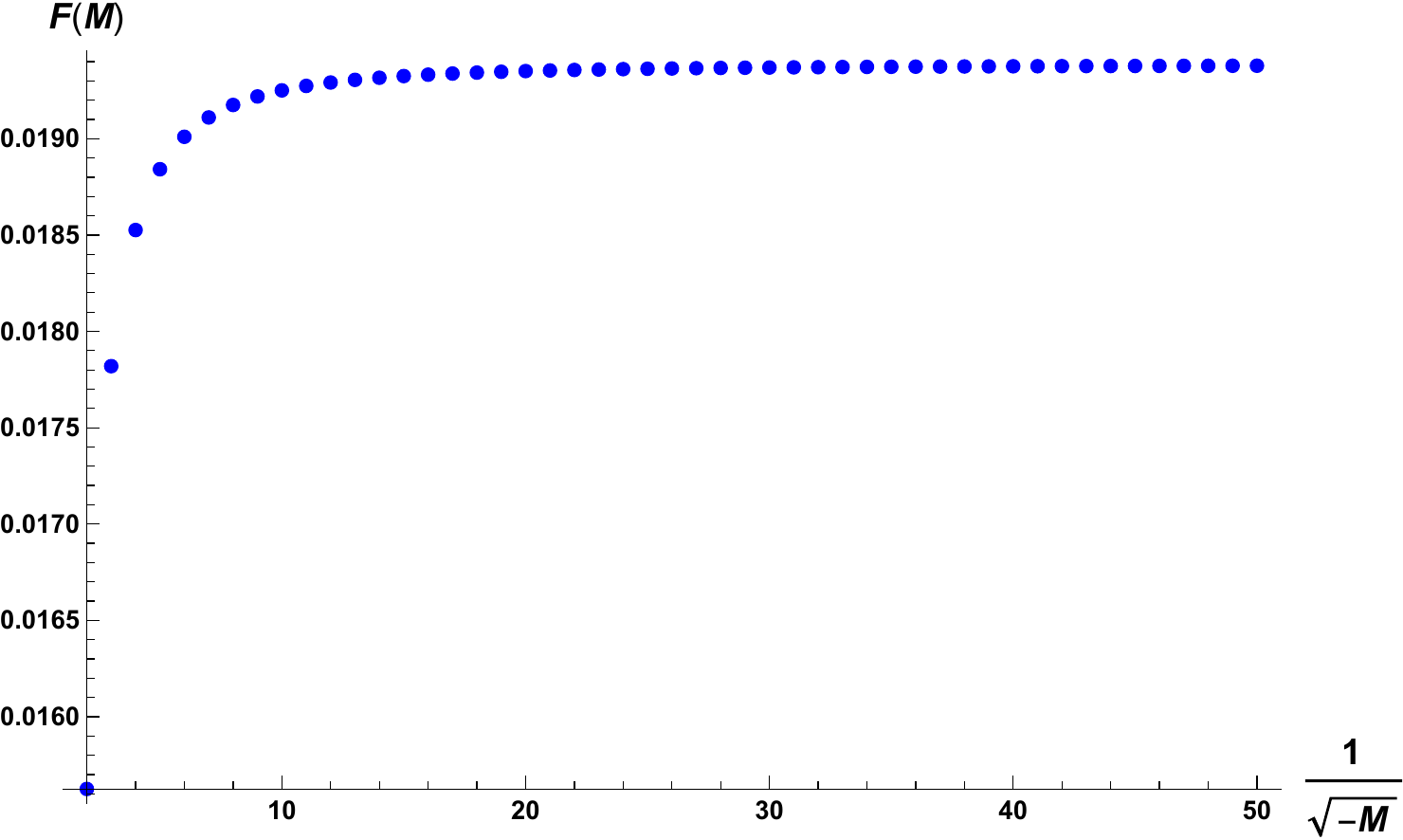}\end{center}
\caption{The function $F(M)$ of Eq. \eqref{eq:FNS} that defines the RSET profile as a function of $1/\sqrt{-M}=N$, where  $N=2,\dots, 50$. The sum in Eq.\eqref{eq:FNS} rapidly approaches the asymptotic value $F(0^-)$ in Eq.\eqref{FNS(0-)} given by the limit $N \to \infty$.}
\label{fig:FBHNS2} 
\end{figure}

We consider static NS space-times with $\beta=2\sqrt{-M}= 2/N$ and $N\in\mathbb{Z}^+$. The RSET on this space-time can then be obtained from Eq.\eqref{eq:RSET S}, the embedding Eq.\eqref{embeddingM-J} and identification matrix in Eq.\eqref{HNS} in the static limit ($\beta_+=\beta$, $\beta_-=0$), where the summation range is $ 1 \le n \le N-1$. As in the static BH case, it is $d_n>0,$ $\forall n>0$, and for any space-time point,  so that $\Theta(d_n)=1$. The result, derived in~\cite{CFMZ1}, is
\footnote{The symbol $N$ is not used for the same quantity here as in~\cite{casals2018quantum}, but the expressions in both places are equivalent. On the other hand, there is a typographical error in Eq.14 in~\cite{CFMZ1} in that a factor of $1/2$ is missing, but the remaining formulas in \cite{CFMZ1} are correct.
}
\begin{equation}\label{eq:RSET BH'}
\kappa\, \rset{T^{ \mu}{}_{ \nu}(x)} =\frac{l_P}{ r^3}  F(M)\, \text{diag}(1,1,-2),
\end{equation}
in $\{t,r,\theta\}$ coordinates, where
\begin{equation}\label{eq:FNS}
F(M)\equiv \frac{(-M)^{3/2}}{4\sqrt{2}} \sum_{n=1}^{N-1} \frac{\cos (2n\pi \sqrt{-M})+3}{\left(1-\cos (2n\pi \sqrt{-M}\right))^{3/2}},
\end{equation}
where we have used Eq.\eqref{eq:symm summand NS}.
The function $F(M)$ is plotted in Fig.\ref{fig:FBHNS2}.

From (\ref{eq:FNS}) it is clear that the result is nontrivial for $N\geq 1$, which in turn implies $-1/4<M<0$. For static conical singularities with $-1<M<-1/4$, the computation requires an integral formula instead of a sum.

The expression for the summand in $F(M)$, including the factor $(-M)^{3/2}$ in Eq.\eqref{eq:FNS} for the NS, can be obtained from the corresponding one for the BH in Eq.\eqref{eq:FBH} by analytic continuation, $M\rightarrow -M$. 
However, in the NS case, the images for $n$ and $n-N$ are repeated, whereas in the BH case all images with different $n$ are distinct, which accounts for the different overall factors in $F(M)$ in the two cases. 
Furthermore, for the NS, unlike the BH case, the sum  runs over a finite range and consequently $F$ is manifestly finite and positive. 

\begin{figure}[h!]
\includegraphics[width=10cm]  {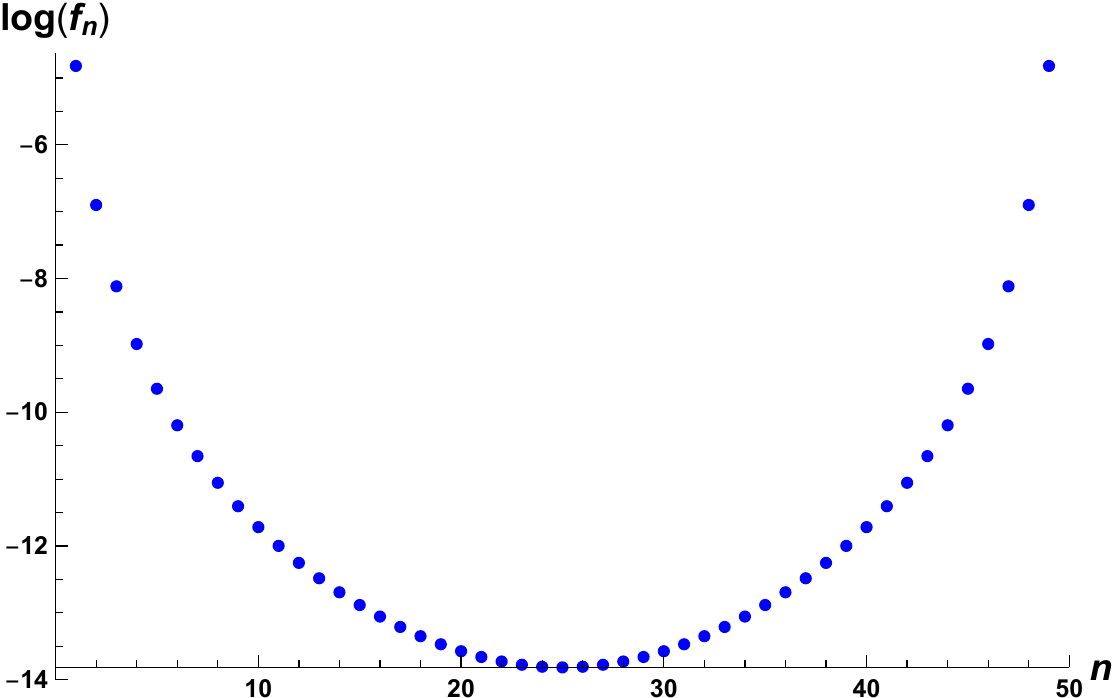}
\caption{Logarithm of the summand $f_n$ in  Eq.\eqref{eq:FNS} as a function of $n$. The plot shows the range $n=1,\dots, N-1=49$ and exhibits the symmetry $n \to N-n$.}
\label{fig:fn} 
\end{figure}

\begin{figure}[h!]
\includegraphics[width=10cm]  {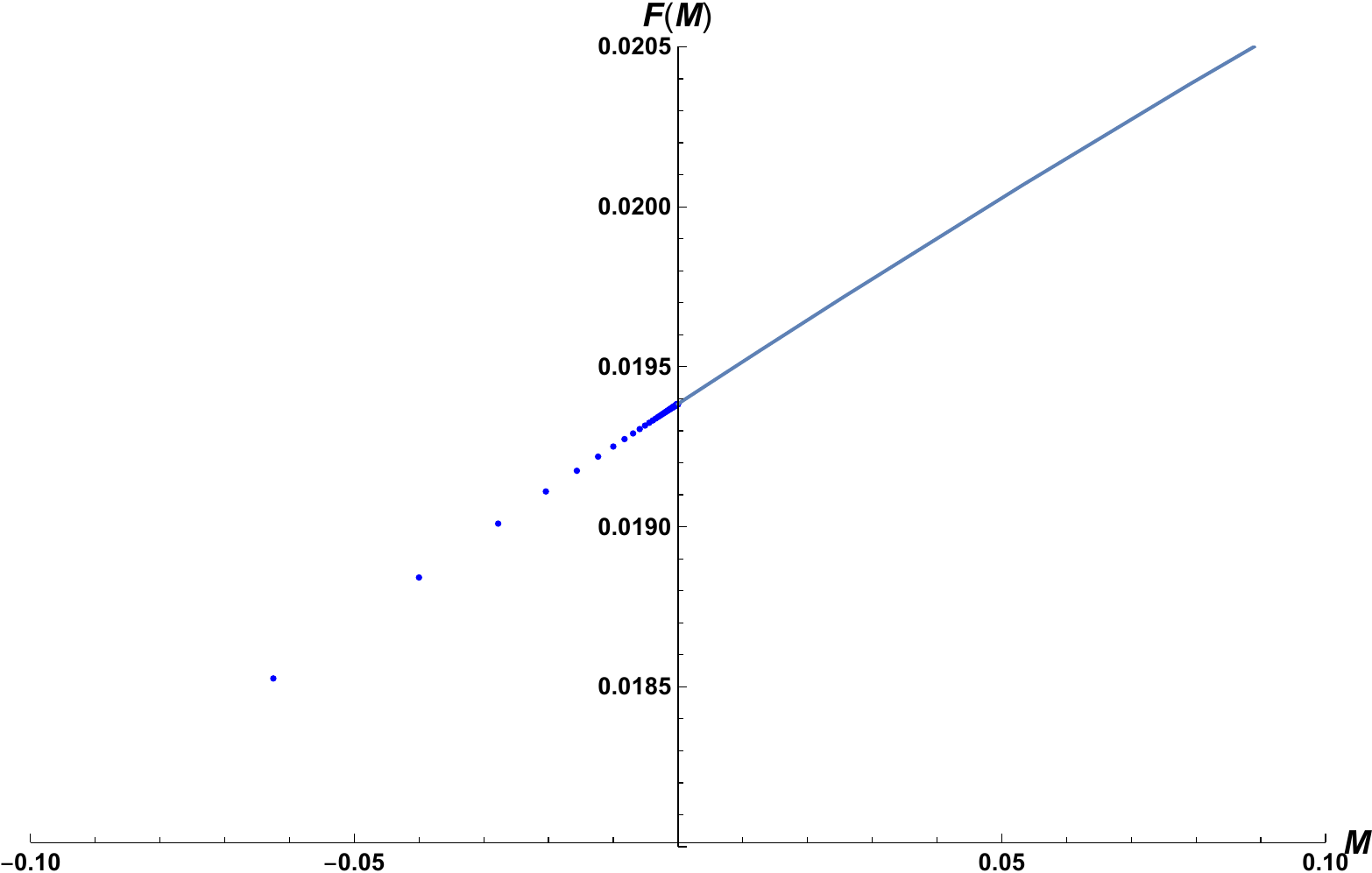}
\caption{The function $F(M)$  as a function of $M$ around $M=0$. For $M<0$, $F(M)$ is given by the finite sum Eq.\eqref{eq:FNS} and it is plotted using  dots. For $M>0$, $F(M)$ is given by the infinite series Eq.\eqref{eq:FBH} and it is represented with a solid line. The figure shows that $F(M)$ and its derivative $dF/dM$ are continuous at $M=0$, which can be also  verified analytically.}
\label{fig:bothsides} 
\end{figure}

The value of $F(0)$ may be obtained  by taking the limit $M=-1/N^2\to 0^-$ (i.e., $N\to \infty$) in Eq.\eqref{eq:FNS}, which is numerically found to be
\begin{equation} \label{FNS(0-)}
F(0^-)=0.0193841 \approx \frac{\zeta(3)}{2\pi^3},
\end{equation}
where $\zeta$ is the Riemann zeta function (see Fig.\ref{fig:FBHNS2}). This value matches the limit $M\to 0^+$ of $F(M)$ in the BH geometry, Eq.\eqref{eq:FBH}, in spite of the fact that there is apparently a mismatch by a factor of two between Eqs.\eqref{eq:FNS} and \eqref{eq:FBH}. The apparent mismatch arises from taking the limit by applying the L'H\^opital rule to the summand and not considering  the symmetry $n \to N-n$ that it has (see Fig.\ref{fig:fn})  -- this symmetry adds a factor of two apparently lost in the sum in Eq.\eqref{eq:FNS}.
The continuity of $F(M)$ and its derivative across $M=0$ is manifest in Fig.\ref{fig:bothsides}.



\vspace{0.5cm}
\textbf{\textit{2B. RSET for the rotating NS}} \\

We consider the case $\beta_{\pm}=2/N_{\pm}$, $N_{\pm}\in\mathbb{Z}^+$, where $N_- > N_+$. The number of distinct images is $N$, the least common multiple of $N_+$ and $N_-$. Then, using Eq.\eqref{eq:RSET S}, the symmetry \eqref{eq:symm summand NS}, as well as the embedding Eq.\eqref{embeddingM-J} and identification matrix in Eq.\eqref{HNS}, we obtain the following components of the RSET for the rotating NS: 
\begin{align}
\kappa \langle T^{t}\,_{t} \rangle=& \frac{ l_{P}}{(\beta_{+}^2-\beta_{-}^2)^{2}} \Sp_{n=1}^{N-1} \frac{2\left(2 r^2 \left(3 {a}_{n}-\left(\beta _+^2-\beta _-^2\right) {b}_{n}\right)+\ell^2 g_n\right) {c}_{n} +3  \beta_{+} \beta_{-}{e}_{n}\left(8 r^2+\beta _-^2 \ell^2+\beta _+^2 \ell^2\right)}{{d}_{n}^{5/2}},
\label{eq:Ttt subext NS}
\\
\kappa\langle T^{r}\,_{r} \rangle=& \frac{l_{P}}{2} \Sp_{n=1}^{N-1} \frac{{c}_{n}}{{d}_{n}^{3/2}},
\\
\kappa \langle T^{\theta}\,_{\theta} \rangle=&- \frac{  l_{P}}{(\beta_{+}^2-\beta_{-}^2)^{2}} \Sp_{n=1}^{N-1} \frac{2\left(2 r^2 \left(3 \bar{a}_{n}+\left(\beta _+^2-\beta _-^2\right) {b}_{n}\right)+\ell^2 \bar{g}_n\right) {c}_{n} +3 \beta_{+} \beta_{-}{e}_{n}\left(8 r^2+\beta _-^2 \ell^2+\beta _+^2 \ell^2\right)}{{d}_{n}^{5/2}},
\\
\kappa\langle T^{t}\,_{\theta} \rangle=& -\frac{3  l_{P}\ell}{ (\beta_{+}^2-\beta_{-}^2)^{2}} \Sp_{n=1}^{N-1} \frac{\left(4 ({c}_{n}-4) r^2-{a}_{n} \ell^2\right){c}_{n}\beta _+\beta _- +{e}_{n} \left(4 r^2\left(\beta _-^2+\beta _+^2\right)+2\beta _-^2 \beta _+^2 \ell^2 \right)}{{d}_{n}^{5/2}},
\\
\kappa\langle T^{\theta}\,_{t} \rangle=&  \frac{3  l_{P}}{ \ell(\beta_{+}^2-\beta_{-}^2)^{2}} 
\Sp_{n=1}^{N-1} \frac{\left(4 ({c}_{n}-4) r^2-{a}_{n}\ell^2\right){c}_{n}\beta _+\beta _-   +{e}_{n}\left(4 r^2\left(\beta _-^2+\beta _+^2\right)+\ell^2(\beta _-^4 +\beta _+^4 )\right)}{{d}_{n}^{5/2}},
\label{eq:Ttht subext NS}
\end{align}
with
\begin{align}
{a}_{n}\equiv &2\beta _+^2 \sin ^2\left(\frac{n\pi  \beta _- }{2}\right)+2\beta _-^2 \sin ^2\left(\frac{n\pi  \beta_+ }{2}\right), \label{anNS}
\\
\bar{a}_{n}\equiv &2\beta _+^2 \sin ^2\left(\frac{n\pi  \beta_+ }{2}\right)+2\beta _-^2 \sin ^2\left(\frac{n\pi  \beta _- }{2}\right), 
\\
{b}_n\equiv &\cos \left(\pi  n \beta_{+} \right)-\cos \left(\pi  n \beta_{-} \right)=2\left( \sin ^2\left(\frac{\pi n \beta _- }{2} \right)-\sin ^2\left(\frac{\pi  n \beta _+ }{2} \right) \right), 
\label{bnNS}
\\
{c}_n\equiv & \cos \left(\pi  n \beta_{+} \right)+\cos \left(\pi  n \beta_{-} \right)+2,
\label{cnNS} \\
{e}_n\equiv & 2 \sin \left(\pi  n \beta_{+} \right) \sin \left(\pi  n \beta_{-} \right),
  \label{enNS}\\
 {g}_n\equiv &\beta _-^2 \left(\beta _+^2+2 \beta _-^2\right) \sin ^2\left(\frac{n\pi  \beta_+ }{2}\right)+\beta _+^2 \left(\beta _-^2+2 \beta _+^2\right) \sin ^2\left(\frac{n\pi  \beta _- }{2}\right),
 \\
\bar{g}_n\equiv &\beta _-^2\left(\beta _-^2+2 \beta _+^2 \right) \sin ^2\left(\frac{n\pi  \beta_+ }{2}\right)+\beta _+^2\left(\beta _+^2+2 \beta _-^2 \right) \sin ^2\left(\frac{n\pi  \beta _- }{2}\right),
\end{align}
and
\begin{equation} \label{dnNS}
d_{n} =2\s(x,H^nx) =4 \ell^2 \frac{\beta_{-}^2 \sin ^2\left(\frac{\pi n \beta _+ }{2} \right)-\beta_{+}^2 \sin ^2\left(\frac{\pi n \beta _- }{2} \right) -2 r^2\ell^{-2} {b}_n}{\beta_{+}^2-\beta_{-}^2}.
\end{equation}

Note that this RSET has the generic form 
\begin{equation} \label{RSETgenNS}
\rset{T^{ \mu}{}_{ \nu}}= \frac{1}{2} \Sp_{n=1}^{N-1} \left[\tau^{ \mu}{}_{ \nu}(r;n,\beta_+,\beta_-)+\tau^{ \mu}{}_{ \nu}(r;-n,\beta_+,\beta_-)\right],
\end{equation}
for some tensor $\tau^{ \mu}{}_{ \nu}$. The components  $\rset{T^{ t}{}_{ r}}$ and  $\rset{T^{ r}{}_{ \theta}}$ vanish because $\tau^{ r}{}_{ t}$ and  $\tau^{ r}{}_{ \theta}$ are antisymmetric under the change $n \to -n$. 
For instance,  the component  $\tau^{ t}{}_{ r}$ given by  $$ \tau^{ t}{}_{ r}=-\frac{48 \ell^3 r z_n \sin \left(\frac{1}{2} \pi  \beta _+ n\right)}{\left(\beta _+^2-\beta _-^2\right) \left(\beta _-^2 \ell^2+4 r^2\right) \left(\beta _+^2 \ell^2+4 r^2\right)d_n^{5/2}}$$ with
$$
z_n\equiv \beta _+(4  r^2+\beta _-^2 \ell^2 ) \sin \left(\pi  \beta _- n\right) \sin \left(\frac{1}{2} \pi  \beta _+ n\right)-2 \beta _- \left(4  r^2+\beta _+^2 \ell^2\right) \sin ^2\left(\frac{1}{2} \pi  \beta _- n\right) \cos \left(\frac{1}{2} \pi  \beta _+ n\right),
$$
is odd under $n  \to -n$. This simplifies the backreaction problem since it is sufficient to consider a solution of the metric semiclassical equations of the stationary form in Eq.\eqref{eq:ansatz} (i.e., with no $g_{tr}$ or $g_{r\theta}$ components).

Note that $b_n>0$ in \eqref{dnNS} implies $d_n<0$ and therefore those terms with $b_n >0 $ do not contribute to the sum defining RSET\footnote{Writing $b_n$ as $2( \sin ^2 y-\sin ^2 x )$ with $y<x$, and imposing $b_n>0$ implies that $\sin ^2 x/x^2-\sin ^2 y/y^2 <0$, from which immediately follows that $d_n$ would be negative.}.
The special case $b_n=0$ would make $d_{n}$ to be independent of $r$, so that the RSET would diverge at radial infinity, thus  leading to a breakdown of the perturbative approximation. It can be seen that $b_n$  vanishes for $n=N$, which is outside the range of the sum $1\le n \le N-1$. In addition, there is a discrete set of pairs of $\beta_{\pm}$ for which this also happens in the range of the sum. This set is given by
\be \label{eq:S}
\mathcal{S}\equiv \{\beta_{\pm} \mid n(\beta_+\pm \beta_-)= 2 k, k\in \mathbb{Z}\}
\ee
and it must be removed from the analysis. For example, the case $\beta_+=2/3$ and $\beta_-=1/3$, yielding $b_2=b_4=0$, belongs to $\mathcal{S}$.

For $b_n < 0$, $d_n$ grows as $r^2$ for sufficiently large $r$ and the above RSET components go as $r^{-3}$ at infinity, which is the same behavior as the RSET in the static case and as the classical stress-energy tensor. 

Finally, and similarly to the BH case, $d_n$ in Eq.\eqref{dnNS} vanishes at some radii $r_{n}$ given by
\begin{equation} \label{r02}
r_{n}^2\equiv \frac{\ell^2}{2 {b}_n} \left[\beta _-^2 \sin ^2\left(\frac{1}{2} \pi n \beta _+ \right)-\beta _+^2 \sin ^2\left(\frac{1}{2} \pi  n \beta _- \right)\right] >0,
\end{equation}
for some $n$. Since $b_n<0$, the numerator of \eqref{r02} must be negative in order for $r_n$ to be real valued.
At each of these zeroes, the RSET blows up and, therefore, the Kretschmann invariant \eqref{eq:Kretschmann} diverges, signaling curvature singularities.

Let us now examine under which conditions one can make sure that $b_{n}<0$ for some $n$ in order for the sum in the RSET to be nonvanishing. For $\beta_-=0$ ($J=0$), ${b}_n$ is negative for all $n$, and since ${b}_n$ is a continuous function of $\beta_{\pm}$, it should still be negative for some range $\beta_-\neq 0$ ($|J|>0$). 

\begin{figure}[h!]
\begin{center}
  \includegraphics[width=12cm]  {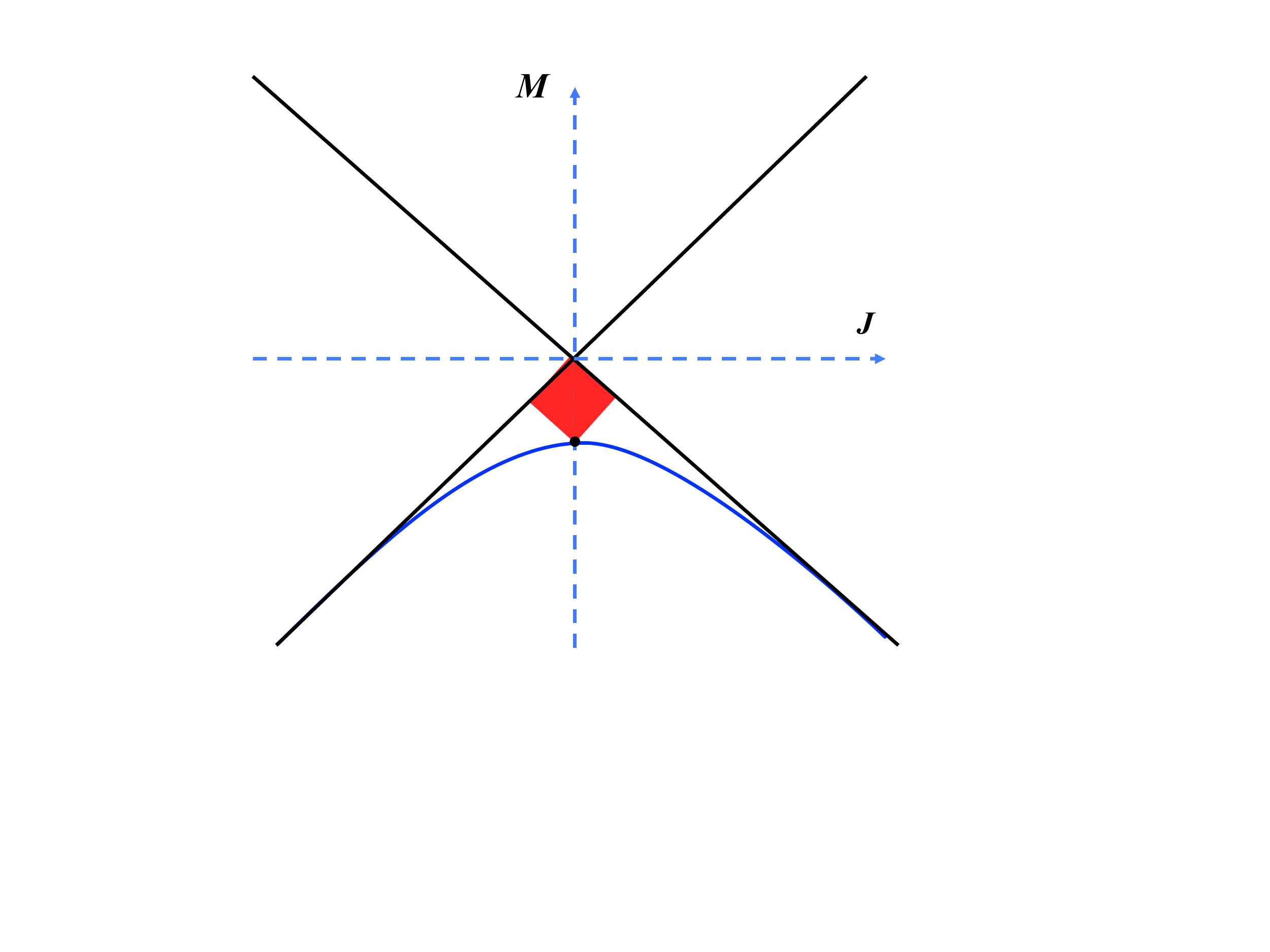}
  \end{center}
\caption{Region $0<\beta_+ + \beta_- <2$ (see Eq.\eqref{range}) in the $M$-$J$ plane corresponds to the central square between the zero mass state ($M=0$, $J=0$) and anti-de Sitter ($M=-1$, $J=0$). In this region $b_1<0$, which guarantees that the RSET contains at least one nonvanishing term in the sum ($n=1$).}
\label{fig:Beta+-} 
\end{figure}

Since $\beta_{\pm}=2/N_{\pm}$, then $0<\beta_+ \pm \beta_- \leq 3$. The largest possible $\beta_{+}=2$ yields $N=|N_-|$ and $b_n \geq 0$ for all $n\in\{1, \cdots,N_- -1\}$, hence this case is excluded. Therefore the only allowed values for $\beta_{\pm}$ are contained in the domain
\begin{equation} \label{range}
0<\beta_+ \pm \beta_- < 2.
\end{equation} 
The region covered by this condition  corresponds to NSs in the square region  $J>M >J -1$ and  $-J>M >-J -1$ as shown in Fig.\ref{fig:Beta+-}. This region includes all the static NSs with masses in the range $0<M<-1$. One can now observe that since $\sin^2x$ grows monotonically for small $x$ and $\beta_+ \geq \beta_-$, at least for $n=1$, ${b}_1<0$, and therefore the sum for RSET contains always the first term. It is also easy to see that, in that same domain, 
the factor in brackets in Eq.\eqref{r02}
is negative for $n=1$,
which renders $r^2_{1}>0$.

\section{Solution of the semiclassical equations}\label{sec:semicl sln}

In this section we solve
analytically
the semiclassical Einstein equations:
\begin{equation} \label{sEFE}
G_{\mu \nu}-\ell^{-2}g_{\mu \nu}=\kappa \langle T_{\mu \nu} \rangle.
\end{equation}
Here, the RSET 
is calculated on a classical BTZ background space-time 
(such as the one in the previous section when using transparent boundary conditions)
and the solution
$g_{\mu \nu}$ corresponds to the quantum-backreacted geometry (that is, $g_{\mu \nu}$ in 
Eq.\eqref{sEFE} is {\it not} the classical BTZ background).

We provide details of the integration 
differentiating between the non-rotating and rotating cases. For the static case, this section contains a review of existing results in the literature \cite{Lifschytz:1993eb, MZ2, CFMZ1} and new observations about the RSET conservation under different boundary conditions. In the rotating case we include a thorough description of the results briefly announced in~\cite{CFMZ2}.

\subsection{Static geometries}

Let us consider a general form for a static and circularly-symmetric three-dimensional line element:
\begin{equation} \label{sa}
ds^2=-A(r)dt^2+\frac{dr^2}{B(r)}+ r^2 d \theta^2.
\end{equation}
The functions $A(r)$ and $B(r)$ are determined so that this metric is a solution of the semiclassical Eq.\eqref{sEFE} with a static RSET of the form $\langle T^{\mu}{}_{\nu}\rangle= \mbox{\textrm{diag}}( \langle T^{t}{}_{t}(r)\rangle,\langle T^{r}{}_{r}(r)\rangle,\langle T^{\theta}{}_{\theta}(r)\rangle)$ as a source. In particular, the RSETs of Eqs.\eqref{eq:RSET BH'} and \eqref{eq:RSET static BH} have this diagonal form; also, the
static BTZ BH  geometries, Eq.\eqref{BTZ} or \eqref{NSline} with $J=0$, have the form in Eq.\eqref{sa}.

The semiclassical Einstein equations containing the above RSET as a source reduce to
\begin{align}
\frac{B'}{2 r}-\frac{1}{\ell^2}&=\kappa  \langle T^{t}{}_{t}\rangle, \label{e1}\\
\frac{B A'}{2 r A}-\frac{1}{\ell^2}&=\kappa  \langle T^{r}{}_{r}\rangle, \label{e2}\\
\frac{ A A' B'-B \left(A'^2-2A A'' \right)}{4A^2 }-\frac{1}{\ell^2}&=\kappa  \langle T^{\theta}{}_{\theta}\rangle, \label{e3}
\end{align}
where a prime on a function denotes derivative with respect to its argument.
Using Eqs.\eqref{e1} and \eqref{e2}, Eq.\eqref{e3} becomes
\begin{equation} \label{e3a}
\langle T^{r}{}_{r}\rangle'+\frac{A'}{2 A}\left(\langle T^{r}{}_{r}\rangle-\langle T^{t}{}_{t}\rangle\right)+\frac{1}{r} \left(\langle T^{r}{}_{r}\rangle-\langle T^{\theta}{}_{\theta}\rangle\right)=0.
\end{equation}
In its turn, the  only nonvanishing  component of $\nabla_{\mu} \langle T^{\mu}{}_{\nu}\rangle $ is
\begin{equation} \label{conser}
\nabla_{\mu} \langle T^{\mu}{}_{r} \rangle=\langle T^{r}{}_{r}\rangle'+\frac{A'}{2 A}(\langle T^{r}{}_{r}\rangle-\langle T^{t}{}_{t}\rangle) +\frac{1}{r} \left(\langle T^{r}{}_{r}\rangle-\langle T^{\theta}{}_{\theta}\rangle\right).
\end{equation}
We note that  Eq.\eqref{e3a} is equivalent to $\nabla_{\mu} \langle T^{\mu}{}_{r} \rangle=0$. As expected, once the three field equations are satisfied, the conservation of $ \langle T^{\mu}{}_{\nu} \rangle$ holds. 

\vspace{0.5cm}
\textbf{1. Dirichlet and Neumann boundary conditions} \\

The RSET that we gave in Sec.\ref{sec:RSET} was for a conformal scalar field satisfying transparent boundary conditions.
From \cite{Lifschytz:1993eb}, one can show that the  RSET components computed using Dirichlet and Neumann boundary conditions
for a conformal scalar field
on the BTZ BH background satisfy the following relation
\begin{equation} \label{rel}
\langle T^{r}{}_{r}\rangle'+\frac{r}{r^2-r_{+}^2}(\langle T^{r}{}_{r}\rangle-\langle T^{t}{}_{t}\rangle) +\frac{1}{r} \left(\langle T^{r}{}_{r}\rangle-\langle T^{\theta}{}_{\theta}\rangle\right) = 0.
\end{equation}
Comparing the above expression with Eq.\eqref{conser}, it is noted that the conservation of the RSET  is guaranteed if 
\begin{equation} \label{Acondition}
\frac{A'}{2 A}=\frac{r}{r^2-r_{+}^2}.
\end{equation}
This condition is exactly verified by the static BTZ BH  geometry, Eq.\eqref{BTZ} with $J=0$. This means that the RSET
for a field satisfying
Dirichlet or Neumann boundary conditions
is conserved on the BTZ BH background. 
\MC{Phrase it better?}
If, on the other hand, $A$ were such that it did not satisfy Eq.\eqref{Acondition}, then the RSET would not be conserved. In that case, the integrability condition for Eqs. \eqref{e1}-\eqref{e3} would not be fulfilled. However, if $A$ satisfied $A\propto (r^2-r_{+}^2)+O(l_P)$, then Eq. \eqref{rel} would be satisfied at order $l_P$. Then the semiclassical equations \eqref{e1}-\eqref{e3} for a  RSET for a field satisfying  Dirichlet or Neumann boundary conditions would only be compatible at linear order in  $l_P$.

\vspace{0.5cm}
\textbf{2. Transparent boundary conditions}
\vspace{0.5cm}

The components of the RSET for a conformal scalar field satisfying transparent boundary conditions on the BTZ background geometries, for either BH (Eq.\eqref{eq:RSET static BH}) or NS (Eq.\eqref{eq:RSET BH'}), satisfy the algebraic relations 
\begin{equation}\label{eq:props diag RSET}
\langle T^{t}{}_{t}\rangle=\langle T^{r}{}_{r}\rangle  \quad  \mbox{and} \quad \langle T^{t}{}_{t}\rangle +\langle T^{r}{}_{r}\rangle+\langle T^{\theta}{}_{\theta}\rangle =0,
\end{equation}
which imply $\langle T^{r}{}_{r}\rangle-\langle T^{\theta}{}_{\theta}\rangle = 3 \langle T^{r}{}_{r}\rangle$. In this case, Eq.\eqref{conser} reduces to
\begin{equation} \label{e3tbc}
\nabla_{\mu} \langle T^{\mu}{}_{r} \rangle= \langle T^{r}{}_{r}\rangle' + \frac{3}{r} \langle T^{r}{}_{r}\rangle,
\end{equation}
whose right hand side vanishes since $\langle T^{r}{}_{r}\rangle$ is proportional to $r^{-3}$ (see Eqs.\eqref{eq:RSET static BH} and \eqref{eq:RSET BH'}).
Note that the term with $ A'/A$ in Eq.\eqref{conser} is absent in Eq.\eqref{e3tbc} because $\langle T^{t}{}_{t}\rangle=\langle T^{r}{}_{r}\rangle$.
This shows that a  RSET calculated on a fixed background space-time and which is of the form 
\begin{equation}\label{eq:RSET form}
 \langle T^{\mu}{}_{\nu}\rangle=\frac{\mbox{constant}}{r^3}\,\textrm{diag}(1,1,-2),
\end{equation}
in $\{t,r,\theta\}$ coordinates is conserved on the general static metric in Eq.\eqref{sa}  and so fulfills the integrability condition for  Eqs. \eqref{e1}-\eqref{e3}. In particular, the form \eqref{eq:RSET form}
is satisfied by the RSET for a field  with transparent boundary conditions on a BTZ background space-time.

Because Eq.\eqref{e3} is satisfied by virtue of \eqref{e3tbc}, it is only necessary to solve  Eqs.\eqref{e1} and \eqref{e2}.    Subtracting Eqs.\eqref{e1} and \eqref{e2},
and using Eq.\eqref{eq:props diag RSET},
we obtain
\begin{equation}
\frac{A'}{A}=\frac{B'}{B}.
\end{equation}
Thus $A=B$ (up to a constant which can be taken equal to 1) and, from Eq.\eqref{e1},  we obtain 
\begin{equation} \label{exactsolutiontbc}
A=B= \frac{r^2}{\ell^2} -c_0 +2 \kappa \int r \langle T^{t}{}_{t}\rangle dr,
\end{equation}
where $c_0$ is an integration constant. 

Note that for the transparent boundary conditions and in the coordinates of Eq. \eqref{sa}, the exact solution given by Eq.\eqref{exactsolutiontbc} of the semiclassical equations \eqref{e1}-\eqref{e3} is a linear function of the source. Thus, if $c_0$ is chosen to be the mass $M$ of the static BTZ geometries, then
the exact (i.e., without expanding for small $l_P$) solution for the metric coefficients $A$ and $B$
coincides with the solution one would obtain if expanding $A$ and $B$ to linear order in $l_P$ around a BTZ static metric.


\vspace{0.5cm}
\textbf{Black hole}
\vspace{0.5cm}

Let us first briefly review the static  ($J=0$) BTZ BH case, which was analyzed in~\cite{MZ2} considering tranparent boundary conditions.  Using Eq.\eqref{eq:RSET static BH}, the integral appearing in Eq.\eqref{exactsolutiontbc} becomes
\begin{equation}
2 \kappa \int r \langle T^{t}{}_{t}\rangle dr=-\frac{2l_P F(M)}{r},
\end{equation}
where $F(M)$ is given in Eq.\eqref{eq:FBH}. The backreacted metric, as given by Eqs.\eqref{sa} and \eqref{exactsolutiontbc}, is then
\be \label{eq:backreacted static BH}
ds^2=-\left(\frac{r^2}{\ell^2}-c_0-\frac{2l_P F(M)}{r}\right)dt^2 + \frac{dr^2}{\displaystyle \left(\frac{r^2}{\ell^2}-c_0 -\frac{2 l_P F(M)}{r}\right)}+r^2d\theta^2
\ .\ee

\textbf{Naked singularity} \\

In the static NS case, the RSET in Eq.(\ref{eq:RSET BH'}) has the same structure as for the static BH case. Therefore, the quantum-backreacted metric has the same form as in Eq.\eqref{eq:backreacted static BH}, but now $F(M)$ is instead given by the finite sum  \eqref{eq:FNS}.

\subsection{Rotating geometries}

In this section we set as a source of the Einstein semiclassical equations \eqref{sEFE} the RSET corresponding to a conformally coupled scalar field on the rotating BTZ background geometries. In order to solve these backreaction equations we consider a general stationary and circularly-symmetric three-dimensional line element: 
\begin{equation} \label{eq:ansatz}
ds^2=-N(r)^{2} f(r)dt^2+\frac{dr^2}{f(r)}+ r^2 \left( d\theta+ k(r) dt \right)^2,
\end{equation} 
for some functions $N(r)$, $f(r)$ and shift function $k(r)$.
We are interested in finding the linear corrections in $l_P$ to the rotating BTZ geometries. For this purpose, we write the metric functions explicitly up to order $O(l_p)$ as
\begin{equation} \label{lpcorrections}
N(r)=N_0(r)+l_{P} N_1(r)+O(l_P^2),  \quad f(r)=f_0(r)+l_{P} f_1(r)+O(l_P^2), \quad k(r)=k_0(r)+l_{P} k_1(r)+O(l_P^2),
\end{equation}
where the functions labeled with a subindex $0$ are the background metric coefficients and those with subindex $1$ correspond to their first-order backreaction corrections in $l_P$. 

The zero-th order field equations provide the equations for the background functions:
\begin{equation} \label{diffeqs}
N_0'=0, \quad k_0''+\frac{3}{r}k_0'=0, \quad f_0''+\frac{3}{r}f_0'=\frac{8}{\ell^2} \quad f_0'+\frac{r^3 k_0'^2}{2 N_0^2}= \frac{2 r }{ \ell^2},
\end{equation}
where a prime means derivative with respect to their argument, $r$. Thus, it is $N_0(r)=\mbox{constant}$, which is taken to be 1. In its turn, the first integral of the equation for $k_0$ gives $r^3 k_0'=J=\mbox{constant}$, so that
\begin{equation}
k_0=-\frac{J}{2 r^2}+k_0(\infty).
\end{equation}
We choose $k_0(\infty)=0$ in order to describe the BTZ geometries in a coordinate frame such that the shift function at infinity vanishes. 
Thus, we have
\begin{equation}\label{eq:N0,k0}
N_0=1, \quad k_0=-\frac{J}{2 r^2}.
\end{equation}
Moreover, from Eq.\eqref{diffeqs} we have
\begin{equation}
f_0'= \frac{2 r }{ \ell^2}-\frac{J^2}{2 r^3},
\end{equation}
and hence,
\begin{equation} \label{b0}
f_0= \frac{ r^2 }{ \ell^2}-M+\frac{J^2}{4 r^2}, 
\end{equation}
where $M$ is a constant of integration.
Eq.\eqref{b0} is the usual expression for the lapse function
of the BTZ geometries with mass $M$ and angular momentum $J$.

The next order of the field equations provides linear differential equations for $N_1(r)$, $k_1(r)$ and $f_1(r)$. 
Explicitly, the $O(l_P)$ semiclassical Einstein equations \eqref{sEFE} read
\begin{align} 
16\rt^6 \frac{\kappa}{l_p} \langle T_{\tit\tit}\rangle &=f_0 \left(r^5 \left(16 J  k_1'-8  f_1'\right)+4 J^2 r^4 N_1''-8 J^2 r^3 N_1'+8 J^2  r^2 N_1+8 J  r^6 k_1''\right)\nonumber \\&+2 J^2 r^4 f_1''-3 J^4r N_1'+6 J^4  N_1-6 J^3  r^3 k_1'+\frac{12}{\ell^2} J^2 r^5 N_1', \label{tt}  \\
 8 \frac{\kappa}{l_p}\langle T_{\tit\tht}\rangle &= -4f_0 \left( J \left(N_1''-\frac{  N_1'}{r}\right)+ r^2 k_1''+3 r k_1'\right) -2 J f_1''+3 J \left(\frac{ J^2 }{r^3}-\frac{4 r }{\ell^2}\right)N_1'-\frac{6 J^3 N_1}{r^4}+\frac{6 J^2 k_1'}{r}, \label{tth} \\
 \frac{\kappa}{l_p} \langle T_{\rt\rt}\rangle &= \frac{r^3 \left(f_1'+J k_1'\right)-J^2 N_1}{2 r^4 f_0}+\frac{N_1'}{r}, \label{rr} \\
 4 \frac{\kappa}{l_p} \langle T_{\tht \tht }\rangle &= 4 r^2 f_0 N_1''+2 r^2 f_1''+\left(\frac{12 r^3}{\ell^2}-\frac{3 J^2}{r}\right) N_1'+\frac{6 J^2 N_1}{r^2}-6 J r k_1'. \label{thth} 
\end{align}
We first isolate the relevant second derivatives appearing in these equations:
\be  \rt^4J k_1''=  (-\rt^3Jk_1' +2\rt^3f_1' +J^2\rt N_1'-2J^2 N_1) + \frac{1}{ f_0}\left(2 J r^2 \frac{\kappa}{l_p} \langle T_{\tit\tht}\rangle + 4 r^4 \frac{\kappa}{l_p} \langle T_{\tit\tit}\rangle \right)\ ,\ee
\be  
4 r^2 f_0 N_1''+2 r^2 f_1'' =4 \frac{\kappa}{l_p} \langle T_{\tht \tht }\rangle -\left(\frac{12 r^3}{\ell^2}+\frac{3 J^2}{r}\right) N_1'-\frac{6 J^2 N_1}{r^2}+6 J r k_1'.
 \ee
 
Substituting these equations into Eq.\eqref{tt}  we obtain
\be  \label{gurb}
\mathcal{A}\equiv
4J^2\rt^2 \frac{\kappa}{l_p} \langle T_{\tht \tht }\rangle +16J \rt^4 \frac{\kappa}{l_p} \langle T_{\tit\tht}\rangle + 16\rt^6\frac{\kappa}{l_p}  \langle T_{\tit\tit}\rangle=-8 r^2 f_0(\rt^3(Jk_1' + f_1') -J^2 N_1)\ , 
\ee  
which combined with Eq.\eqref{rr}  gives 
\be  \label{z} 
N_1' = \rt \frac{\kappa}{l_p}  \langle T_{\rt\rt}\rangle  + \frac{\mathcal{A}}{16\rt^5f_0^2}\ .
\ee
This last equation determines $N_1$. Since the RSET is traceless, we obtain
\begin{equation} \label{TR}
  \langle T_{\tit\tit}\rangle=f_0^2\langle T_{r r}\rangle+\left(\frac{1 }{ \ell^2}-\frac{M}{r^2}\right) \langle T_{\tht \tht }\rangle-\frac{J}{r^2} \langle T_{t \tht }\rangle,
\end{equation}
so that $\mathcal{A}$ becomes
\begin{equation}
\mathcal{A}=  16\rt^6\frac{\kappa}{l_p} f_0 \left( f_0 \langle T_{r r}\rangle +\frac{1}{r^2} \langle T_{\tht \tht }\rangle \right).
\end{equation}
We now use the combination of Eqs.\eqref{rr} and \eqref{thth} that eliminates $k_1'$, together with  
Eqs. \eqref{TR} and (\ref{z}), and obtain
\be  \label{b12p}
(\rt^3 f_1')' = -\frac{\mathcal{A}}{8\rt^3 f_0}-2\rt^2N_1'\left(\frac{4\rt^2}{\ell^2}-M-\frac{J^2}{2\rt^2}\right) -2\rt^3 f_0 N_1''.  \ee
This equation allows to solve for $f_1$. Finally, the equation determining $k_1$ comes from  Eq.\eqref{gurb}:
\be \label{yy} Jk_1' = -f_1' + \frac{J^2 N_1}{\rt^3} - \frac{
\mathcal{A}
}{8\rt^5f_0}.   
\ee
In this way, we have decoupled the semiclassical Einstein equations at the linear approximation in $l_P$. We note that the components of the RSET  satisfy the integrabity condition of these equations,
\begin{align}
\frac{d \langle T_{r r} \rangle}{dr}&=
  \left(\frac{J^2 \ell^2-2 \ell^2 r^2 f_0-4 r^4}{\ell^2 r^3 f_0}\right)\langle T_{r r} \rangle +\frac{\ell^2 \langle T_{t t} \rangle-\langle T_{\theta \theta} \rangle}{\ell^2 r f_0^2},
\end{align}
which corresponds to the  covariant conservation (at first order in $l_P$) of the RSET.
 
The integration of Eqs.\eqref{z}, \eqref{b12p} and \eqref{yy} give
\begin{align}
N_1(r)&=\frac{\kappa}{l_P} \int dr  \left(2 r \langle T_{r r} \rangle  +\frac{\langle T_{\theta \theta} \rangle }{r f_0(r)} \right) + c_{1},
\label{eq:N1}
\\
f_1(r)&= \int dr \left[ -2 f_0(r) N_1'(r)+ N_1(r)\left(-\frac{2M}{r}+\frac{J^2}{r^3} \right) \right.  \nonumber \\ 
& \left.+\frac{2}{r^3}\int dr \left( 2 M r N_1(r) +\frac{ \kappa}{ l_P} r^3 f_0(r)\langle T_{r r} \rangle \right)  \right]  +\frac{c_2}{r^2} +c_3,  \\
J k_1(r)&=-f_1(r)-2 f_0(r) N_1(r) +2 \int r dr \left( \frac{2 N_1(r)}{\ell^2} +f_0(r) l_P^{-1} \kappa \langle T_{r r} \rangle \right) + c_4.
\label{eq:k1}
\end{align}
The integration constants $c_1, c_2, c_3$ and $c_4$ are set to zero 
so that $N_1$, $f_1$ and
$k_1$ vanish for vanishing RSET.
Note that $f_1$ and $k_1$ are determined by $N_1$ and $\langle T^\rt_\rt \rangle=f_0(r)\langle T_{rr}\rangle$. 

We next give the explicit form of the semiclassical corrections for the different BTZ backgrounds.

\vspace{0.5cm}
\textbf{1. Semiclassical corrections to the nonextremal black hole} 
\vspace{0.5cm}

The backreaction corrections to the nonextremal rotating BTZ BH are determined using Eqs.\eqref{eq:RSET subext BH tt}--\eqref{eq:RSET subext BH tht} together with Eqs.\eqref{eq:N1}--\eqref{eq:k1}, yielding
\begin{align}
N_{1}(r)=&  \frac{\ell^2}{\left(\alpha_{+}^2-\alpha_{-}^2\right) }
\Sp_{n=1}^{\infty}
\frac{a_n c_n  -2 \alpha_{-} \alpha_{+} e_n}{ b_n d_n^{3/2}},\label{a1bhnext}
\\
f_1(r)=&-
\frac{1}{32r^2 \left(\alpha_{+}^2-\alpha_{-}^2\right)}
\Sp_{n=1}^{\infty}
\frac{4h_n (a_n c_n-2 \alpha_{-}\alpha_{+}e_n)+c_n d_n^2\left(\alpha_{+}^2-\alpha_{-}^2\right)^3}{  b_n^2 d_n^{3/2} },
\label{b1bhnext}  \\
k_1(r)=&\frac{\ell}{4 r^2 }
\Sp_{n=1}^{\infty}
\frac{\left(\alpha_{-}^2+\alpha_{+}^2\right)e_n-\alpha_{-} \alpha_{+} \left(c_n-4 \right)c_n}{b_n^2 d_n^{1/2} }, \label{k1bhnext}
\end{align}
where $a_n$, $b_n$, $c_n$, $e_n$ and $d_n$ are given in Eqs.\eqref{anBH}, \eqref{bnBH}, \eqref{cnBH},
\eqref{enBH}
and \eqref{dnBH}, respectively,  with
\begin{align}
h_n\equiv &\left(4 r^2-\ell^2\alpha_{+}^2\right) \left(4 r^2-\ell^2\alpha_{-}^2\right) b_n+ \left(\alpha_{+}^2-\alpha_{-}^2\right) \left(4 r^2-\frac{(\alpha_{+}^2+\alpha_{-}^2)\ell^2}{2}\right)d_n. \label{hnBH}
\end{align}

For plots of the (equivalent of the) functions $N_1(r)$, $f_1(r)$ and $k_1(r)$ in
the backreacted metric, we refer
the reader to~\cite{casals2018quantum}.

\vspace{0.5cm}
\textbf{2. Semiclassical corrections to the extremal black hole} \\
\vspace{0.5cm}

Using Eqs.\eqref{eq:RSET ext BH tt}--\eqref{eq:RSET ext BH tht} together with Eqs.\eqref{eq:N1}--\eqref{eq:k1} we obtain the semiclassical corrections to the extremal BTZ BH case:
\begin{align}
 N_1(r)=&
 \Sp_{n=1}^{\infty}
 \frac{  A_n  }{2 \pi n  \alpha  \sinh \left(2 \pi n \alpha \right)d_n ^{3/2}},
\label{eq:N1 ext BH}
\\
f_1(r)=& -\ell^4
\Sp_{n=1}^{\infty}
\frac{\left(\frac{r^2}{\ell^2}-\alpha ^2\right)^2 B_n+2 \alpha  \left(\frac{r^2}{\ell^2}-\alpha ^2\right) \sinh (2 \pi  \alpha  n)C_n +D_n}{r^2 \pi ^2 n^2 \alpha    \sinh (2 \pi  \alpha  n)d_n^{3/2}}, 
\label{eq:f1 BH ext}
\\
 k_1(r)=&-\gamma \ell
 \Sp_{n=1}^{\infty}
 \frac{  2 \pi ^2\alpha^2 n^2 \left(\cosh^2 \left(2 \pi n \alpha\right)+1\right) -\sinh^2 \left(2 \pi n \alpha\right) }{2 r^2 \pi ^2   n^2 \sinh^2 \left(2 \pi n \alpha\right)d_n^{1/2}},
 \label{eq:k1 ext BH}
\end{align}
where
\begin{align}
A_n&\equiv \ell^2 \left( \sinh^2 \left(2 \pi n \alpha\right)+2 \pi ^2 n^2\alpha^2  \left(\cosh^2 \left(2 \pi n \alpha\right)+1\right)  -2 \pi n \alpha  \left(\cosh\left(2 \pi n \alpha\right)+1\right) \sinh \left(2 \pi n \alpha\right) \right), \\
B_n&\equiv 6 n^3\pi ^3 \alpha ^2  \left(\cosh ^2(2 \pi  \alpha  n)+1\right)-4 n^2 \pi ^2 \alpha   \sinh (2 \pi  \alpha  n) \cosh ^2(\pi  \alpha  n)+3 n \pi   \sinh ^2(2 \pi  \alpha  n), \\
C_n &\equiv n^2\pi ^2 \alpha ^2  \left(\cosh ^2(2 \pi  \alpha  n)+1\right) \textrm{sech}^2(\pi  \alpha  n)+2 n \pi  \alpha   \sinh (2 \pi  \alpha  n)+2 \sinh ^2(\pi  \alpha  n), \\
D_n & \equiv 8 \alpha ^3 \sinh ^2(\pi  \alpha  n) \sinh (2 \pi  \alpha  n),
\end{align}
with $d_n$ given in Eq.\eqref{eq:dn ext BH}. 
We remind the reader that $\alpha = r_+/\ell =\sqrt{M/2}>0$ and the angular momentum $J= \gamma M \ell$ with $\gamma=\pm 1$.

\vspace{0.5cm}
\textbf{3. Semiclassical corrections to the nonextremal naked singularity} 
\vspace{0.5cm}

In the case of the nonextremal NS, using Eqs.\eqref{eq:Ttt subext NS}--\eqref{eq:Ttht subext NS} together with Eqs.\eqref{eq:N1}--\eqref{eq:k1}, the following backreaction corrections are found:
\begin{align}
N_{1}(r)=&  -\frac{\ell^2}{2\left(\beta_{+}^2-\beta_{-}^2\right) }
\Sp_{n=1}^{N-1}
\frac{a_n c_n-2\beta_{+} \beta_{-}  e_n}{b_n d_n^{3/2}},
\label{eq:N1 subext NS}
\\
f_1(r)=&
\Sp_{n=1}^{N-1}
\frac{4h_n (a_n c_n-2 \beta_{+} \beta_{-} e_n)-c_n d_n^2 \left(\beta_{+}^2-\beta_{-}^2\right)^3}{64 r^2 \left(\beta_{+}^2-\beta_{-}^2\right)  b_n^2  d_n^{3/2}},
\label{eq:f1 subext NS}
\\
k_1(r)=&-\frac{\ell}{ 8 r^2 }
\Sp_{n=1}^{N-1}
\frac{\left(\beta_{-}^2+\beta_{+}^2\right)e_n+\beta_{+}  \beta_{-}  \left(c_n-4 \right)c_n}{b_n^2  d_n^{1/2}},
\label{eq:k1 subext NS}
\end{align}
where $a_n$, $b_n$, $c_n$, $e_n$ and $d_n$ are given in Eqs.\eqref{anNS}, \eqref{bnNS}, \eqref{cnNS}, \eqref{enNS}
and \eqref{dnNS}, respectively,  with
\begin{align}
h_n=&\left(4 r^2+\ell^2\beta_{+}^2\right) \left(4 r^2+\ell^2\beta_{-}^2\right) b_n- \left(\beta_{+}^2-\beta_{-}^2\right) \left(4 r^2+\frac{(\beta_{+}^2+\beta_{-}^2)\ell^2}{2}\right)d_n.
\label{eq:hn subext NS}
\end{align}

The integrals involving the RSET components were computed assuming $b_n \neq 0$, which is indeed the case.
 
Aside from the limits in the sums, the above expressions 
\eqref{eq:N1 subext NS}--\eqref{eq:k1 subext NS}
are equal to the nonextremal rotating BH corrections 
\eqref{a1bhnext}--\eqref{k1bhnext}
by means of the replacements $\alpha_{\pm}^2 \to -\beta_{\pm}^2,  \cosh \left(\pi n \alpha_{\pm} \right) \to \cos \left(\pi  n \beta_{\pm} \right)$, and $\alpha_{+} \alpha_{-}\sinh \left(\pi  n \alpha_{+} \right) \sinh \left(\pi  n \alpha_{-} \right)  \to  \beta_{+} \beta_{-}\sin \left(\pi  n \beta_{+} \right) \sin \left(\pi  n \beta_{-} \right)$.

\section{Analysis of the semiclassical-backreacted geometries}\label{sec:analysis}

In this section we shall investigate the physical properties of the geometries given by the semiclassical-backreacted  metrics which we have obtained in the previous section. As usual, we shall split this investigation between the different background space-times.

\subsection{Static black hole}

As shown in \cite{MZ2} for the static BH case, by setting $c_0$ in Eq.\eqref{eq:backreacted static BH}
to be equal to the mass $M>0$ of the background BH, the quantum corrections lead to a growth of order $l_P$ of the event horizon and to the formation of a curvature singularity at $r=0$. 

\subsection{Static naked singularity}

The static solution of semiclassical Einstein equation given in \eqref{exactsolutiontbc} has an arbitrary integration constant $c_0$ whose choice corresponds to the freedom of describing different physical setups. 
The analysis of the space-time structure, performed in \cite{CFMZ1} considering $c_0=M \le 0$ in Eq.\eqref{exactsolutiontbc} corresponds to the study of quantum corrections on the classical conical singularities of mass $M$. For finite $M$ a horizon forms at  the radius
\be
r^{(q)}_+(l_P\to 0) =  \frac{2F(M)}{-M} l_P + O(l_P^2),
\ee
while for $M\to 0^{-}$, the horizon is at 
\be \label{smallmass}
 r^{(q)}_+(M\to  0^{-}) =  \left[2\frac{l_p}{\ell}F(0) \right]^{1/3}\ell + O(M).
\ee
This horizon hides a curvature singularity inside (at $r=0$). In the background space-time, the (naked) singularity was a  causal singularity and of timelike character. On the other hand, in the backreacted space-time, the (horizon-hidden) singularity  is a curvature singularity and of spacelike character (as in Schwarzschild space-time).

The fact that the backreacted metric corresponds to a BH prompts the question of whether there is a classical solution of Einstein equations that corresponds to this metric. We examine this possibility by choosing $c_0= 3 (F(M)l_P/\ell )^{2/3}$ in Eq.\eqref{exactsolutiontbc}, so as to match the BH classical solution \eqref{ds}, which exhibits an event horizon with radius 
\begin{equation}\label{eq:r+ static NS}
r_+= 2 \ell \left(\frac{c_0}{3}\right)^{1/2}=2 \ell \left( \frac{l_p}{\ell} F(M) \right)^{1/3}.
\end{equation}
This horizon is of the same order in $l_P$ as the result in Eq.\eqref{smallmass}. This classical solution for the metric extremizes the one-loop effective action where the r\^ole of the classical scalar field is played by $\sqrt{\frac{8C}{\kappa(r+C)}}$ (see Eq.\eqref{psie}), with  $C=\ell\left(F(M)l_p/\ell\right)^{1/3}$.
This dressed BH has a mass  --the conserved charge associated with the time translation symmetry at infinity-- given by \cite{MZ1}
\be \label{calM}
\mathcal{M}= 3 \left(\frac{l_P F(M)}{\ell}\right)^{2/3}. 
\ee
The corresponding temperature and entropy of this black hole are~\cite{MZ1}
\begin{equation}\label{eq:T,S static NS}
T= \frac{3 \sqrt{3}}{ \pi}  \left( \frac{l_P}{\ell}  \right) \mathcal{M}^{1/2}, \qquad S= \frac{2 \pi }{3 \sqrt{3}}  \left(\frac{\ell}{l_P}  \right)\mathcal{M}^{1/2},
\end{equation}
respectively\footnote{In Eqs. \eqref{calM} and \eqref{eq:T,S static NS} we have set $\kappa=\pi$. }. The first law of thermodynamics $d \mathcal{M}=T dS$ is directly verified from Eqs. \eqref{calM} and \eqref{eq:T,S static NS}. As noted in \cite{MZ1}, due to the conformal coupling, the area law for entropy is corrected as
\be
S=\left(1-\frac{\kappa}{8}\phi^2(r_+)\right) \frac{\pi r_+}{2 l_p}= \frac{\pi r_+}{3 l_p}.
\ee

\subsection{Rotating black hole}
From the analytical solution of the backreaction equations given in the previous section, we shall now investigate how the quantum corrections modify the background BH geometry. 
It is important to stress that our results are valid for nonextremal  BHs as well as for  extremal ones.

\vspace{0.5cm}
\textit{Asymptotic structure}
\vspace{0.5cm}

At infinity the corrections are negligible, since $N_1\to \frac{1}{r^3}$, $f_1\sim \frac{1}{r}$ and $k_1\sim \frac{1}{r^3}$ as $r\to \infty$
(we remind the reader that $N_0(r)=1$, $k_0(r)= O(\frac{1}{r^2})$ and $f_0(r)\to r^2$). Therefore, the quantum corrections do not modify the asymptotic structure of the BTZ background space-time.

\vspace{0.5cm}
\textit{Horizons} 
\vspace{0.5cm}

Let us now study the quantum backreaction on the Cauchy and event horizons.
In order to study their stability properties it is useful to compute the curvature invariants  of the quantum-backreacted space-time.
The correction
``$-2 \kappa   \langle T^{\mu}{}_{\mu}\rangle$"
(which is $O(l_P)$)
to the background Ricci scalar $R= 6 \Lambda$ 
 is zero
because the RSET, which is the source of the semiclassical Einstein equations, is traceless.
In its turn, the
Kretschmann of the backreacted metric is
\be \label{eq:Kretschmann}
R_{\mu\nu\rho\sigma}R^{\mu\nu\rho\sigma}=4R_{\mu\nu}R^{\mu\nu}-R^2=12\Lambda^2+4\kappa^2\langle T^{\mu}{}_{\nu}\rangle \langle T^{\nu}{}_{\mu}\rangle \ ,
\ee 
where the semiclassical Einstein equations and the tracelessness of the RSET have been used.

With regards to the event horizon, we first note that the RSET is regular at the classical event horizon.
At $r=r_+=\ell \alpha_{+}/2$, we have from Eqs.\eqref{eq:RSET subext BH tt}--\eqref{eq:RSET subext BH tht} that
\begin{equation} \label{T2}
\left.\kappa^2\langle T^{\mu}_{\, \,\nu}\rangle \langle T^{\nu}{}_{\mu}\rangle \right|_{r=r_+}=\frac{3 l_{P}^2}{4 \ell^6 }\sum_{n,m =1}^{\infty} 
\frac{ c_n \, c_m}{(\cosh (n\pi  \alpha_{+} )-1)^{3/2} (\cosh (m\pi  \alpha_{+} )-1)^{3/2}},
\end{equation}
with $c_n=\cosh (n\pi  \alpha_{+} )+\cosh (n\pi  \alpha_{-} )+2$, already defined in \eqref{cnBH}. The invariant \eqref{T2} is regular and so, from Eq.\eqref{eq:Kretschmann},
it follows that the Kretschmann scalar at the event horizon (of the background space-time) is also regular.

In order to find the event horizon of the quantum-corrected solution, we look for the largest root of 
\be\label{zzp}   g^{rr}=f(r)=\frac{r^2}{\ell^2}-M+\frac{J^2}{4r^2}+l_P f_1(r)=0\, \ee
where we have used Eqs.\eqref{lpcorrections}, \eqref{eq:N0,k0} and \eqref{b0}.
Working at $O(l_P)$, it is enough to replace $f_1(r)$ with $f_1(r_+)$, provided $r_+ \gg l_P$, and 
consider the largest solution of the resulting  quartic equation\footnote{Clearly, this procedure would not work if we wanted to analytically extend our solution to the NS regime, where there is no horizon at the classical level.}. The radius $r_{+}^{(q)}$ of the event horizon of the backreacted metric is then given by
\be \label{solhor} \left(\frac{r_{+}^{(q)}}{\ell}\right)^2=\frac{M-l_P f_1(r_+)}{2}+\frac{1}{2} \sqrt{\left(M-l_P f_1(r_+)\right)^2-\frac{J^2}{\ell^2}}\ . \ee
It is understood that the above expression must be expanded at leading order in $l_P$, which in the non-extremal case yields
\be \label{lalu} 
r_+^{(q)}=r_+\left(1- \frac{2 l_P f_1(r_+)}{\alpha_+^2-\alpha_-^2}\right),\ee
with $\alpha_+^2-\alpha_-^2=4\sqrt{M^2-J^2\ell^{-2}}$.

From Eq.\eqref{b1bhnext} we arrive at 
\be  
 f_1(r_+)  = -
\frac{\sqrt{2}(\alpha_+^2-\alpha_-^2)}{8 \ell \alpha _+^2}
\sum_{n=1}^\infty \frac{\alpha_+^2 c_n\left( c_n-4\right) -2\alpha_+\alpha_-e_n}{ b_n^2 \sqrt{\cosh(n\pi\alpha_+)-1}  }.\label{bune}
\ee
One can proof that $f_1(r_+)<0$. In fact, by writing $\alpha_{\pm}\equiv s\pm\Delta$,
where $s>0$ and $\Delta>0$, we have that
the numerator of the summand in Eq.\eqref{bune} is
\begin{align}
&
\alpha_+^2\left( (\cosh(n\pi\alpha_+) +\cosh(n\pi\alpha_-))^2-4\right) -4\alpha_+\alpha_-\sinh(n\pi\alpha_+)\sinh(n\pi\alpha_-)=
\nonumber \\ &
4(s+\Delta)[s\sinh^2(n\pi\Delta)(1+ \cosh^2(n\pi s)) + \Delta\sinh^2(n\pi s)(1+\cosh^2(n\pi\Delta))]>0\ .
\end{align}
Thus,  Eq. \eqref{lalu} implies $r_+^{(q)} >r_+$.
That is, the radius of the quantum-corrected event horizon is larger than the classical one. 

The event horizon of the extreme BTZ black hole is located at $r_+^{\textrm{ext}}= \ell \sqrt{M/2}\equiv \ell \alpha$. From  Eq.\eqref{eq:f1 BH ext} we obtain
\be \label{eq:b1ext hor}
 f_1^{\textrm{ext}}(\ell\alpha)\equiv
f_1(r_+^{\textrm{ext}})=-\frac{1}{\ell\pi^2}\sum_{n=1}^{\infty} \frac{1}{n^2\sinh(n\pi\alpha)}  < 0.\ee
 Following the same procedure for obtaining the radius of the event horizon in the nonextremal case, we obtain the corrected horizon radius
\be \label{laluu} r_+^{\textrm{ext}(q)}=r_+^{\textrm{ext}} +\frac{\ell}{2}\sqrt{-l_P f_1^{\textrm{ext}}(\ell\alpha)},\ee
where the leading order correction term is now $O(\sqrt{l_P})$, instead of $O(l_P)$ as in the non-extremal case shown in Eq. \eqref{lalu}. Note that  the corrected horizon $r_+^{\textrm{ext}(q)}$ is greater than the classical one $r_+^{\textrm{ext}}$. Moreover, 
from Eqs.\eqref{bune} and \eqref{eq:b1ext hor}, it is easy to see that $f_1^{\textrm{ext}}(\ell\alpha)=\lim_{J\to M \ell}f_1(r_+)$.

Let us now consider the quantum corrections at the inner (Cauchy) horizon
$r_-$. As already remarked in \cite{Steif}
and as we mentioned in 
Sec.\ref{sec:RSET}, the RSET in Eqs.\eqref{eq:RSET subext BH tt}--\eqref{eq:RSET subext BH tht} is divergent at a series of circles $r=r_n$ for which $d_n$ vanishes,
i.e., at
\be \frac{r_n^2}{\ell^2}=\frac{\alpha_-^2(\cosh(n\pi\alpha_+)-1)-\alpha_+^2(\cosh(n\pi\alpha_-)-1)}{4(\cosh(n\pi\alpha_+)-\cosh(n\pi\alpha_-))}\ .
\ee
As $n\to \infty$, $r_n$ approaches $r_-$ from the inside, i.e., $\frac{r_n^2}{\ell^2}\to \frac{\alpha_-^2}{4}$. 
This accumulation produces an essential  singularity at $r_-$.
We see via Eq.\eqref{eq:Kretschmann} that the divergence of $\langle T^{\mu}{}_{\nu}\rangle \langle T^{\nu}{}_{\mu}\rangle$ at $r_-$ produces a curvature singularity (in the Kretschmann scalar) there.
As mentioned, the singularity at the Cauchy horizon arises when approaching it from its inside, i.e., as $r\to r_-^-$.
In Kerr, it has been shown that 
classical field perturbations
in the region inside the Cauchy horizon
possess unstable modes~\cite{dotti2008gravitational}.
However, near the singularity in Kerr there exist closed timelike curves and so the initial value problem is in principle not well-posed
there (even after requiring specific boundary conditions on the singularity).
The rotating BTZ BH case here, on the other hand, possesses no closed timelike curves anywhere and so we are free from their troubles.

A singularity at the Cauchy horizon is not unexpected. In ($3+1$)-D,   classical perturbations of the external region
of  Reissner-Nordstr\"om  and Kerr space-times
grow without bound at the inner (Cauchy) horizon, thus producing a
``mass inflation"
curvature singularity there \cite{poisson1989inner,Ori:1991zz,ori1992structure,DafermosLuk2017}. It was shown in \cite{Chan:1994rs} that a similar unbounded growth of the perturbations (and of the local mass function) happens in ($2+1$)-D for the rotating BTZ BH at $r_-$.
Furthermore, at a quantum level, there are indications that the RSET diverges in at least a part of the CH in Reissner-Nordstr\"om and Kerr(-Newman) background space-times~\cite{hiscock1980quantum,birrell1978falling,Ottewill:2000qh,Lanir:2018vgb}.

Within our linear perturbative analysis, and 
following the same reasoning adopted in \cite{bapo}, we can  study the quantum corrections to the inner horizon.  As we did for the event horizon, provided $r_- \gg l_P$, we  can replace $f_1(r)$ with $f_1(r_-)$ in Eq.\ref{zzp} and consider its smallest positive root. The radius $r_{-}^{(q)}$ of the Cauchy horizon of the backreacted metric is then given via
\be \label{solhorintquartic} 
\left(\frac{\rt_{-}^{(q)}}{\ell}\right)^2=\frac{M-l_P f_1(r_-)}{2}-\frac{1}{2} \sqrt{\left(M-l_P f_1(r_-)\right)^2-\frac{J^2}{\ell^2}}, \ee
which at $O(l_P)$ reduces to
\be \label{solhorint}
r_-^{(q)}=r_-\left(1+ \frac{2 l_P f_1(r_-)}{\alpha_+^2-\alpha_-^2}\right).
\ee

It turns out that the sign of the quantum correction to the 
radius of the
inner horizon is given by the sign of $f_1(r_-)$, where
\be \label{eq:b1(r_-)}
f_1(r_-)=\frac{\sqrt{2}(\alpha_+^2-\alpha_-^2)}{8 \ell \alpha _-^2}
\sum_{n=1}^\infty \frac{\alpha_-^2 c_n\left( c_n-4\right) -2\alpha_+\alpha_-e_n}{b_n^2  \sqrt{\cosh(n\pi\alpha_-)-1}
}.
\ee
One can show, that close to extremality, $f_1(r_-)$ is finite and negative.
This means that the inner horizon is pushed inwards (i.e. $\rt_{-}^{(q)}<r_-$) and 'disappears from the space-time', which ends up in the future at a spacelike curvature singularity at $r_-$.
Thus, the causal structure
of the backreacted rotating black hole
is essentially that of the static black hole in Fig.\ref{fig:Penrose}(a).
This means that, in this case, quantum effects act to preserve strong CCH. The same considerations apply to the extremal case, where $f_1(r_-)$ is given by \eqref{eq:b1ext hor} and the corrected inner horizon radius takes the form \eqref{laluu} but with $r_+$ replaced by $r_-$ and a minus sign in front of the correction term.

\begin{figure}[h!]
\begin{center}
  \includegraphics[width=8cm] 
  {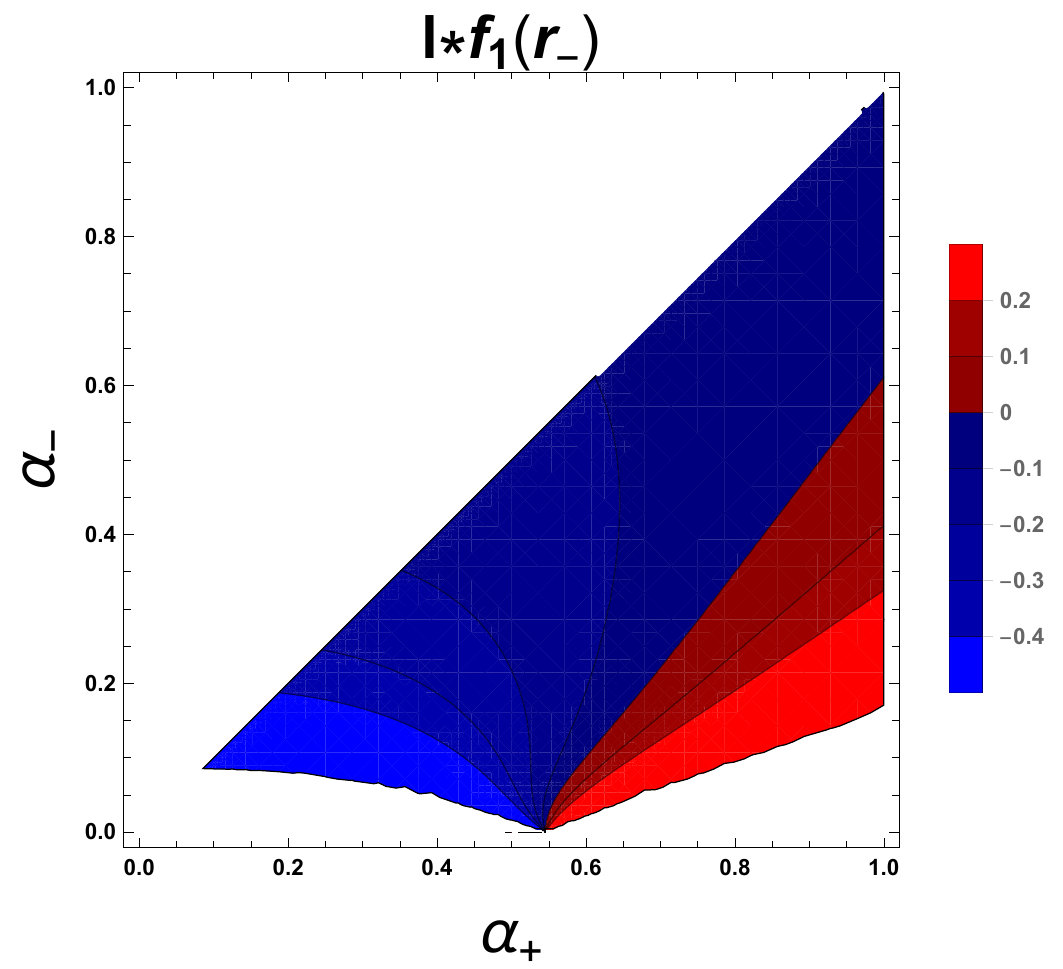}
  \end{center}
\caption{
This is $f_1(r_-)$ in Eq.\eqref{eq:b1(r_-)}
as a function of 
$\alpha_-$ and $\alpha_+ (>\alpha_-)$.
The red and blue shades correspond to, respectively, positive and negative values of $f_1(r_-)$.
Eq.\eqref{eq:b1(r_-)} shows that $f_1(r_-)$ diverges in the static limit $\alpha_-\to 0$, as the plot indicates (N.B.: the numerical calculation was not accurate enough for $\alpha_-$ very close to zero). 
}
\label{fig:ContourPlots} 
\end{figure}

We cannot check, with our method, what happens in the opposite regime (i.e., in the weak rotating limit) since, in this regime, the inner horizon is close to $r=0$ and so $f_1(r)$ cannot be replaced with $f_1(r_-)$ in Eq.(\ref{zzp}).
We plot $f_1(r_-)$ 
in Fig.\ref{fig:ContourPlots}.
This plot shows 
that the sign of $f_1(r_-)$ changes when varying $M$ and $J$.  When it is negative,  the radius of the Cauchy horizon diminishes and is smaller than the radius where the curvature singularity is. Therefore, in this case, the singularity is also spacelike and strong CCH is preserved.
On the other hand, when $f_1(r_-)$ is positive, the radius of the Cauchy horizon increases and is larger than the radius where the singularity is. Therefore, in this case, the singularity is timelike and strong CCH continues to be violated even in the backreacted space-time.
Fig.\ref{fig:ContourPlots} also indicates a divergence in $f_1(r_-)$  in the static limit. This divergence seems to come from the fact that
the image of the point $r=0$ in the static case is itself, and so the
the chordal distance in Eq.\eqref{dnBH} is equal to zero in the static
case for the point $r=0$.

\vspace{0.5cm}
\textit{Hypersurfaces outside rotating black holes: ergosphere and absence of superradiant instability} 
\vspace{0.5cm}
 
Another surface of interest in the rotating BTZ space-time is the static limit surface, defined by $g_{tt}=0$.
In order to find the radius of the static limit surface of the quantum-corrected space-time, we solve 
\be g_{tt}=-N^2(r) f(\rt)+\rt^2k^2(\rt)=0\ .\ee 
Working at $O(l_P)$, the equation to solve is
\be 
\label{solerg} - \left(\frac{\rt^2}{\ell^2}-M\right)-l_P\left(2 f_0 N_1+ f_1+J k_1\right) =0. \ee
Using the results in Eqs.\eqref{a1bhnext}, \eqref{b1bhnext} and \eqref{k1bhnext},
we see that the three last terms take a rather simple form: 
\be \label{3rt}  
2 f_0 N_1+ f_1+ J k_1 =-
\Sp_{n=1}^\infty \frac{(\alpha_+^2+\alpha_-^2)c_n\left( c_n-4\right) -4\alpha_+\alpha_-e_n}{4 b_n^2 d_n^{1/2}},
\ee
which is shown to be negative\footnote{\label{fn:sign}This statement can be checked by writing $\alpha_\pm=s\pm\Delta$, and so we have that the numerator of the summand in  Eq.\eqref{3rt}  is
$(\alpha_+^2+\alpha_-^2)\left((\cosh(n\pi\alpha_+)+\cosh(n\pi\alpha_-))^2-4\right)-8\alpha_+\alpha_-\sinh(n\pi\alpha_+)\sinh(n\pi\alpha_-)=8[s^2\sinh^2(n\pi\Delta)(1+\cosh^2(n\pi s))+\Delta^2\sinh^2(n\pi s)(1+\cosh^2(n\pi\Delta))] >0$.} for all $r$ such that $d_n >0$.

In order to solve Eq.\eqref{solerg}, for large enough 
radius
in comparison with $l_P$, we evaluate the terms in Eq.\eqref{3rt} (which are multiplied by $l_P$ when appearing in Eq.\eqref{solerg}) on the classical static limit surface $\rt_{SL}^2=\ell^2 M=\ell^2(\alpha_+^2+\alpha_-^2)/4$.
Let us denote by $r_{SL}^{(q)}$
and $r_{SL}^{\textrm{ext}(q)}$ the radii of the static limit surface of the quantum-backreacted  nonextremal and extremal BH geometries,  respectively.
For the nonextremal geometry, we obtain
\be \label{qerg} 
\frac{r_{SL}^{(q)2}}{\ell^2}-\frac{r_{SL}^2}{\ell^2}=-l_P\left.\left(2 f_0 N_1+ f_1+ J k_1\right)\right|_{r=r_{SL}} > 0.
\ee
 Therefore, like for the event horizon, the quantum corrections increase the radius of the static limit surface.
 In the extremal case we obtain, from Eq.\eqref{solerg}  and using Eqs.\eqref{eq:N1 ext BH}--\eqref{eq:k1 ext BH},
\be \label{qergext} \left(r_{SL}^{\textrm{ext}(q)}\right)^2-\left(r_{SL}^{\textrm{ext}}\right)^2=l_P \ell\sum_{n=1}^\infty \frac{\sinh^2(2 n \pi  \alpha  )+n^2 \pi ^2 \alpha ^2  (\cosh (4 n \pi  \alpha  )+3) }{2 \pi ^2  n^2 \sinh^2(2 n \pi  \alpha  )\sqrt{\sinh (n\pi  \alpha  ) (\sinh (n \pi  \alpha  )+n \pi  \alpha  \cosh (n \pi  \alpha  ))}},\ee 
which is also positive.
At this point, it is interesting to evaluate the quantum correction to the ``size" $r_{SL}-r_+$ 
of the ergoregion by computing, from Eqs.\eqref{lalu} and \eqref{qerg},
\begin{align} \label{sizeerg} \left[\left(r_{SL}^{(q)}\right)^2-\left(r_+^{(q)}\right)^2\right] - \left[ r_{SL}^{2}-r_+^{2 }\right]= \frac{\sqrt{2}l_P \ell }{8 }\sum_{n=1}^\infty &\left[ \frac{\sqrt{\alpha_+^2-\alpha_-^2}\left((\alpha_+^2+\alpha_-^2)c_n\left( c_n-4\right) -4\alpha_+\alpha_-e_n\right)}{ b_n^2 \sqrt{\alpha_+^2(\cosh(n\pi\alpha_+)-1)-\alpha_-^2(\cosh(n\pi\alpha_-)-1)}} \nonumber \right. \\ &\left. - \frac{\alpha_+^2 c_n\left( c_n-4\right) -2\alpha_+\alpha_-e_n}{ b_n^2 \sqrt{\cosh(n\pi\alpha_+)-1}}\right]. 
\end{align}
We could not determine the sign of the right hand side of Eq.\eqref{sizeerg} analytically. However, we carried out a numerical evaluation and this sign seems to be  always negative (although for $\alpha_-$ very close to zero the numerics were not reliable).   

In the extremal case, where the $O(\sqrt{l_P})$ correction to $r_+$ is larger than the $O(l_P)$ correction to $r_{SL}$, we have 
\be  \label{sizeergext} \left[\left(r_{SL}^{\textrm{ext}(q)}\right)^2-\left(r_+^{\textrm{ext}(q)}\right)^2\right] - \left[ \left(r_{SL}^{\textrm{ext} }\right)^2-\left(r_+^{\textrm{ext} }\right)^2\right]=-\ell r_+^{\textrm{ext}} \sqrt{-l_P f_1^{\textrm{ext}}(\ell\alpha)}<0,\ee
where $r_+^{\textrm{ext}(q)}$
denotes the radius of the event horizon of the backreacted extremal BH geometry.

We shall now turn to the evaluation of the quantum corrections to the angular velocity of the BH
using Eq.\eqref{eq:N0,k0}. We find that the angular velocity of the quantum-corrected BH is
\be \label{qrbh} \Omega_H^{(q)}= \left.\frac{g^{\tit\tht}}{g^{\tit\tit}}\right|_{r=r_{+}^{(q)}}=
-\left. k\right|_{r=r_{+}^{(q)}}=\frac{J}{2\left(r_{+}^{(q)}\right)^2}-l_P k_1(r_+). \ee
In the nonextremal case, combining the two contributions to $\Omega_H^{(q)}$, we find at $O(l_P)$,
\be\label{zz}  \Omega_H^{(q)}-\Omega_H = 
-\frac{\sqrt{2}l_P(\alpha_+^2-\alpha_-^2)}{\ell^2 \alpha_+^2}\sum_{n=1}^\infty\frac{\sinh(n\pi\alpha_+)\sinh(n\pi\alpha_-)}{b_n^2
(\cosh(n\pi\alpha_+)-1)^{1/2}}, \ee 
where $\Omega_H=J/(2 r_+^2)$. Note that the right-hand side of \eqref{zz} has a sign opposite to that of $\Omega_H$ because $J \sinh(n\pi\alpha_-)>0$. Therefore, the quantum corrections to the angular velocity reduce its absolute value. The same effect occurs in the extremal case. 
We denote by $\Omega_H^{\textrm{ext}(q)}$ and $\Omega_H^{\textrm{ext}}$ the angular velocity of the black hole in, respectively, the backreacted and background geometries.
From Eq.\eqref{qrbh}  and  Eqs.\eqref{eq:N1 ext BH}--\eqref{eq:k1 ext BH}, we obtain
\be \Omega_H^{\textrm{ext}(q)}-\Omega_H^{\textrm{ext}}= -\frac{\gamma  \sqrt{-f_1^{\textrm{ext}}(\ell \alpha})}{2 \sqrt{2} \alpha  \ell} \sqrt{l_P},\ee
with $\Omega_H^{\textrm{ext}}=\gamma/(2 \ell)$. Since  the quantum correction to $r_+^{\textrm{ext}}$ is of the order $O(l_P^{1/2})$ (see Eq. \eqref{laluu}), we obtain the same leading order for the correction  to the angular velocity.

Finally, we inspect the possible appearance of a speed of light surface, which would -- likely -- make the space-time superradiantly unstable~\cite{harhaw}\footnote{Although the BTZ BH is unstable under {\it massive} scalar field perturbations due to modes whose frequency has a real part that lies within the superradiant regime~\cite{Iizuka:2015vsa}, this is not considered the ``standard" superradiant instability, which refers to a massless scalar field.}.
For this purpose, we consider the quantum-corrected Killing vector $\chi^{(q)}=\partial/\partial t+ \Omega_H^{(q)}\partial/\partial \theta$.
This vector has squared norm 
\begin{align}\label{norm}  
\chi^{(q)^2}=&
g_{\mu\nu}\chi^{(q)^\mu}\chi^{(q)^\nu} 
=
-N^2 f+\rt^2 k^2+2\rt^2 k \Omega_H^{(q)}+\rt^2\Omega_H^{2(q)} \nonumber \\
=&
-(f_0(r)+l_P f_1(\rt)) -2l_P f_0(r) N_1
+
\nonumber \\ &
\frac{J^2}{4\rt^2} \left(\frac{\rt^2}{\rt_+^{2(q)}}-1\right)^2 
-J l_P \left(\frac{\rt^2}{\rt_+^2}-1 \right)(k_1(\rt_+)-k_1(\rt))+O(l_P^2).
\end{align}
Classically, 
\be \label{chicext}\chi^2=-\frac{(r_+^2-r_-^2)(r^2-r_+^2)}{l^2r_+^2}\ . 
\ee
The vector $\chi^{(q)}$ is timelike in the near-horizon region (where the terms in the second line of (\ref{norm}) go like ``$-A(\rt ^2- \rt_+^{(q)^2})$", with $A$ a positive constant, and the terms in the third line go like $\sim (\rt ^2- \rt_+^{(q)^2})^2$)  and becomes null on the horizon. At radial infinity, where 
$N_1$, $f_1$, $k_1 \to 0$, we have that 
\be \label{norminf} \chi^{(q)^2} \sim -\frac{\rt^2}{\ell^2}\left( 1-\ell^2\Omega_H^{(q)2}\right ),
\quad r\to \infty.
\ee
The condition for it to be spacelike, and (likely) for the space-time to develop a superradiant instability  is \be \label{cone} \ell\Omega_H^{(q)}
>1\ .\ee Classically, this condition is not met, i.e., $\ell\Omega_H\leq 1$ (the equality being realized in the extremal case).   Eq.\eqref{zz} implies
that, in the non extremal case, it is $\ell\Omega_H^{(q)}<\ell\Omega_H<1$, and, in the extremal case, $\ell\Omega_H^{ext(q)}<\ell\Omega_H^{ext}=1$. This suggests that the quantum effects do not change the superradiant stability property of the BTZ BH.



\begin{figure}[h!]
\begin{center}
  \includegraphics[width=9cm]  
  {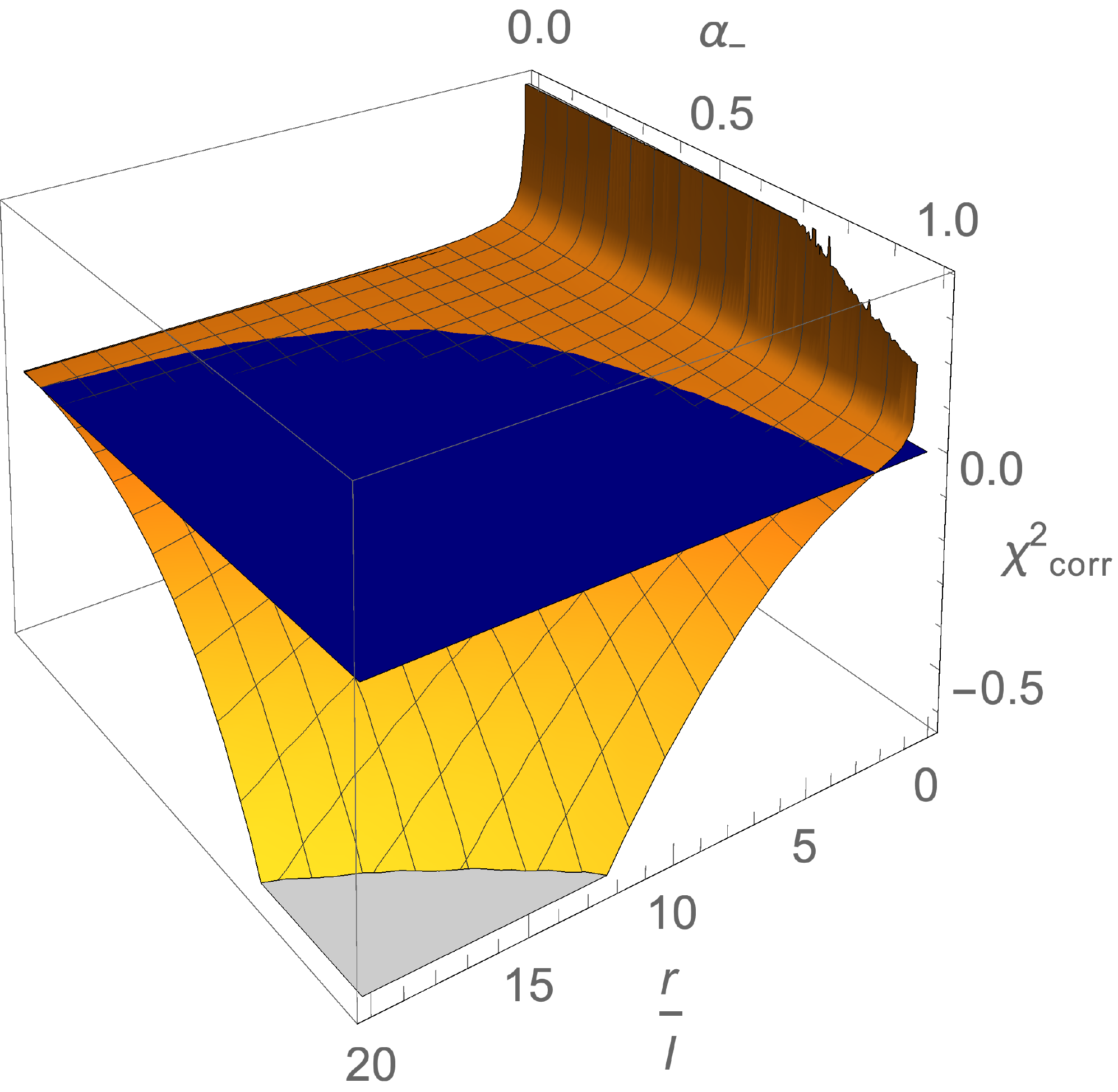}
  \end{center}
\caption{Plot of the $O(l_P)$ correction
($\chi^2_{\text{corr}}$)
to $\chi^2$ in 
Eq.\eqref{chiSq sub}
as a function of $r$ and $\alpha_-$ for the fixed value of $\alpha_+=(\sqrt{3}+1)/\sqrt{2}\approx 1.93$.
The blue horizontal plane corresponds to 
zero.
}
\label{fig:chiSq} 
\end{figure}

In order to investigate the norm of  $\chi^2$ more widely, we first obtain explicitly the $O(l_P)$ correction to $\chi^2$ in the subextremal case from Eqs.\eqref{norm} and \eqref{lalu}:
\begin{align}\label{chiSq sub}
\chi^{(q)^2}=
\chi^2+l_P \left(
- f_1(\rt) -2f_0(r) N_1
+
\frac{J^2 \ell^2 f_1(r_+)}{2r_+^4} \frac{(\rt^2-r_+^2)}{(r_+^2-r_-^2)} 
-J \left(\frac{\rt^2}{\rt_+^2}-1 \right)(k_1(\rt_+)-k_1(\rt))
\right)
+O(l_P^2).
\end{align}
We plot
this $O(l_P)$ correction to $\chi^2$
in Fig.\ref{fig:chiSq}.
[N.B.: for the large-$r$ behaviour in Eq.\eqref{norminf}
to be seen in Fig.\ref{fig:chiSq} for small $\alpha_-$, the plot should be performed to larger values of $r$.]

In the extremal case, where $\chi^2=0$ identically (see \eqref{chicext}), 
the quantum-corrected 
$\chi^{ext (q)^2}$
is entirely given by the quantum corrections, whose leading term, $O(\sqrt{l_p})$, comes from the first term  in the last line of Eq.\eqref{norm} and
from $r_+^{(q)}$ (see Eq.\eqref{laluu}). We then obtain
\begin{equation}
\chi^{ext (q)^2}=-\frac{J^2\ell}{2\left(r_+^{ext}\right)^3}\left(\frac{\rt^2}{\rt_+^2}-1 \right)
\sqrt{-l_pf_1(l\alpha)}<0\ . 
\end{equation} 
Thus the quantum corrections turn the classically identically-null $\chi^{ext}$  timelike.

\subsection{Rotating naked singularity }   

\vspace{0.5cm}
\textit{Emergence of an event horizon} 
\vspace{0.5cm}

The first-order quantum correction to the metric component $g^{rr}$ of the NS geometry, $f_1(r)$ in Eq.\eqref{lpcorrections}, is responsible for the formation of a horizon. In order to see this, we note that $f_1(r)$ has a finite number of poles at radii $r_n$ where $d_n$ vanishes [see Eq.\eqref{r02}]. As we will shown below at these poles  $f_1 \to -\infty$, turning the otherwise positive definite  $g^{rr}_{\text{classical}}=f_0(r)$, into a function that vanishes at some finite radii. The largest radius at which $g^{rr}$ vanishes is the event horizon of the quantum-backreacted metric, $r_{+}^{(q)}$. 

The zeroes of $d_n$ form a finite set, the largest of them, which we denote by $r_*$, occurs at a certain value $n=n_{*}$,
\begin{equation} \label{r04}
r_*^2= \frac{\ell^2}{2 {b}_{n_{*}}} \left(\beta _-^2 \sin ^2\left(\frac{1}{2} \pi n_{*} \beta _+ \right)-\beta _+^2 \sin ^2\left(\frac{1}{2} \pi n_{*} \beta _- \right)\right).
\end{equation}
This zero appears twice in the sum that defines $f_1(r)$ due the symmetry of the summand in Eq.\eqref{eq:f1 subext NS} under $n \to N-n$. At $r=r_*$ the geometry has a curvature singularity (since the Kretschmann invariant \eqref{eq:Kretschmann} diverges) and therefore the spacetime cannot be extended to $r<r_*$.

From \eqref{eq:f1 subext NS} the correction $f_1(r)$ can be seen to diverge as $(r-r_*)^{3/2}$ near $r_{*}$,
\be\label{eq:f1,r*}
f_1(r)= \frac{\Xi f_0(r_{*})}{ (r-r_*)^{3/2}}+ C\,,
\quad r\to r_*,
\ee
where $C$ is a finite constant and
\be \label{gamma}
\Xi\equiv \frac{\sqrt{\beta _+^2-\beta _-^2} \ell^2 \left(a_{n^{*}} c_{n^{*}} -2 \beta _- \beta _+ e_{n^{*}}\right)}{32 b_{n^{*}} (-b_{n^{*}} r^{*})^{3/2}}.
\ee

First, we note that the combination $a_{n_{*}} c_{n_{*}} -2 \beta _- \beta _+ e_{n_{*}}$  in the numerator of the above equation is positive definite. Moreover, since $b_{n_{*}}<0$,   $\Xi<0$. Then, using Eq.\eqref{eq:f1,r*}, the condition $g^{rr}|_{r_{+}^{(q)}}=0$ that defines the quantum-corrected horizon can be written as 
\be \label{horNS1}
(f_0(r_{+}^{(q)})+l_P C)(r_{+}^{(q)}-r_{*})^{3/2}+l_P \Xi f_0(r_{*})=0.
\ee
Since $f_0(r)$ is an analytic function for $r\neq0$, one can write  $f_0(r_{+}^{(q)})=f_0(r_{*})+f_0'(r_{*})(r_{+}^{(q)}-r_{*})+O(r_{+}^{(q)}-r_{*})^2$ near $r_{*}$. Replacing this Taylor expansion in   Eq.\eqref{horNS1}, one finds that: (i) $r_{+}^{(q)}-r_*$ must be of the order $l_P^{2/3}$, (ii) $C$ can be ignored and consequently,
\be \label{solhorNS} r_{+}^{(q)}=r_{*}+(-\Xi \, l_P)^{2/3}+O(l_P^{7/3}).
\ee
Thus, the existence of a horizon and its radius have been established for the backreacted spacetime. The classical NS has been replaced by a rotating black hole whose horizon encloses a curvature singularity. 
This singularity at $r=r_*$ is spacelike since $g^{rr}$ has no zero within $[r_*, r_{+}^{(q)})$.

Thus, in the cases that satisfy Eq.\eqref{range}, except for the set $\mathcal{S}$ defined in Eq.\eqref{eq:S}, an event horizon forms; the other cases would have to be investigated separately.

\vspace{0.5cm}
\textit{Ergosphere} 
\vspace{0.5cm}

The radius of the static limit surface, which is the boundary of the ergosphere, is determined by Eq. \eqref{solerg}. This equation can be solved near the singularity $r=r_{*}$, yielding
\be \label{rSLNS}
r_{SL}^{(q)}=r_{*}+\mu\, l_P^2,
\ee
with
\be \label{muRSLNS}
\mu\equiv \frac{(\beta _+^2-\beta _-^2) \left(\left(\beta _-^2+\beta _+^2\right) (4-c_{n_*}) c_{n_*}-4 \beta _- \beta _+ e_{n_*}\right)^2}{16 (-b_{n_*})^{5} r_{*} \left( \left(\beta _-^2+\beta _+^2\right)+\left(\frac{2r_*}{\ell}\right)^2\right)^2}.
\ee
It follows that the right-hand side of the above equation is positive because $b_{n_*}<0$.
Since the distance $r_{SL}^{(q)}-r_{*}$ is of order $O(l_P^2)$ and  $r_{+}^{(q)}-r_{*}$ is of order $O(l_P^{2/3})$, as shown in Eq.\eqref{solhorNS}, the static limit surface is located behind the event horizon.

\section{Summary and discussion }\label{sec:discussion}
In this paper we have considered the $O(l_P)$ RSET for a conformally coupled massless scalar field in a background ($2+1$)-dimensional BTZ geometry. This background corresponds to a black hole ($M>0$) or to a naked conical singularity ($M<0$). Using this RSET as an effective source for the Einstein equations, we have computed the quantum corrections to the original background metric (backreaction) both in the static and  rotating cases. Our findings can be summarized as follows:\\

\textbf{Static Black Hole}

$\bullet$ 
The RSET given in Eqs.\eqref{eq:RSET static BH} is diagonal, traceless and conserved with respect to the background black hole geometry. For a fixed $M$, the backreacted metric has a quantum-corrected horizon with a radius larger than the classical one,
\begin{equation}
r^{(q)}_+ = r_+ + \frac{F(M)}{M} l_P + O(l_P^2) > r_+,
\end{equation}
where $F(M)$ is given in Eq.\eqref{eq:FBH} and $r_+=\sqrt{M}\ell$. For very small mass, 
\begin{equation}
    r^{(q)}_+ =  \left(2\frac{l_p}{\ell}F(0^+) \right)^{1/3}\ell + O(M),
\end{equation}\\
where $F(0^+)=\zeta(3)/(2 \pi^3)\approx 0.0193841$.\\

$\bullet$
A {\it curvature} spacelike singularity is formed at
$r=0$.\\

\vspace{0.5cm}
\textbf{Rotating Black Hole}

$\bullet$ 
The RSET given in Eq.\eqref{eq:RSET subext BH tt}-\eqref{eq:RSET subext BH tht} is traceless and conserved with respect to background black hole geometry \eqref{eq:ansatz}. Its only off-diagonal $t-\theta$ components are compatible with the stationary rotating black hole solution. Again, the non-extremal backreacted metric has a quantum-corrected event horizon with a radius larger than the classical one,
\begin{equation}
r^{(q)}_+ = r_+ - \frac{2 f_1(r_+) r_+}{\alpha_+^2-\alpha_-^2}l_P  + O(l_P)^2 > r_+,
\end{equation}
where $f_1(r_+)<0$. \\

$\bullet$
The radius  $r_-$ (which is the inner -- Cauchy -- horizon of the classical background space-time) becomes an accumulation surface for divergent contributions to the RSET at which the Kretschmann invariant blows up. In the quantum-corrected space-time, the curvature singularity at $r_-$  can be either spacelike ($r_-^{(q)}<r_-$; this is the case close to, and at, extremality)  or, depending on the values of $M$ and $J$, timelike ($r_-^{(q)}>r_-$).
In the former case, quantum mechanics provides a mechanism for strong  cosmic censorship.\\

$\bullet$ 
Similarly to the event horizon, the ergosphere  is also pushed outwards (the quantum correction to its radius is always positive), while the black hole angular velocity generically diminishes.\\

$\bullet$ 
In the extremal limit, our results could be interpreted by saying that the quantum corrections take the solution away from extremality. 

\vspace{0.5cm}
\textbf{Static Naked Singularity}

$\bullet$ 
The RSET given in Eq.\eqref{eq:RSET BH'} is diagonal, traceless and conserved with respect to the background static conical geometry. The backreacted metric presents a horizon of non-vanishing radius,
\begin{equation}
    r^{(q)}_+ =  \frac{2F(M)}{-M} l_P + O(l_P^2),
\end{equation}
where $F(M)$ is given in Eq.\eqref{eq:FNS}. This result is valid for a finite mass $M$. In the limit $M \to 0^-$, the horizon radius is given by
\begin{equation}
    r^{(q)}_+ =  \left[2\frac{l_p}{\ell}F(0) \right]^{1/3}\ell + O(M).
\end{equation}
Fig.\ref{fig:bothsides} shows the continuity at $M=0$ between the radius of the event horizon of the quantum-backreacted  black hole and the radius of the newly-formed event horizon of the quantum-backreacted naked singularity. \\

$\bullet$
A {\it spacelike curvature} singularity is formed at
$r=0$.
The appearance of a horizon  around the classical naked singularity, and the fact that the timelike singularity of the background spacetime has become spacelike in the backreacted spacetime,
means that, at least in this simplified setting, quantum mechanics provides a mechanism for strong cosmic censorship.\\

$\bullet$
The backreacted geometry is obtained as a classical solution of the Einstein equations in the presence of the RSET given in Eq.\eqref{eq:RSET BH'}. This stress-energy tensor happens to be the same as that for the Einstein-Hilbert action conformally coupled to a scalar field,  Eq.\eqref{eq:SET scalar} with $C= \ell[F(M)l_P/\ell]^{1/3}$. Hence, the backreacted metric can be interpreted as a classical solution of the form 
\begin{equation} \label{ds'}
ds^2= -f(r)dt^2+f^{-1}(r)dr^2+r^2 d\theta^2,
\end{equation}
with $f(r)\equiv \frac{1}{ \ell^2}\left(r^2-3C^2- \frac{2C^3}{r}\right)$. In this interpretation, the geometry is that of a black hole with a tiny positive mass, $\mathcal{M}= 3[F(M)l_P /\ell]^{2/3}$, and a horizon radius $\tilde{r}_+$ of order $l_P^{1/3}$, $\tilde{r}_+^2 = (4/3)\mathcal{M}\ell^2$.\\


$\bullet$ In \cite{efk}, black holes localised on the brane in 3+1-dimensional Randall-Sundrum braneworlds \cite{ehm1, ehm2} were interpreted, via the AdS/Conformal Field Theory (CFT) correspondence, as
static quantum-corrected BTZ black holes and naked singularities. In particular, and despite the fact that the dual quantum theory (CFT) living on the brane is poorly known, use of the AdS/CFT dictionary gives a brane black hole metric that has the same form as ours:
\be ds^2=-\left(\frac{r^2}{\ell^2}-M-\frac{r_1(M)}{r}\right)dt^2 + 
\frac{dr^2}{\left(\frac{r^2}{\ell^2}-M-\frac{r_1(M)}{r}\right)}
+r^2d\theta^2\ ,
\ee
for some function $r_1(M)$.
For a slightly curved brane, 
it is $r_1(M)\sim N l_p f(M)$ ($N$ being the (large) number of degrees of freedom of the CFT on the brane)  and $f$ is a function that depends on the mass $M$. For zero-mass black holes, where $f(0)\sim O(1)$, as well as for naked singularities ($M<0$), 
the correction term $\frac{r_1(M)}{r}$ leads to the formation of a horizon, in agreement with our results.

\vspace{0.5cm}
\textbf{Rotating Naked singularity}

$\bullet$ 
The RSET given in Eqs.\eqref{eq:Ttt subext NS}-\eqref{eq:Ttht subext NS} is traceless and conserved with respect to the background rotating conical geometry. Similarly to the rotating black hole background above, its only off-diagonal $t-\theta$ components are compatible with the stationary rotating solution \eqref{eq:ansatz}. Again, the backreacted metric also has an event horizon of radius
\begin{equation}
    r_{+}^{(q)} = r_{*}+(-\Xi \, l_P)^{2/3},
\end{equation}
where $r_{*}$ is the largest zero of $d_n$, and $\Xi$ is the finite expression given in Eq.\eqref{gamma}.\\ 

$\bullet$ 
A {\it spacelike curvature} singularity is formed at the radius $r_*$ given by Eq.\eqref{r04}.
As in the static case, the appearance of a horizon and the spacelike character of the singularity in the backreacted spacetime
mean that quantum mechanics acts as a strong cosmic censor.\\

$\bullet$ 
A legitimate concern is about the validity of the perturbative approximation in powers of $l_P$ for the geometry in view of the fact that $l_P f_1$ diverges at some finite $r$. This divergence is responsible for the formation of a horizon, which implies a change of topology and of the causal structure of the spacetime. The point is that for $r>>l_P$ the geometry receives a very small correction of order $\hbar$ (as clearly seen in the static case, with a small horizon of the same order, $r_+\sim l_P$). This is not too different from a perturbation of the Schwarzschild geometry by the addition of a small electric charge or angular momentum: the appearance of a (small) second horizon produces a small correction to the exterior metric. Depending on the experimental resolution, it might be irrelevant for an external observer whether the geometry has a second horizon or not, even if the topology and the causal structure both suffer major changes. For a small $M<0$, the perturbative approximation is certainly more reliable and $r_+^{(q)}\sim (l_P)^{1/3}$ for $J=0$, and $r_+^{(q)}\sim (l_P)^{2/3}$ for $J\neq 0$.

\vspace{0.5cm}
\textbf{Extensions and open questions}

$\bullet$ 
For rotating naked singularities we have assumed $\beta_{\pm}=2/N_{\pm}$, with integer $N_{\pm}$. The method of images can also be extended to arbitrary rational values of $\beta_{\pm}$, but we did not consider this case in order to keep the discussion as simple as possible and to be able to make definite claims. Despite of the restriction on the values of $\beta_{\pm}$, our results are sufficient to explore rotating geometries for small angular momentum to claim that the conclusions drawn for the static case are generic and not an accidental consequence of the static symmetry. In the case of static flat space, the authors of \cite{Souradeep-Sahni} have shown that for continuous values of the angular deficit the resulting RSET interpolates between  the discrete values obtained for $\beta=2/N$. This suggests that a similar extension to arbitrary real values of $\beta_{\pm}$ is possibly doable in the spirit of \cite{Cheeger}. \\

$\bullet$ 
Another direction in which this work can be extended is the inclusion of quantum matter to examine  backreaction on other spacetimes. For example, other (2+1)-D geometries with naked singularities like BTZ spacetimes with $M<-1$ (angular excesses), or with $M\ell<|J|$; spacetimes with closed timelike curves \cite{deDeo-Gott}, etc.\\

$\bullet$ Quantum matter was also shown to form a horizon around conical singularities in asymptotically {\it flat} three-dimensional spacetimes in \cite{Souradeep-Sahni, Soleng}. Although those papers did not identify the backreacted geometry as a black hole --perhaps because the existence of black holes in 2+1 dimensions was not widespread at the time--, they suggest that results similar to ours could be found for naked singularities in flat and de Sitter three-dimensional spacetimes.\\

$\bullet$ The existence of locally propagating \MC{gravitational?} degrees of freedom in higher dimensions means that, without a quantum theory of gravity, the cosmic censorship hypothesis for $D \geq 4$ could only be tested semiclassically: quantum effects on cosmic strings, the Big Bang or Big Crunch singularities in 3+1 dimensions could only be examined with quantum matter on a classical background\MC{But semiclassical doesn't necessarily have to be on a classical {\it background}, it could also be a self-consistent solution?}.\\



\begin{acknowledgments}
We are thankful to Adrian C.~Ottewill and Elizabeth Winstanley for useful discussions.
M.C. acknowledges partial financial support by
CNPq (Brazil), process number 
310200/2017-2. 
This work has been partially funded by the Fondecyt
grants  1161311 and 1180368. The Centro de Estudios Cient\'{\i}ficos (CECs) is funded by the Chilean Government through the Centers of Excellence Base Financing Program of Conicyt. A.F. acknowledges partial financial support by the Spanish Ministerio de Economía, Industria y Competitividad Grants No. FIS2014-57387-C3-1-P and No. FIS2017-84440-C2-1-P, the Generalitat Valenciana Project No. SEJI/2017/042 and the Severo Ochoa Excellence Center Project No. SEV-2014-0398.

\end{acknowledgments}

\appendix
\section{Black holes and naked singularities as identifications in $\mathbb{R}^{2,2}$}\label{sec:BTZ as ids} \label{identifications}      

A general Killing vector $\Kill$
for the pseudosphere 
Eq.\eqref{eq:pseudo}
embedded in 
$\mathbb{R}^{2,2}$
can be written in terms of the $so(2,2)$ generators $J_{a b}:=X_b \partial_a-X_a \partial_b$.
Let us parametrize the 
pseudosphere 
with coordinates $(t, r, \theta)$.
The BTZ geometries are then obtained by identifying points in the pseudosphere 
via the Killing vector
\begin{equation}
\Kill= \partial_{\theta}=\frac{\partial X^a}{\partial \theta}  \partial_a=\frac{1}{2} \omega^{ a b} J_{a b},
\end{equation}
where the antisymmetric matrix $\omega^{ a b}$ characterizes the identification.

Since $\Kill= \partial_{\theta}$, the identification corresponding to the action of the vector $2\pi \Kill$ means that the geometry is periodic in $\theta$ with period $2\pi$. Identifying a point in the manifold with itself rotated by $2\pi$ can also be represented by the action of the matrix $H:= e^{2\pi \Kill}$ in the embedding space, such that $H^{a}\,_{b} \, X^b(\theta) = X^a(\theta +2\pi)$. 

\vspace{0.5cm}
\textbf{1. Rotating nonextremal BTZ black hole} 
\vspace{0.5cm}

The rotating BTZ BH with mass $M$ and angular momentum $J$ is described by the line element in Eq.\eqref{BTZ}.
It may be expressed in terms of $\alpha_{\pm}$ 
in Eq.\eqref{alfa+-}
as
\begin{equation} \label{BTZline}
ds^2= -\left(\frac{r^2}{\ell^2}-\frac{ \alpha_{+}^2+\alpha_{-}^2}{4} \right) dt^2+\frac{dr^2 \ell ^2 r^2}{\left(r^2-\frac{\ell^2  \alpha_{+}^2}{4}\right) \left(r^2-\frac{\ell^2  \alpha_{-}^2}{4}\right)}-\frac{\ell \alpha_{+} \alpha_{-}}{2}d t d\theta+r^2 d\theta^2.
\end{equation}

The various BTZ BH regions can be parametrized in terms of $(t,r,\theta)$ coordinates in the following way: 

Region I:  $r>r_{+}$.
\begin{align} \label{regionI}
&X^0=\sqrt{A_{-}} \cosh \left(\frac{\alpha_{+} \ell \theta -\alpha_{-} t}{2 \ell}\right),  & X^{1}=\sqrt{A_{-}} \sinh \left(\frac{\alpha_{+} \ell \theta -\alpha_{-} t}{2 \ell}\right), \nonumber \\
&X^{2}=\sqrt{A_{+}}\cosh \left(\frac{\alpha_{+} t-\alpha_{-} \ell \theta }{2 \ell}\right),  & X^3=\sqrt{A_{+}} \sinh \left(\frac{\alpha_{+} t-\alpha_{-} \ell \theta }{2 \ell}\right).
\end{align}
Region II: $r_{-}< r <r_{+}$.
\begin{align} \label{regionII}
&X^0=\sqrt{A_{-}} \cosh \left(\frac{\alpha_{+} \ell \theta -\alpha_{-} t}{2 \ell}\right),  & X^{1}=\sqrt{A_{-}} \sinh \left(\frac{\alpha_{+} \ell \theta -\alpha_{-} t}{2 \ell}\right), \nonumber \\
&X^{2}=-\sqrt{A_{+}}\sinh \left(\frac{\alpha_{+} t-\alpha_{-} \ell \theta }{2 \ell}\right),  & X^3=-\sqrt{A_{+}} \cosh \left(\frac{\alpha_{+} t-\alpha_{-} \ell \theta }{2 \ell}\right).
\end{align}
Region III: $0<r<r_{-}$.
\begin{align} \label{regionIII}
&X^0=\sqrt{-A_{-}} \sinh \left(\frac{\alpha_{+} \ell \theta -\alpha_{-} t}{2 \ell}\right), & X^{1}=\sqrt{-A_{-}} \cosh \left(\frac{\alpha_{+} \ell \theta -\alpha_{-} t}{2 \ell}\right), \nonumber \\
&X^{2}=-\sqrt{-A_{+}} \sinh \left(\frac{\alpha_{+} t-\alpha_{-} \ell \theta }{2 \ell}\right),  & X^3=-\sqrt{-A_{+}} \cosh \left(\frac{\alpha_{+} t-\alpha_{-} \ell \theta }{2 \ell}\right),
\end{align}
where 
\begin{equation}
A_{\pm}\equiv \frac{4 r^2-\alpha _{\pm}^2 \ell^2}{\alpha _+^2-\alpha _-^2}.
\end{equation}

The rotating BTZ space-time is obtained through identifications generated by the Killing vector 
\begin{equation} \label{xiBTZ}
\Kill=\frac{\alpha_+}{2} J_{0 1}+\frac{\alpha_-}{2} J_{23}.
\end{equation}
The identification matrix $H= e^{2\pi \Kill}$ then takes the form
\begin{equation}\label{HBTZ}
H=
 \left( \begin{array}{cccc}
\cosh(\pi\alpha_{+})  & \sinh(\pi\alpha_{+})  & 0 & 0 \\
\sinh(\pi\alpha_{+})  & \cosh(\pi\alpha_{+})  & 0 & 0 \\
0 & 0 & \cosh(\pi\alpha_{-})  & -\sinh(\pi\alpha_{-})  \\
0 & 0 &-\sinh(\pi\alpha_{-})  & \cosh(\pi\alpha_{-})
 \end{array} \right).
\end{equation}

Using coordinates $(t,r,\theta)$, the chordal distance $\sigma(x,x')$ (Eq. \eqref{eq:world func R22}) for each region of the BTZ BH spacetime is given by:

Region I:  $r>r_{+}$,
\begin{align}
\sigma(x,x')&=\sqrt{\frac{4 r^2-\alpha _-^2 \ell^2}{\alpha _+^2-\alpha _-^2}} \sqrt{\frac{4 r'^{\,2}-\alpha _-^2 \ell^2}{\alpha _+^2-\alpha _-^2}} \cosh \left(\frac{\alpha _+ \ell \left(\theta ' -\theta \right)+\alpha _- \left(t-t'\right)}{2 \ell}\right)\nonumber \\&-\sqrt{\frac{4 r^2-\alpha _+^2 \ell^2}{\alpha _+^2-\alpha _-^2}} \sqrt{\frac{4 r'^{\,2}-\alpha _+^2 \ell^2}{\alpha _+^2-\alpha _-^2}} \cosh \left(\frac{\alpha _- \ell \left(\theta -\theta '\right)+\alpha _+ \left(t'-t\right)}{2 \ell}\right)- \ell^2.
\end{align}
Region II: $r_{-}< r <r_{+}$,
\begin{align}
\sigma(x,x')&=
\sqrt{\frac{4 r^2-\alpha _-^2 \ell^2}{\alpha _+^2-\alpha _-^2}} \sqrt{\frac{4 r'^{\,2}-\alpha _-^2 \ell^2}{\alpha _+^2-\alpha _-^2}} \cosh \left(\frac{\alpha _+ \ell \left(\theta ' -\theta \right)+\alpha _- \left(t-t'\right)}{2 \ell}\right)\nonumber \\&+\sqrt{\frac{\alpha _+^2 \ell^2-4 r^2}{\alpha _+^2-\alpha _-^2}} \sqrt{\frac{\alpha _+^2 \ell^2-4 r'^{\,2}}{\alpha _+^2-\alpha _-^2}} \cosh \left(\frac{\alpha _- \ell \left(\theta -\theta '\right)+\alpha _+ \left(t'-t\right)}{2 \ell}\right)- \ell^2.
\end{align}
Region III: $0<r<r_{-}$,
\begin{align}
\sigma(x,x')&=
-\sqrt{\frac{\alpha _-^2 \ell^2-4 r^2}{\alpha _+^2-\alpha _-^2}} \sqrt{\frac{\alpha _-^2 \ell^2-4 r'^{\,2}}{\alpha _+^2-\alpha _-^2}} \cosh  \left(\frac{\alpha _+ \ell \left(\theta ' -\theta \right)+\alpha _- \left(t-t'\right)}{2 \ell}\right)\nonumber \\&+\sqrt{\frac{\alpha _+^2 \ell^2-4 r^2}{\alpha _+^2-\alpha _-^2}} \sqrt{\frac{\alpha _+^2 \ell^2-4 r'^{\,2}}{\alpha _+^2-\alpha _-^2}} \cosh \left(\frac{\alpha _- \ell \left(\theta -\theta '\right)+\alpha _+ \left(t'-t\right)}{2 \ell}\right)- \ell^2.
\end{align}
\textbf{2. Extremal BTZ black hole} \\

The extremal BTZ BH of mass $M$ is described by the line element
\begin{equation}\label{mext}
ds^2=-dt^2 \left(\frac{r^2}{\ell^2}-2  \alpha^2\right)+\frac{dr^2 \ell^2 r^2}{\left(r^2-\ell^2 \alpha^2\right)^2}-2\gamma \ell \alpha^2 d t d\theta+r^2 d\theta^2,
\end{equation}
where $\alpha\equiv r_+/\ell =\sqrt{M/2}>0$ and the angular momentum is  $J= \gamma M \ell$ with $\gamma=\pm 1$. The coordinate ranges are $-\infty< t < \infty, \;  0<r<\infty, \; 0\leq \theta <2\pi$ (periodic). We note that line element for the extremal black hole is equal to the extremal limit of the line element  for the nonextremal rotating black hole, Eq.\eqref{BTZline}.

The extremal BTZ BH can be embedded in  $\mathbb{R}^{2,2}$ in the following way. 
\noindent  For the region $r>r_+$,
\begin{align}\label{eq:X ext BH,r>rH}
X^0=\frac{\ell \left(\left(\sqrt{A(r)} \left(u-1\right)+\frac{1}{\sqrt{A(r)}}\right) \sinh \alpha v+\left(\sqrt{A(r)} \left(u+1\right)+\frac{1}{\sqrt{A(r)}}\right) \cosh \alpha v\right)}{2 \sqrt{2}}, \nonumber\\
X^1=\frac{\ell \left(\left(\sqrt{A(r)} \left(u+1\right)+\frac{1}{\sqrt{A(r)}}\right) \sinh \alpha v+\left(\sqrt{A(r)} \left(u-1\right)+\frac{1}{\sqrt{A(r)}}\right) \cosh \alpha v\right)}{2 \sqrt{2}},\nonumber\\
X^2=\frac{\ell \left(\left(\frac{1}{\sqrt{A(r)}}-\sqrt{A(r)} \left(u-1\right)\right) \sinh \alpha v+\left(\sqrt{A(r)} \left(u+1\right)-\frac{1}{\sqrt{A(r)}}\right) \cosh \alpha v\right)}{2 \sqrt{2}}, \nonumber\\
X^3=\frac{\ell \left(\left(\frac{1}{\sqrt{A(r)}}-\sqrt{A(r)} \left(u+1\right)\right) \sinh \alpha v+\left(\sqrt{A(r)} \left(u-1\right)-\frac{1}{\sqrt{A(r)}}\right) \cosh \alpha v\right)}{2 \sqrt{2}},
\end{align}
and for the region $r<r_+$ we have
\begin{align}\label{eq:X ext BH,r<rH}
X^0=&-\frac{\ell \left(\left(\frac{1}{\sqrt{-A(r)}}-(u+1) \sqrt{-A(r)}\right) \sinh (\alpha  v)+\left((1-u) \sqrt{-A(r)}+\frac{1}{\sqrt{-A(r)}}\right) \cosh (\alpha  v)\right)}{2 \sqrt{2}}, \nonumber\\
X^1=&-\frac{\ell \left(\left((1-u) \sqrt{-A(r)}+\frac{1}{\sqrt{-A(r)}}\right) \sinh (\alpha  v)+\left(\frac{1}{\sqrt{-A(r)}}-(u+1) \sqrt{-A(r)}\right) \cosh (\alpha  v)\right)}{2 \sqrt{2}},\nonumber\\
X^2=&\frac{\ell \left(\left((u+1) \sqrt{-A(r)}+\frac{1}{\sqrt{-A(r)}}\right) \sinh (\alpha  v)+\left((1-u) \sqrt{-A(r)}-\frac{1}{\sqrt{-A(r)}}\right) \cosh (\alpha  v)\right)}{2 \sqrt{2}}, \nonumber\\
X^3=&\frac{\ell \left(\left(\frac{1}{\sqrt{-A(r)}}-(1-u) \sqrt{-A(r)}\right) \sinh (\alpha  v)+\left(-(u+1) \sqrt{-A(r)}-\frac{1}{\sqrt{-A(r)}}\right) \cosh (\alpha  v)\right)}{2 \sqrt{2}}.
\end{align}
Here,
\begin{equation}
A(r)\equiv \frac{r^2-\ell^2\alpha^2}{\ell^2 \alpha }, \quad u\equiv \theta +\frac{\gamma  t}{\ell},  \quad v\equiv \theta -\frac{\gamma  t}{\ell}.
\end{equation}

The extremal BH  is obtained through identifications generated by the Killing vector 
\begin{equation} \label{KillingNSext}
\Kill=\alpha  (J_{01}+J_{23})+\frac{1}{2} (J_{02}+J_{03}+J_{12}+J_{13}),
\end{equation}
so that the identification matrix $H= e^{2\pi \Kill}$ takes the form 
\begin{equation} \label{HBTZext}
H =\left(
\begin{array}{cccc}
\cosh (2  \pi  \alpha ) & \sinh (2 \pi  \alpha ) &  e^{2  \pi  \alpha } \pi  & - e^{2  \pi  \alpha } \pi  \\
\sinh (2  \pi  \alpha ) & \cosh (2 \pi  \alpha ) & e^{2  \pi  \alpha } \pi  & - e^{2  \pi  \alpha } \pi  \\
e^{-2  \pi  \alpha } \pi  & -e^{-2  \pi  \alpha } \pi  & \cosh (2 \pi  \alpha ) & -\sinh (2  \pi  \alpha ) \\ e^{-2 \pi  \alpha } \pi  & -e^{-2  \pi  \alpha } \pi  & -\sinh (2  \pi  \alpha ) & \cosh (2  \pi  \alpha ) \\
\end{array}
\right).
\end{equation}
The $n$-th power of $H $ is
\begin{equation}\label{eq:Hn ext BH,r>rH}
H^n=\left(
\begin{array}{cccc}
 \cosh (2 n \pi  \alpha ) & \sinh (2 n \pi  \alpha ) & n e^{2 n \pi  \alpha } \pi  & -n e^{2 n \pi  \alpha } \pi  \\
 \sinh (2 n \pi  \alpha ) & \cosh (2 n \pi  \alpha ) & n e^{2 n \pi  \alpha } \pi  & -n e^{2 n \pi  \alpha } \pi  \\
 n e^{-2 n \pi  \alpha } \pi  & -n e^{-2 n \pi  \alpha } \pi  & \cosh (2 n \pi  \alpha ) & -\sinh (2 n \pi  \alpha ) \\
 n e^{-2 n \pi  \alpha } \pi  & -n e^{-2 n \pi  \alpha } \pi  & -\sinh (2 n \pi  \alpha ) & \cosh (2 n \pi  \alpha ) \\
\end{array}
\right).
\end{equation}

In terms of the coordinates $(t,r,\theta)$, the chordal distance $\sigma(x,x')$ (Eq. \eqref{eq:world func R22}) for the extremal BTZ BH is
\begin{align}
\sigma(x,x')&=
\frac{ \ell \sqrt{A(r) A\left(r'\right)}}{2} \left(\ell(\theta-\theta ')+\gamma( t-t')\right) \sinh \left(\frac{\alpha  \left( \ell(\theta-\theta ')+\gamma ( t'-t)\right)}{\ell}\right)\nonumber\\&+\frac{\ell^2 \left(A(r)+A(r')\right) }{2 \sqrt{A(r) A\left(r'\right)}}\cosh \left(\frac{\alpha  \left( \ell(\theta-\theta ')+\gamma ( t'-t)\right)}{\ell}\right)-1,
\end{align}
in the region $r>r_{+}$, and 
\begin{align}
\sigma(x,x')&=
\frac{ \ell \sqrt{A(r) A\left(r'\right)}}{2}  \left(\ell (\theta '-\theta)+\gamma ( t'-t)\right) \sinh \left(\frac{\alpha  \left( \ell(\theta-\theta ')+\gamma ( t'-t)\right)}{\ell}\right)\nonumber\\&-\frac{\ell^2 \left(A(r)+A(r')\right) }{ 2\sqrt{A(r) A\left(r'\right)}}\cosh \left(\frac{\alpha  \left( \ell(\theta-\theta ')+\gamma ( t'-t)\right)}{\ell}\right)-1,
\end{align}
in the region $r<r_{+}$.
\vspace{0.5cm}

\textbf{3. Rotating nonextremal naked singularity} \\

The spinning NS with mass $M<0$ and angular momentum $J$ ($M\neq -1$ if $J=0$) is described by the line element \begin{equation} \label{NSline}
ds^2= -\left(\frac{r^2}{\ell^2}+\frac{ \beta_{+}^2+\beta_{-}^2}{4} \right) dt^2+\frac{dr^2 \ell ^2 r^2}{\left(r^2+\frac{\ell^2  \beta_{+}^2}{4}\right) \left(r^2+\frac{\ell^2  \beta_{-}^2}{4}\right)}-\frac{\ell \beta_{+} \beta_{-}}{2}d t d\theta+r^2 d\theta^2,
\end{equation} 
with $-\infty< t < \infty, \;  0<r<\infty, \; 0\leq \theta <2\pi$ (periodic).
In this case, the embedding is given by
\begin{align} \label{embeddingM-J}
&X^0=A_{+}(r)  \cos \left(\frac{\ell \theta  \beta _-+t \beta _+}{2 \ell}\right), \, & X^{1}=A_{-}(r)\cos \left(\frac{t \beta _-+\ell \theta  \beta _+}{2 \ell}\right), \nonumber \\
&X^{2}=A_{-}(r)\sin \left(\frac{t \beta _-+\ell \theta  \beta _+}{2 \ell}\right), \, & X^3=A_{+}(r) \sin \left(\frac{\ell \theta  \beta _-+t \beta _+}{2 \ell}\right),
\end{align}
with
\begin{equation}
A_{\pm}(r) \equiv \sqrt{\frac{4 r^2+\ell^2 \beta _{\pm}^2}{\beta _+^2-\beta _-^2}}.
\end{equation}

This geometry is obtained through identifications generated by the Killing vector 
\begin{equation} \label{KillingNS}
\Kill=\frac{\beta_+}{2} J_{2 1}+\frac{\beta_-}{2} J_{30}.
\end{equation}
This Killing vector is spacelike, 
\begin{equation}\label{xiNS}
\Kill^2=\frac{\beta_{+}^2}{4} \left[(X_1)^2+ (X_2)^2\right] - \frac{\beta_{-}^2}{4}\left[(X_0)^2+ (X_3)^2\right] = r^2 >0.
\end{equation} 

Exponentiating (\ref{xiNS}) yields the identification matrix
\begin{equation}\label{HNS}
H = 
 \left( \begin{array}{cccc}
\cos \left(\pi  \beta _-\right)   & 0 & 0 &  -\sin \left(\pi  \beta _-\right)  \\
0 &\cos \left(\pi  \beta _+\right) &  -\sin \left(\pi  \beta _+\right)  & 0 \\
0 & \sin \left(\pi  \beta _+\right) & \cos \left(\pi  \beta _+\right)   & 0  \\
\sin \left(\pi  \beta _-\right)  & 0 & 0 & \cos \left(\pi  \beta _-\right) 
 \end{array} \right).
\end{equation}

Using the coordinates $(t,r,\theta)$, the chordal distance $\sigma(x,x')$ (Eq. \eqref{eq:world func R22}) for the nonextremal NS reads
\begin{align}\label{eq:sigma subext NS}
\sigma(x,x')&=
-\sqrt{\frac{\beta _-^2 \ell^2+4 r^2}{\beta _+^2-\beta _-^2}} \sqrt{\frac{\beta _-^2 \ell^2+4 r'^{\,2}}{\beta _+^2-\beta _-^2}} \cos \left(
\frac{\beta _+ \ell \left(\theta -\theta '\right)+\beta _- \left(t-t'\right)}{2 \ell} 
\right)\nonumber\\&+\sqrt{\frac{\beta _+^2 \ell^2+4 r^2}{\beta _+^2-\beta _-^2}} \sqrt{\frac{\beta _+^2 \ell^2+4 r'^{\,2}}{\beta _+^2-\beta _-^2}} \cos \left(\frac{\beta _- \ell \left(\theta -\theta '\right)+\beta _+ \left(t-t'\right)}{2 \ell}\right)-\ell^2.
\end{align}

The embedding (\ref{embeddingM-J}) breaks down for $\beta _{+} = |\beta _{-}|$, which corresponds to the extremal case $M\ell=-|J|$. This means that the Killing vector for the identification that gives rise to the extremal NS cannot be obtained by just taking the limit $\beta_+=|\beta_-|$ in (\ref{xiNS}) and the matrix $H$ needs to be recalculated for this case as well -- we give it in the next subsection. 

Note that $H(\beta_+, \beta_-)$ remains unchanged if either $\beta_+$ or $\beta_-$ are shifted by even integer numbers. In addition, if $n$ and $m$ are two integers such that $n(\beta_+ - \beta_-) =2m$, then 
\begin{equation} \label{resonance}
H^n(\beta_+, \beta_-)=H(n\beta_+, n\beta_-)=H(n\beta_-+2m, n\beta_-)= H(n\beta_-, n\beta_-) \,,
\end{equation}
which is the form of the naive extremal limit. As we see in Sec.\ref{sec:RSET}, this feature leads to a singularity in the RSET and to a breakdown of the perturbative regime for the system.

\vspace{0.5cm}
\textbf{4. Extremal naked singularity}
\vspace{0.5cm}

Although the line-element of  the extremal NS  coincides with the  line-element in Eq.\eqref{NSline} when taking the 
extremal limit $M\ell=-|J|$, the extremal NS space-time
cannot be obtained by taking the limit $\beta_+=|\beta_-|= 2\beta$ in the embedding (\ref{embeddingM-J}), Killing vector (\ref{xiNS}) or the identification matrix (\ref{HNS}). In fact, the extremal metric
\begin{equation} \label{Extreme-g}
 ds^2 = - \left(\frac{r^2}{\ell^2}+2  \beta^2\right)dt^2+\frac{ \ell^2 r^2 dr^2}{\left(r^2+\ell^2 \beta^2\right)^2}-2\gamma \ell \beta^2 d t d\theta+r^2 d\theta^2, 
 \end{equation} 
  with $-\infty< t <\infty, \;  0<r<\infty, \; 0\leq \theta <2\pi$ (periodic),
is obtained via the embedding
\begin{align} \label{embeddingM=J}
X^0&=\frac{\ell \left(\left(\frac{1}{B(r)}-(v-1) B(r)\right) \sin (\beta  u)+\left(-(v+1) B(r)-\frac{1}{B(r)}\right) \cos (\beta  u)\right)}{2 \sqrt{2}}, \nonumber\\
X^1&=\frac{\ell \left(\left(-(v-1) B(r)-\frac{1}{B(r)}\right) \sin (\beta  u)+\left((v+1) B(r)-\frac{1}{B(r)}\right) \cos (\beta  u)\right)}{2 \sqrt{2}},\nonumber\\
X^2&=\frac{\ell \left(\left((v+1) B(r)-\frac{1}{B(r)}\right) \sin (\beta  u)+\left((v-1) B(r)+\frac{1}{B(r)}\right) \cos (\beta  u)\right)}{2 \sqrt{2}}, \nonumber\\
X^3&=\frac{\ell \left(\left(-(v+1) B(r)-\frac{1}{B(r)}\right) \sin (\beta  u)+\left((v-1) B(r)-\frac{1}{B(r)}\right) \cos (\beta  u)\right)}{2 \sqrt{2}}.
\end{align}
 Here $\beta\equiv \sqrt{-M/2}>0$ and the angular momentum is  $J= -\gamma M \ell$ with $\gamma=\pm 1$, where
\begin{equation}
B(r)\equiv \sqrt{\frac{r^2+\ell^2\beta^2}{\ell^2 \beta }}, \quad u\equiv \theta +\frac{\gamma  t}{\ell},  \quad v\equiv \theta -\frac{\gamma  t}{\ell}.
\end{equation}

This extremal NS  is obtained through identifications generated by the Killing vector 
\begin{equation} \label{KillingNSext}
\Kill=\beta  (J_{03}-J_{12})-\frac{1}{2} (J_{01}+J_{03}+J_{12}-J_{23}).
\end{equation}
The  identification matrix in this case is given by
\begin{align}\label{HNSext}
&H = e^{2\pi \Kill}= \nonumber \\
& \left(\begin{array}{cccc}
 \cos(2\pi \beta) + \pi \sin(2\pi \beta) & -\pi \cos(2\pi \beta) &
  \pi \sin(2\pi \beta) & \pi \cos(2\pi \beta) -\sin(2\pi \beta)  \\
 -\pi \cos(2\pi \beta) & \cos(2\pi \beta) -\pi \sin(2\pi \beta ) &
 -\pi \cos(2\pi \beta) -\sin(2\pi \beta) & \pi \sin(2\pi \beta ) \\
 -\pi \sin(2\pi \beta) & \pi \cos(2\pi \beta) +\sin(2\pi \beta ) &
 \cos(2\pi \beta)-\pi  \sin (2 \pi  \beta) & -\pi  \cos (2\pi \beta) \\
 \sin (2 \pi  \beta )-\pi \cos(2\pi \beta) & -\pi \sin (2\pi \beta) &
 -\pi \cos(2\pi \beta) & \cos(2\pi \beta) + \pi \sin(2\pi \beta ) 
\end{array}\right),
\end{align}
and the $n$-th power of $H$ is obtained replacing $\pi$ by $n\pi$ in the above expression.

For the extremal NS, the chordal distance $\sigma(x,x')$ (Eq. \eqref{eq:world func R22}) is given by 
\begin{align}
\sigma(x,x')&=
\frac{ \ell B(r) B\left(r'\right)}{2}  \left( \ell(\theta-\theta ') +\gamma ( t'-t)\right) \sin \left(\frac{\beta  \left( \ell(\theta-\theta ')+\gamma  (t-t')\right)}{\ell}\right)\nonumber\\&+\frac{\ell^2 \left(B(r)^{2}+B(r')^2\right) }{2 B(r) B\left(r'\right)}\cos \left(\frac{\beta  \left( \ell(\theta-\theta ')+\gamma  (t-t')\right)}{\ell}\right)-1.
\end{align}

\section{Two-point function in CAdS$_3$}\label{sec:G_A}  

In this appendix we derive the anti-commutator in CAdS$_3$, Eq.\eqref{eq:G^1A as sigma_M}.
Let us  consider the line element in the covering space of AdS$_3$
in coordinates
$\rho\in [0,\pi/2]$, $\theta\in (0,2\pi]$, $\tau \in \mathbb{R}$~\cite{Lifschytz:1993eb}:
\begin{align}\label{eq:AdS metric}
& ds^2 = 
 \ell^2\sec^2\rho \left(-d\tau^2+d\rho^2+\sin^2\rho\ d\theta^2\right).
\end{align}

The transformation between these coordinates in AdS$_3$ and those in
Eq.\eqref{eq:R22} in
$\mathbb{R}^{(2,2)}$ is~\cite{Lifschytz:1993eb}:
\begin{equation}
X^0=\ell\frac{\cos\tau}{\cos\rho},\quad
X^1=\ell\tan\rho\cos\theta,\quad
X^2=\ell\tan\rho\sin\theta,\quad
X^3=\ell\frac{\sin\tau}{\cos\rho}.
\end{equation}

This transformation allows us to write the function $\s$ in Eq.\eqref{eq:world func R22}
in the AdS$_3$ coordinates of Eq.\eqref{eq:AdS metric} as
\begin{equation}\label{eq:sigma_M-ap}
\s
(x,x')
=\ell^2\left(\cos(\Delta\tau)\sec\rho\sec\rho'-1-\tan\rho\tan\rho'\cos\Delta\theta\right).
\end{equation}

The metric Eq.\eqref{eq:AdS metric} is manifestly conformal to half of the Einstein Universe $\mathbb{R}\times S^2$ with a conformal factor $\Omega(x)=\ell/\cos\rho$, 
and, therefore (see, e.g., Eq.3.154~\cite{Birrell:Davies}), $G^+_A(x,x')=\sqrt{\cos\rho\cos\rho'}G^+_E(x,x')/\ell$, where $G^+_A$ is the Wightman function 
in AdS$_3$ and  $G^+_E$ is the Wightman function in the Einstein Universe. 
By using this fact and explicitly calculating $G^+_E$, App.A~\cite{Lifschytz:1993eb} finds
\begin{equation}\label{eq:G^+A}
G^+_A(x,x')=\lim_{\epsilon\to 0^+}\frac{1}{4\sqrt{2}\pi\ell}\left(\cos(\Delta\tau-i\epsilon)\sec\rho\sec\rho'-1-\tan\rho\tan\rho'\cos\Delta\theta\right)^{-1/2},
\end{equation}
where $\Delta\tau\equiv \tau-\tau'$ and $\Delta\theta\equiv \theta-\theta'$,
for a quantum state which corresponds to imposing transparent b.c.; the `$i\epsilon$' corresponds to the Feynman prescription.
Now, by using
\begin{align}
& \cos(\Delta\tau-i\epsilon) \approx \cos(\Delta\tau)+i\sin\left(\Delta\tau\right) \epsilon, \quad \epsilon \rightarrow 0^+,
\\&
\lim_{\epsilon \rightarrow 0^+}\left(x\pm i\epsilon\right)^{-1/2} =
|x|^{-1/2}e^{\mp i\pi\Theta(-x)/2},
\quad x\in \mathbb{R},
\end{align}
together with Eqs.\eqref{eq:elem func} and \eqref{eq:sigma_M-ap},
it readily follows that the anti-commutator corresponding to Eq.\eqref{eq:G^+A} is
given by Eq.\eqref{eq:G^1A as sigma_M}.

\section{Two-point function in static BTZ naked singularity via mode sums}\label{sec:GF in BTZ modesum}

In Sec.\ref{sec:two-pt func}, we gave the two-point function on a static BTZ NS space-time as derived by applying the method of images
on the two-point function in AdS$_3$.
Specifically, the two-point function on a static BTZ NS
is given by Eq.\eqref{Green-NS2} with $N=1/\sqrt{-M}$.
In this appendix, we are going to rederive that expression by instead using  mode sums over homogeneous solutions of the field equation, Eq.\eqref{wave-eq}.
This alternative derivation will enable us to: (i) clarify the boundary conditions used in obtaining the two-point function; 
and (ii) relate the upper summation index (i.e., $N-1$) in the sum over the two-point functions in AdS$_3$ to null geodesics.
Point (ii) is useful, not only for corroborating the result in Eq.\eqref{Green-NS2} in the static case,
but also for acquiring a direct geometrical understanding of its sum. This understanding could be used as a guide for finding what the upper summation index should be for a rotating NS space-time,
in a more general case than that done in Sec.\ref{sec:two-pt func}.

We start with the homogeneous field Eq.\eqref{wave-eq}
and write a field mode solution as 
\bea
\phi_{\ind}(x)=N_{\ind}e^{-i\omega t+im\theta}R_{\ind}(r),
\eea
where $N_{\ind}$ is a normalization constant,
$\omega\in\mathbb{C}$ and $m\in\mathbb{Z}$.
The radial function $R_{\ind}(r)$ is found to satisfy the ordinary differential equation
\bea\label{eq:radial ODE static NS}
\left(\frac{1}{r}\frac{d}{dr}\left(r(r^2-M)\frac{d}{dr}\right)
-\frac{m^2}{r^2}+\frac{\omega^2}{r^2-M}+\frac{3}{4}\right)R_{\ind}(r)=0,
\eea
where $r\in (0,\infty)$.
We need to choose boundary conditions for the solutions
of this radial equation at the singularity ``$r=0$" and at the AdS boundary $r=\infty$.

Let us now define normalize quantities as:
$\rb\equiv r/\sqrt{-M}\in (0,\infty)$, $\ob\equiv \omega/\sqrt{-M}$ and $\mb\equiv m/\sqrt{-M}$, where we remind the reader that $M<0$ for a static NS.
From now on we restrict ourselves to the case that $M=-1/N^2$, with $N\in\mathbb{Z}^+$, so that, in particular,
$\mb=m\cdot N\in\mathbb{Z}$.

The solutions of Eq.\eqref{eq:radial ODE static NS} can be expressed in terms of associated Legendre functions.
In particular, we choose the following two linearly independent solutions:
\bea\label{eq:slns radial ODE static NS}
{}_1R_{\ind}(\rb)\equiv \left(1+\rb^2\right)^{-1/4}
P^{-\mb}_{-1/2+\ob}\left(\frac{1}{\sqrt{1+\rb^2}}\right),
\quad
{}_2R_{\ind}(\rb)\equiv \left(1+\rb^2\right)^{-1/4}
P^{\mb}_{-1/2+\ob}\left(-\frac{1}{\sqrt{1+\rb^2}}\right).
\eea

It can be easily checked that the functions in Eq.\eqref{eq:slns radial ODE static NS} satisfy Eq.\eqref{eq:radial ODE static NS}.
In what follows, it will be convenient to use the 
coordinate $\rho\in (0,\pi/2)$ defined via $\cos\rho \equiv \left(1+\rb^2\right)^{-1/2}$.
In terms of this new coordinate, the solutions
read
\bea\label{eq:slns radial ODE static NS rho}
{}_1R_{\ind}(\rho)= \sqrt{\cos\rho}\,
P^{-\mb}_{-1/2+\ob}\left(\cos\rho\right),
\quad
{}_2R_{\ind}(\rho)= \sqrt{\cos\rho}\,
P^{\mb}_{-1/2+\ob}\left(-\cos\rho\right).
\eea
Let us now turn to their boundary conditions.
They behave as 
${}_1R_{\ind}=O(\rb^{|\mb|})$ (for which 
$\mb\in\mathbb{Z}$ is needed) and
${}_2R_{\ind}=O(\rb^{-|\mb|})$ as $\rb\to 0^+$~\cite{Olver:1974}.
That is, ${}_1R_{\ind}$ is regular as $r\to 0^+$ and
square integrable near $r=0$; on the other hand,
${}_2R_{\ind}$ is irregular as $r\to 0^+$ 
and is not square integrable near $r=0$ (except for $m=0$).
Near the AdS boundary, ${}_2R_{\ind}$ obeys transparent boundary conditions~\cite{Avis1978quantum,Shiraishi:1993ti}.
Therefore, ${}_1R_{\ind}$ is the appropriate solution near the singularity $r=0$ and ${}_2R_{\ind}$ is the appropriate one near the AdS
boundary $r=\infty$.

We now proceed to write the two-point function similarly to the way that it is done 
in~\cite{shiraishi1994vacuum,Shiraishi:1993ti} for the BH case.
The idea is that one first Euclideanizes the space-time.
The field equation becomes elliptic in the Euclidean manifold and so there is a unique Green function (under the conditions of square-integrability near the origin and transparent boundary conditions at infinity), which is the so-called Euclidean Green function.
The Euclidean Green function may be constructed in the usual way that one constructs a Green function: with the radial part of the modes given by   the radial solution which satisfies the desired boundary condition near $r=0$ evaluated at the point with the smallest radius (i.e., $r_<\equiv \min(r,r')$),
times the
radial solution which satisfies the desired boundary condition near $r=\infty$ evaluated at the point with the largest radius (i.e., $r_>\equiv \max(r,r')$).
The Euclidean Green function is obtained as a frequency-integral of the modes constructed as per above (it is a frequency instead of a discrete sum since, in this case,
the corresponding Euclidean manifold contains no conical singularity and so no periodicity is required in the Euclideanized time).
One then de-Euclideanizes and obtains the Feynman Green function from the Euclidean Green function by the corresponding analytic continuation.
When de-Euclideanizing, the integration contour over the purely-imaginary frequencies in  the Euclidean Green function is deformed to an integral over just below the real axis for $Re(\omega)<0$ and just above the real axis for $Re(\omega)>0$; we denote such contour by $\mathcal{C}$ (see, e.g., Fig.1 in~\cite{DeWitt:1965}).
Specifically, the Feynman Green function in the static NS space-time when the field satisfies transparent boundary conditions is given by
\bea\label{eq:FGF}
G_F(x,x')=
-\frac{N^2}{(2\pi)^2}
\int_{\mathcal{C}}
d\omega\,  \sum_{m=-\infty}^{\infty}
e^{-i\omega t+im\theta}
\frac{{}_1R_{\ind}(\rho_<){}_2R_{\ind}(\rho_>)}{\tan\rho\cdot W\left[{}_1R_{\ind},{}_2R_{\ind}\right]}.
\eea
Here we have taken $t'=0$ and $\theta'=0$ without loss of generality (due to the stationarity and circular symmetry of the space-time).
The factor ``$\tan\rho$" is required so that the denominator is constant.
The Wronskian is given by
\bea\label{eq:Wronskian}
W\left[{}_1R_{\ind},{}_2R_{\ind}\right]\equiv
{}_1R_{\ind}\frac{d{}_2R_{\ind}}{d\rho}-
{}_2R_{\ind}\frac{d{}_1R_{\ind}}{d\rho}
=-\frac{2}{\pi}\cot\rho \cos\left(\pi\left(\mb-\ob\right)\right).
\eea
The Feynman propagator \eqref{eq:FGF} satisfies the Green function equation \eqref{eq:GF eq}.

From Eqs.\eqref{eq:FGF} and \eqref{eq:Wronskian} it readily follows that
\bea\label{eq:FGF simp}
G_F(x,x')=\frac{N}{8\pi}\left(\cos\rho\cos\rho'\right)^{1/2}
\int_{\mathcal{C}}
d\ob\,  \sum_{m=-\infty}^{\infty} (-1)^{\mb}
e^{-i\ob \bar t+im\theta}
\frac{P_{-1/2+\ob}^{-\mb}(\cos\rho_<)P_{-1/2+\ob}^{\mb}(-\cos\rho_>)}{\cos(\pi\ob)},
\eea
where we have defined $\bar t\equiv t \sqrt{-M}$.
We note that the only singularities in the complex-$\omega$ plane of the integrand (for $r,r'\in (0,\infty)$) in Eq.\eqref{eq:FGF simp}
are the poles  which correspond to the zeros of the denominator, i.e., $\ob=n+1/2$ with $n\in \mathbb{Z}$.
We next wish to perform the infinite-sum and integral in Eq.\eqref{eq:FGF simp}.

For the sum, we will mirrror a similar calculation in the Appendix of~\cite{Davies:1987th}.
We start with Eq.8.794 of~\cite{GradRyz}.
After basic operations and the use of the property
\bea
P_{\nu}^{k}\left(\cos\psi_1\right)P_{\nu}^{-k}\left(\cos\psi_2\right)=
P_{\nu}^{-k}\left(\cos\psi_1\right)P_{\nu}^{k}\left(\cos\psi_2\right)
\eea
for $k\in\mathbb{Z}$, $\psi_1,\psi_2\in\mathbb{R}$,
$\nu\in\mathbb{C}$,
we obtain
\bea\label{eq:prop Leg}
\sum_{k=-\infty}^{\infty}(-1)^k P_{\nu}^{-k}\left(\cos\psi_1\right)P_{\nu}^{k}\left(\cos\psi_2\right)e^{ik\varphi}=
P_\nu\left(\cos\psi_1\cos\psi_2+\sin\psi_1\sin\psi_2\cos\varphi\right),
\eea
with $\varphi\in\mathbb{R}$.
We now take $\varphi\to \varphi+2n\pi/N$ onto
Eq.\eqref{eq:prop Leg} and take a sum over $n$ from $0$ to $N-1$.

After some more basic operations and the use of the distributional identity~\cite{Davies:1987th}
\bea \label{eq:distributional id}
\sum_{n=0}^{N-1}e^{ik\left(\varphi+2n\pi/N\right)}=
Ne^{ik\varphi}\sum_{m=-\infty}^{\infty}\delta_{k,m\cdot N},
\eea
we obtain the useful identity
\bea\label{eq:sum Leg}
\sum_{m=-\infty}^{\infty}(-1)^{m\cdot N} P_{\nu}^{-m\cdot N}\left(\cos\psi_1\right)P_{\nu}^{m\cdot N}\left(\cos\psi_2\right)e^{imN\varphi}=
\frac{1}{N}\sum_{n=0}^{N-1}
P_\nu\left(\cos\psi_1\cos\psi_2+\sin\psi_1\sin\psi_2\cos\left(\varphi+\frac{2n\pi}{N}\right)\right).
\eea
We can now apply Eq.\eqref{eq:sum Leg} to the infinite-sum in Eq.\eqref{eq:FGF simp}:
\bea\label{eq:sum Leg final}
\sum_{m=-\infty}^{\infty} (-1)^{\mb}
e^{im\theta}P_{-1/2+\ob}^{-\mb}(\cos\rho_<)P_{-1/2+\ob}^{\mb}(-\cos\rho_>)=
\frac{1}{N}\sum_{k=0}^{N-1}
P_{-1/2+\ob}\left(\cos\beta_k\right),
\eea
where
\bea
\cos\beta_k\equiv
-\cos\rho\cos\rho'-\sin\rho\sin\rho'\cos\left(\frac{\theta+2k\pi}{N}\right) \ .
\eea
Therefore, we have, from Eqs.\eqref{eq:FGF simp} and \eqref{eq:sum Leg final},
\bea\label{eq:FGF after sum}
G_F(x,x')=\frac{1}{8\pi}\left(\cos\rho\cos\rho'\right)^{1/2}
\sum_{k=0}^{N-1}
\int_{\mathcal{C}}d\ob\,  
e^{-i\ob \bar t}
\frac{P_{-1/2+\ob}\left(\cos\beta_k\right)}{\cos(\pi\ob)}.
\eea
In order to evaluate this contour integral we shall use the residue theorem.
For this purpose, we choose to close the contour $\mathcal C$ in the lower $\omega$-plane. When $t>0$, the integral along the arc at infinite radius in the lower plane vanishes and
so from now on we consider $t>0$.
Then, taking into account the poles of the integrand for 
$\ob>0$ 
(i.e., $\ob=n+1/2$ with $n\in \mathbb{Z}^+\cup {0}$)
when using the residue theorem,
we obtain
\bea\label{eq:FGF after sum&int}
G_F(x,x')=\frac{i}{4\pi}\left(\cos\rho\cos\rho'\right)^{1/2}
\lim_{\epsilon\to 0^+}
\sum_{k=0}^{N-1}
\sum_{n=0}^{\infty}
e^{-i(n+1/2) (\bar t-i\epsilon)}(-1)^{n}P_{n}\left(\cos\beta_k\right),
\eea
where we have introduced a small-$\epsilon$ prescription for convergence.
In order to carry out the $n$-sum, we use the fact that 
$(-1)^{n}P_{n}\left(\cos\beta_k\right)=P_{n}\left(-\cos\beta_k\right)$ for $n\in \mathbb{Z}^+\cup {0}$,
together with Eq.8.921~\cite{GradRyz}, which requires 
that $\left| e^{-i(\bar t-i\epsilon)} \right|=e^{-\epsilon}<1$, which is satisfied for $\epsilon>0$.
As a result, we obtain
\bea
G_F(x,x')=\frac{i}{4\sqrt{2}\pi}\left(\cos\rho\cos\rho'\right)^{1/2}
\lim_{\epsilon\to 0^+}
\sum_{k=0}^{N-1}
\frac{1}{\left(\cos(\bar t-i\epsilon)+\cos\beta_k\right)^{1/2}}.
\eea
By writing it in the original BTZ coordinates,
it is easy to check that (restoring $\ell$)
\bea\label{eq:summand GF}
\frac{\cos(\bar t-i\epsilon)+\cos\beta_k}{\cos\rho\cos\rho'}=
\left(\left(N^2\rt^2+\ell^2\right)\left(N^2\rt'^2+\ell^2\right)\right)^{1/2}\cos\left(\frac{\tit-i\epsilon}{N\ell}\right)-N^2\rt\rt'\cos\left(\frac{\tht+2k\pi}{N}\right)-\ell^2,
\eea

Comparing with Eq.\eqref{eq:sigma subext NS}, we can see that, 
for $k=0$ and $\epsilon=0$,
the right hand side of \eqref{eq:summand GF} is equal to the world function
$\s(x,x')$ with $M=-1/N^2$.
The expression for $k\neq 0$ (and $\epsilon=0$) simply corresponds to
$\s(x,H^kx')$.

In its turn, the $\epsilon$-dependence can be separated out so that:
\bea\
\frac{\cos(\bar t-i\epsilon)+\cos\beta_k}{\cos\rho\cos\rho'} \sim
\sigma_{\epsilon}(x,H^kx')\equiv
\s(x,H^kx')+\sin \bar t\cdot i\epsilon,
\quad \epsilon\to 0^+.
\eea
The final expression for the Feynman Green function is thus
\bea\label{eq:GF NS static sigma}
G_F(x,x')=\frac{i}{4\sqrt{2}\pi}
\lim_{\epsilon\to 0^+}
\sum_{n=0}^{N-1}
\frac{1}{\sqrt{\sigma_{\epsilon}(x,H^nx')}}.
\eea
In its turn, the final expression for the  anti-commutator is thus, from Eqs.\eqref{eq:elem func} and
\eqref{eq:GF NS static sigma},
\bea\label{eq:G1 NS static sigma}
G^{(1)}_{NS}(x,x')=
2\, \text{Im}(G_F(x,x'))=
\frac{1}{2\sqrt{2}\pi}
\sum_{n=0}^{N-1}
\frac{\Theta(\s(x,H^nx'))}{\sqrt{
\s(x,H^nx')
}
}.
\eea
Even though this expression has been derived assuming $t>0$, since the expression for the anticommutator $G^{(1)}(x,x')$ is the same for $t>0$ as for $t<0$\MC{please check this statement}, this expression is actually valid for all $t\in\mathbb{R}$.
Eqs.\eqref{eq:G1 NS static sigma}  and Eq.\eqref{Green-NS2} agree while they have been derived in completely different ways: as a mode-sum here whereas using the method of images there.
We note that the ``sum over caustics"
(or, in other words, the generalization $\theta\to \theta+2n\pi/N$ with the associated sum over $n$ or,
in other words, the ``sum over images" within the method of images) has arisen naturally here from the 
distributional identity Eq.\eqref{eq:distributional id}.




\end{document}